\documentclass[12pt]{article}
\pdfoutput=1
\usepackage{amsmath}
\usepackage{amssymb}
\usepackage{graphicx,bbm,mathrsfs}
\usepackage{color}

\newcommand{\lsim}{\raisebox{-0.13cm}{~\shortstack{$<$ \\[-0.07cm] $\sim$}}~} 
\newcommand{\gsim}{\raisebox{-0.13cm}{~\shortstack{$>$ \\[-0.07cm] $\sim$}}~} 
\newcommand{\beq}{\begin{eqnarray}} 
\newcommand{\eeq}{\end{eqnarray}} 
\newcommand{\tb}{\tan\beta} 
 
\usepackage{listings}
\usepackage{caption}
\usepackage{cite}

\newcommand{\bea}{\begin{align}}
\newcommand{\eea}{\end{align}}
\newcommand{\nbea}{\begin{align*}}
\newcommand{\neea}{\end{align*}}
\newcommand{\nbeq}{\begin{equation*}}
\newcommand{\neeq}{\end{equation*}}
\newcommand{\bear}{\begin{eqnarray}}  
\newcommand{\eear}{\end{eqnarray}}

 \newcommand{\comment}[1]{}
\usepackage{array}
\newcolumntype{M}[1]{>{\centering\arraybackslash}m{#1}}
\newcolumntype{N}{@{}m{0pt}@{}}

\numberwithin{equation}{section}

\begin{document}

\begin{flushright}\small{CERN-TH/2016-095, KCL-PH-TH/2016-20, LPT-Orsay--16--39~~~~~~~~~} \end{flushright} 

\vspace*{5mm} 

\begin{center}
 
\mbox{\large\bf Interference Effects in the Decays of Spin-Zero Resonances} \\
\vspace{2mm}
\mbox{\large\bf into $\mathbf{\gamma \gamma}$
and $\mathbf{t\bar t}$}

\vspace*{8mm}

{\sc Abdelhak~Djouadi$^{1,2}$, John Ellis$^{2,3}$} and {\sc J\'er\'emie Quevillon$^3$ } 

\vspace{6mm}

{\small 
$^1$ Laboratoire de Physique Th\'eorique,  CNRS and Universit\'e Paris-Sud, \\  
B\^at. 210, F--91405 Orsay Cedex, France \\
\vspace{2mm}
$^2$ Theoretical Physics Department, CERN, CH 1211 Geneva 23, Switzerland \\
\vspace{2mm}
$^3$ Theoretical Particle Physics \& Cosmology Group, Department of Physics, \\ 
King's College,  Strand, London WC2R 2LS, United Kingdom
}
\end{center}

\vspace*{8mm}

\begin{abstract}

We consider interference effects in the production via gluon fusion in LHC collisions at 13~TeV and decays into $\gamma \gamma$ and $t {\bar t}$ final states of one or two putative new resonant states $\Phi$, assumed here to be scalar and/or pseudoscalar particles. 
Although our approach is general, we 
use for our numerical analysis the example of the putative $750$ GeV 
state for which a slight excess was observed in the initial 
LHC $13$ TeV data. We revisit previous calculations of the interferences between the heavy-fermion loop-induced  $gg \to \Phi \to \gamma \gamma$ signal and the continuum $gg\to \gamma\gamma$ QCD background,  which can alter the production rate as well as modify the line-shape and apparent mass. We find a modest enhancement by $\sim 20$\% under favorable circumstances, for a large $\Phi$ width. The effect of interference on the apparent scalar-pseudoscalar mass difference in a two-Higgs-doublet model is found to be also modest. An exploratory study indicates that similar effects are to be expected in the $gg \to \Phi \to Z \gamma$ channel. In this and other models with a large $\Phi$ total width, the dominant $\Phi$ decays are expected to be into $t \bar t$ final states. We therefore also study the effects of interference of the $gg\to \Phi \to t\bar t$ signal with the $gg\to t \bar t$ continuum QCD background and show that in the presence of standard fermions only in the $gg\to \Phi$ loops, it is  {\it destructive} causing a {\it dip} in the $t \bar t$ mass distribution. Including additional vector--like quarks leads to a different picture as peaks followed by dips can then occur.  We use the absence of such effects in ATLAS and CMS data to constrain models of the production and decays of the $\Phi$ state(s).

\end{abstract}
\vspace*{1cm}
\begin{flushleft}\small{May 2016} \end{flushleft} 

\newpage

\section{Introduction}

The reports in December 2015 by the CMS~\cite{CMS-diphoton} and ATLAS~\cite{ATLAS-diphoton} Collaborations of possible enhancements in their initial 13-TeV data in the $\gamma \gamma$ invariant-mass spectra near 750~GeV, which might be the first indications of one or more possible new heavy particles $\Phi$, have triggered a frenzy of model-building and theoretical interpretations \cite{full-list}. These studies/speculations have not been discouraged by the updated analyses released by ATLAS~\cite{ATLASMoriond} and CMS~\cite{CMSMoriond} at the Moriond meeting in March 2016, which confirmed the previous enhancements, and included 8-TeV data from both experiments and CMS data taken with the magnet off. The (non-)existence of the $\Phi$ state(s) will presumably be settled by data to be collected at the LHC during 2016 (weasels permitting). 

What information might these data provide, beyond the confirmation of $\gamma \gamma$ invariant-mass peak(s) and clarification of its/their width(s)? Many authors have highlighted the importance of searches for other diboson $\Phi$ decay modes such as $Z\gamma, ZZ$ and $WW$, which already impose relevant constraints on some models~\cite{EEQSY}. If the total $\Phi$ decay width is much larger than the minimal width given by anomalous triangle diagrams, the bulk of its decays may be into $t \bar t$ final states, which are dominant in two--Higgs doublet models, see for example~\cite{DEGQ}.  These decays, which have received scant attention (but see also~\cite{ADM,othertt}), are also potentially observable. 

The $\gamma \gamma$ and $t {\bar t}$ final states both have significant continuum backgrounds, which present opportunities as well as problems. As we discuss in this paper, interference effects on the $\Phi$ line-shape may be able to provide information on both the real and imaginary parts of the $gg \to \Phi \to \gamma \gamma$ and $gg \to \Phi \to t {\bar t}$ amplitudes, providing supplementary constraints on the properties of one or two new state(s), exemplified by the recent 750~GeV excess. There is an extensive literature on interference effects on the corresponding signals of the standard-like 125 GeV Higgs boson, $h$, in the $\gamma \gamma$ and $h\to ZZ^*$ final states, which may generate an observable difference between the apparent masses measured in these final states \cite{int-gamma,Martin} and/or provide loose constraints on the total $h$ width \cite{hgamgam}. There have also been pioneering studies of possible interference effects in the decays of a heavy Higgs boson into $t\bar t$ final states,  in both the standard~\cite{Htt0} and  two--Higgs doublet \cite{Htt1} models. 

In the context of the $\Phi (750)$, an analysis of interference effects between the $gg \! \to \! \Phi \! \to \! \gamma \gamma$ signal and the $gg\to \gamma\gamma$ QCD background has been performed in~\cite{KIAS}, and significant effects have been shown to occur~\footnote{See also the recent analysis~\cite{Trieste} of the spin--2 case.}. As it is natural to consider the ``observed" $\gamma \gamma$ final state before going on to consider possible effects in other channels, we use the analysis of Ref.~\cite{KIAS} as a starting-point and extend it to various scenarios for the $\Phi$ state(s), including a broad or narrow single scalar or pseudoscalar resonance and a possible near-degenerate pair of CP--even $H$ and CP--odd $A$ states as can appear in two--Higgs doublet models~\cite{DEGQ}. 

We assess how large the interference effects could be, depending on the number and masses of the particles in the quantum loops generating the $gg \to \Phi$ and $\Phi \to \gamma\gamma$ amplitudes. We find that interference effects in the imaginary part of the amplitude could enhance the resonance peak only slightly, 
whereas interference effects in the real part (which changes sign at the nominal position of a particle pole) would shift the maxima of the signal cross sections by amounts of $\lesssim {\cal O}(\Gamma_\Phi)$ - which is large for a broad resonance, $\Gamma_\Phi \approx 45$ GeV - rendering the interpretation of the mass peak more complicated. This is especially the case if two $H$ and $A$ states are involved and are almost degenerate in mass, as is the case in supersymmetric models, for instance~\cite{DEGQ}. 

These analyses may be extended to other possible bosonic final states of the $\Phi$ resonance, namely the decays $\Phi \to \gamma Z, ZZ$ and $W^+W^-$. If the $\Phi ZZ$ and $\Phi WW$ couplings are also generated by loops of heavy fermions only (which might not be entirely the case for the scalar $H$ state in two-Higgs-doublet models, for instance),  the situation is qualitatively similar to that of the two--photon and photon--$Z$ decays, with an interference of the signal $gg \to \Phi \to VV$ amplitude with that of the $gg\to VV$ QCD background (but where the longitudinal components of the  vector bosons have to be taken into account). Significant numerical differences should occur because of the different couplings of the $\gamma, Z,W$ bosons to fermions. For the same reason, these diboson final states could provide additional information on the properties of the $\Phi$ resonance and on the additional matter particles that are involved in the quantum loops that generate the $\Phi VV$ couplings. We give one example of possible effects in the $Z \gamma$ final state, leaving a detailed study of the effects in the other channels to future work~\cite{DEQ2}. 

Instead, we focus in the rest of this paper on interference effects between the $gg \!  \to \! \Phi \! \to \!  t \bar t$ signal and the QCD process $gg\to t\bar t$ that generates the major part of the $t\bar t$ background at LHC energies. If the $gg\to \Phi$ cross section is generated by the top quark loops only,  we find the interference to be {\it destructive} with the net effect of a {\it dip} in the measured $t {\bar t}$ cross section beyond the nominal position of the resonance peak. In contrast, if additional heavy quarks contribute to the production amplitude, the interference can become destructive before and constructive after the mass peaks. The magnitudes of these dips and peaks depend on the masses and couplings of the particles mediating the production and decay mechanisms. 

The ATLAS and CMS collaborations have published analyses of $t {\bar t}$ production at the LHC at 8~TeV or 13 TeV \cite{ATLAS-tt,CMS-tt} which give no indication of any structure around 750~GeV, setting limits on any upward or downward deviations of the cross sections from the background that can be used to constrain the properties of possible mediating particles. Since $\Phi \to  t \bar t$ decay is the dominant mode in many scenarios, including that in which the $\Phi$ state is a superposition of the broad $H$ and $A$ states, future LHC data could allow any new state to be observed in this channel, and these interference effects should be included in order to interpret correctly any signal, or its absence. 

The structure of this paper is as follows: in the next Section, we describe briefly the
two benchmark scenarios that we will use for the $\Phi$ resonance, first a singlet $\Phi$ scenario, 
in which it may be narrow or wide, scalar or pseudoscalar, and then a two-Higgs-doublet model 
in which $\Phi$ is a combination of the heavier CP--even scalar state $H$ and the CP--odd 
pseudoscalar state $A$.  In Section 3, we consider interference effects in the $gg\to \gamma \gamma$ process, in both the imaginary part that modifies the signal cross section and the real part that shifts the position of the peak. We also comment on the $\gamma Z$ final state in which the situation is qualitatively similar. Section 4 is devoted to interference in the $gg\to \Phi \to t \bar t$ process with the leading order $gg\to t\bar t$ QCD background amplitudes. In all cases~\footnote{Other additional interesting final states for the $\Phi$ particles would be $\Phi \to \tau^+ \tau^-$ and $\Phi \to gg, b \bar b$. The main background for the former process comes from a source that is not gluon fusion, so there is no signal--background interference. In the later two cases,  the interferences  with the huge two gluon--jet or two $b$--jet backgrounds are rather involved and their treatment is beyond our scope here.},  the impact of the interference and its importance are discussed in various illustrative cases, for singlet and doublet scalar and pseudoscalar resonances that may be narrow or broad. Section 5 summarises our conclusions. 

\section{Benchmarks for the $\Phi (750)$ State(s)}

In this section, we describe two benchmark scenarios that we will use to illustrate our results. The first is a minimal scenario in which the $\Phi$ state is an single scalar or pseudoscalar state \cite{CERN-pp,EEQSY} with no other companion, except for heavy fermions that generate the two--photon and two--gluon couplings. The other benchmark is a two-Higgs-doublet model (2HDM) \cite{2HDM} in which the $\Phi$ state could be either the heavier CP--even $H$ or CP--odd $A$ or a combination of the two states~\cite{DEGQ,ADM}.  

In all the scenarios studied, in which $\Phi$ is a scalar $H$ or pseudoscalar $A$ singlet that is not  accompanied by any bosonic partner particles, the $\Phi$ couplings to  photon and gluon pairs are described via dimension-five operators in an effective field theory:
\beq
{\cal L}_{\rm eff}^H &=& \frac{e}{v } c_{H\gamma\gamma} \, H F_{\mu\nu} F^{\mu\nu} 
                 + \frac{g_s}{v}       c_{Hgg} \, H G_{\mu\nu}  G^{\mu\nu} \nonumber \, , \\
{\cal L}_{\rm eff}^A &=& \frac{e}{v } c_{A\gamma\gamma} \, A F_{\mu\nu} \tilde F^{\mu\nu} 
                 + \frac{g_s}{v}   c_{Agg}  \, A G_{\mu\nu} \tilde G^{\mu\nu} \, ,
\label{eq:lag-eff}
\eeq
with $F_{\mu \nu}= (\partial_\mu A_\nu - \partial_\nu A_\mu)$ the field strength of the
electromagnetic field, $\tilde{F}_{\mu\nu}=\epsilon_{\mu \nu \rho \sigma} F^{\rho \sigma}$ 
and likewise for the SU(3) gauge fields $G_{\mu \nu}$, and $v\approx 246$ GeV is the standard Higgs vacuum expectation value. In addition to Standard Model particles,
the $\Phi \gamma \gamma$ and $\Phi gg$ couplings are induced by new massive particles, which we assume to be vector--like quarks and leptons that couple to the $\Phi=H/A$ resonances according to (we take the Standard Model--like Higgs Yukawa coupling as a reference)
\beq 
\lambda_{\Phi F F}= m_F / v \times \hat g_{\Phi FF}    
\label{eq:yukawa}
\eeq 
Couplings of the singlet states $\Phi$ to standard fermions could also be generated through 
the effective Lagrangians ${\cal A}_m^H = c_f (m_f/\Lambda) \Phi \bar f f$ and ${\cal A}_m^A = i c_f (m_f/\Lambda) \Phi \bar f \gamma^5 f$ in the scalar and pseudoscalar cases, respectively,
with $\Lambda$ some new physics scale in the multi-TeV range~\cite{MAD}. As the Yukawa coupling is proportional to the fermion mass, the top quark should be then the particle that couples most strongly to the $\Phi$ states. The couplings $c_f$ and $\hat g_{\Phi FF}$ 
(\ref{eq:yukawa}) are related by $c_f= (\Lambda / v ) \times \hat g_{\Phi ff}$.

The second benchmark that we consider is a 2HDM in which there are five physical states: two CP--even neutral $h$ and $H$ bosons,  a  CP--odd $A$ and two charged $H^\pm$ bosons. In the general case, the masses $M_h, M_H, M_A$ and $M_{H^\pm}$ are free parameters and one assumes that $h$ is the observed Higgs boson with mass $M_h=125$ GeV. At least two additional mixing parameters $\beta$ and $\alpha$ are needed to characterize fully the model: $\tb = v_2/v_1$ is the ratio of the vacuum expectation values of the two fields with $v_1^2\!+\!v_2^2 \! = \! v^2 \! = \! {\rm (246~GeV)^2}$, and  $\alpha$ is the angle that diagonalises the CP--even $h$ and $H$ mass matrix~\cite{2HDM}.  

The $\Phi$ state will be identified with a neutral Higgs boson, $\Phi=H,A$ or a superposition $H+A$.  There is no coupling of the CP--odd $A$ to the vector bosons $V=W,Z$ by virtue of CP invariance, but the CP--even $h$ and $H$ states share the coupling of the standard Higgs particle and, in units of this coupling, one has $\hat g_{hVV}= \sin(\beta-\alpha)$ and $\hat g_{HVV}= \cos(\beta-\alpha)$. One must take into account the fact that the couplings of the $h$ boson have been rather precisely measured at the LHC, and found to agree with those of a standard Higgs boson within 10\% accuracy overall~\cite{combo}.  This constraint can be accommodated naturally by postulating the alignment limit \cite{align}, in which one has $\alpha=\beta- \frac{\pi}{2}$ and the $h$ couplings are exactly Standard Model--like. Here we adopt this limit, which leads to a  simplified picture, as the couplings of the $\Phi=H/A$ states to massive $V=W,Z$ bosons are then both absent, $\hat g_{\Phi VV}=0$. 

In contrast, the Higgs interactions with fermions are model--dependent in a 2HDM, and two options are generally discussed \cite{2HDM}: Type--I, in which one field generates the masses of all fermions, and Type--II, in which one field generates the masses of isospin down--type fermions and the other the masses of up--type quarks. In the alignment limit $\alpha=\beta- \frac{\pi}{2}$,  the $h$ couplings to a given fermion are again standard, 
while the $H$ and $A$ couplings have the same magnitude.  In the case of third-generation fermions, they are given  by
\begin{eqnarray}
{\rm Type\!-\!I}~:~  | \hat g_{\Phi tt } | = \cot\beta \, , \ | \hat  g_{\Phi bb } | =  | \hat  g_{\Phi \tau\tau } | =  \cot\beta  \, ,\\
{\rm Type\!-\!II}~:~  | \hat  g_{\Phi tt } | = \cot\beta \, , \ | \hat  g_{\Phi bb } | =  | \hat  g_{\Phi \tau\tau } | =  \tan\beta \, ,
\label{Phiff}
\end{eqnarray}
when normalized to the standard Higgs coupling, $g_{Hff}^{\rm SM}= m_f/v$. The absolute values 
of the couplings are given as there is a sign ambiguity that depends on the isospin and the 
model type. In the Type-II case, there is a relative minus sign between 
the $At \bar t$ and $H t\bar t$ couplings with the latter having the opposite sign to the $h t {\bar t}$ coupling, for instance,. 

In the case of the bottom  quarks and and tau leptons, their couplings are significant only in Type-II models and for large $\tb$ values, $\tb \gsim 20$, which are excluded by LHC $\Phi \to \tau\tau$  searches \cite{LHC-tautau}. In both model types, the $\Phi$ couplings to top quarks are large for low values of $\tb$. Nevertheless, $\tb$ values less than unity must be avoided not only for perturbativity reasons but also  because of the ATLAS and CMS limits from searches for $t\bar t $ production~\cite{ATLAS-tt,CMS-tt}. We therefore assume $\tb = 1$ in our studies, in which case both the Type--I and Type--II models lead to similar phenomenology. 

All these features appear in the context of the Minimal Supersymmetric extension of the Standard Model (MSSM),
which is essentially a Type--II 2HDM with the additional restriction of
near-degeneracy between the heavier Higgs states $M_A \approx M_H \approx M_{H^\pm}$ in the so--called decoupling limit in which $\alpha=\beta -\frac{\pi}{2}$ and, hence, the light $h$ state
is automatically Standard Model-like. We adopt the  assumption of approximately equal Higgs masses in our 2HDM scenario, in particular because this constraint is favored by high-precision electroweak data \cite{gfitter}. In our analyses, we use as a basic input $M_A=750$~GeV, which then leads to $M_H= 766$ GeV for the heavy CP--even Higgs mass~\footnote{These values are obtained in the context of the so--called $h$MSSM scenario \cite{hMSSM} in which the constraint $M_h=125$~GeV has been enforced, and which allows one to consider low values of $\tb$.} when $\tb =1$. 

\begin{figure}[!h]
\vspace*{-1.7cm}
\centerline{ \hspace*{-7.5cm}
\includegraphics[scale=0.58]{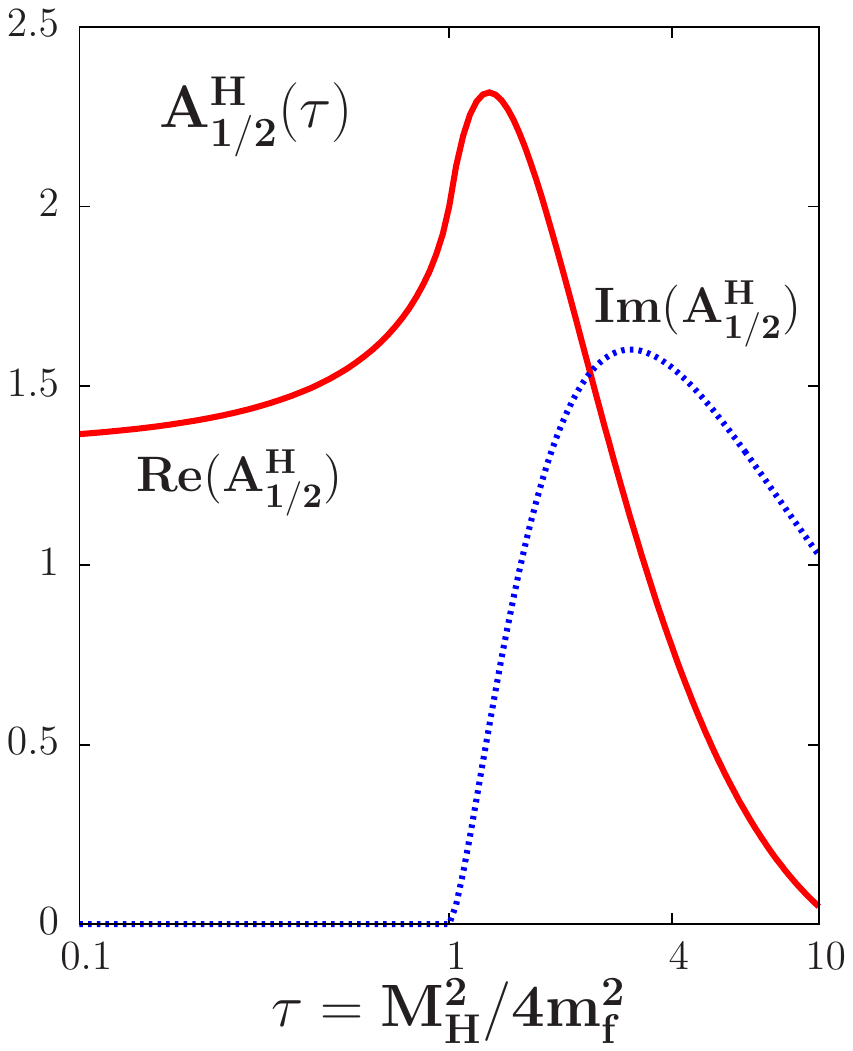} \hspace*{-6.2cm}
\includegraphics[scale=0.58]{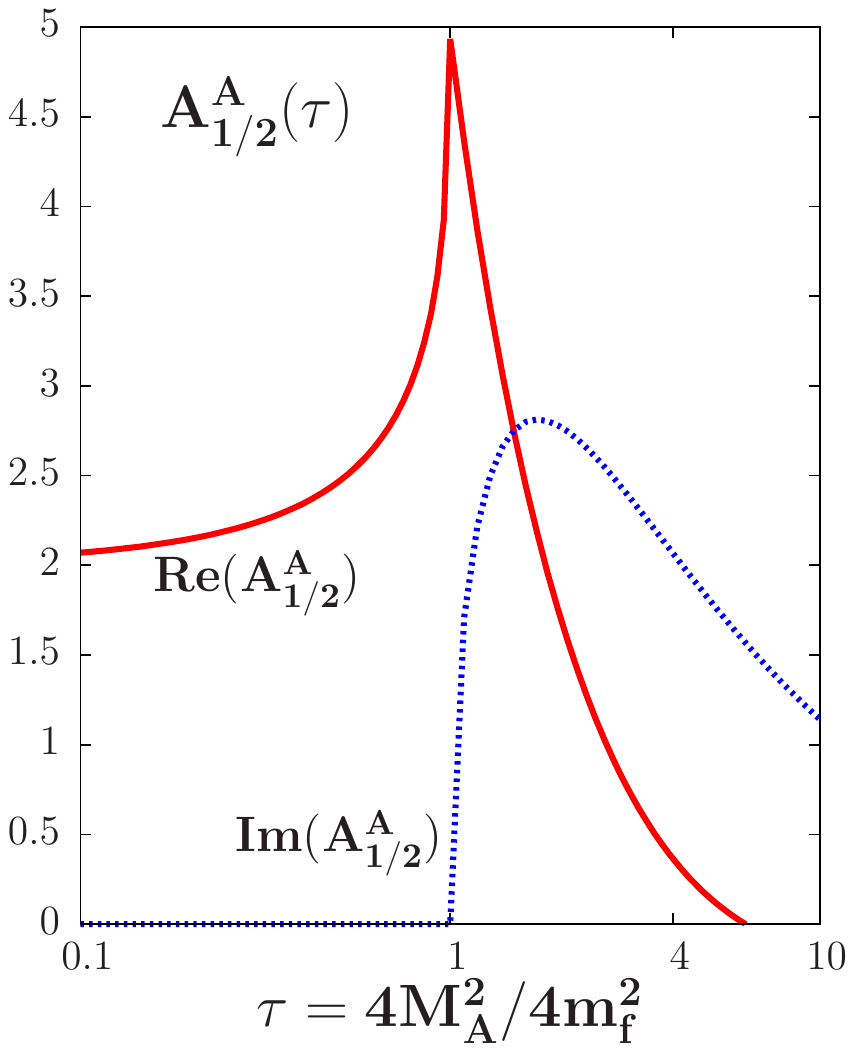} \hspace*{-7.2cm}
}
\vspace*{-9.3cm}
\caption{\it The real and imaginary parts of the form factors $A^\Phi_{1/2}$ with fermion loops in the case of a CP--even state $H$ (left panel) and of a CP--odd $A$ state (right panel) as functions of the variable $\tau=M_\Phi^2/4m_F^2$.} 
\label{fig:A12}
\vspace*{-1mm}
\end{figure}

As discussed above, the couplings of the $\Phi$ states to gluons and photons are assumed to be generated  by loops of heavy fermions $F$, which can be either third-generation Standard Model
fermions or new vector--like fermions, in which case the partial decay width into the $gg$ and $\gamma\gamma$ final states are given by \cite{Anatomy} (see also~\cite{venerable}):
\begin{eqnarray}
\Gamma(\Phi  \to gg) & = &  \frac{G_\mu\alpha_s^2 M_\Phi^3}
{64\sqrt{2}\pi^3} \bigg| \sum_Q  \hat g_{\Phi QQ} A_{1/2}^\Phi  (\tau_Q) 
\bigg|^2 \, , \nonumber \\
\Gamma(\Phi  \to \gamma\gamma) & = &  \frac{G_\mu\alpha^2 M_\Phi^3}
{128\sqrt{2}\pi^3} \bigg| \sum_F  \hat g_{\Phi FF} N_c e_F^2 A_{1/2}^\Phi  (\tau_F) 
\bigg|^2 \, ,
\label{eq:Gammagg}
\end{eqnarray}
with $N_c$ a color factor, $e_F$ the electric charge of the fermions $F$, and  $\hat g_{\Phi FF}$
the reduced Yukawa coupling in units of $m_F/v$. The quantities $A_{1/2}^{\Phi}$ are the usual form factors for the contributions of spin--$\frac12$ fermions that, in terms of the variable 
$\tau_F \equiv M_\Phi^2 / 4 m_F^2$, are given in the CP--even $H$ and CP--odd $A$ cases by 
\begin{eqnarray}
& A_{1/2}^{H}(\tau)  = 2 \left[  \tau +( \tau -1) f(\tau)\right]  \tau^{-2} \, , \ \  
  A_{1/2}^{A} (\tau) =  2 \tau^{-1} f(\tau) \, ,  \label{eq:Af} \\ 
& f(\tau)=\left\{ \begin{array}{ll}  \displaystyle \arcsin^2\sqrt{\tau } & 
{\rm for} \;  \tau \leq 1 \, , \\
\displaystyle -\frac{1}{4}\left[ \log\frac{1+\sqrt{1-\tau^{-1} }} 
{1-\sqrt{1-\tau^{-1}}}-i\pi \right]^2 \hspace{0.5cm} & {\rm for} \; \tau > 1 \, .
\end{array} \right.
\label{eq:formfactors}
\end{eqnarray}
These are displayed in Fig.~\ref{fig:A12} for the CP--even (left panel) and CP--odd (right panel) cases as functions of the loop variable $\tau=M_\Phi^2/4m_F^2$. The form factors vanish in the zero--mass limit for the fermions, while in the infinite-mass limit they reach constant values $A_{1/2}^H \to \frac43$ and $A_{1/2}^A \to 2$. They are real below the kinematical threshold $M_\Phi = 2m_F$
and develop an imaginary part above, reaching their maximum values near the threshold.  

There are, in principle, also $W$ boson loop contributions to the $H\to \gamma \gamma$ decay mode~\cite{EGN}. However, as we are assuming the alignment limit of the 2HDM (or the decoupling limit  of the MSSM), there is no $HWW$ coupling, $\hat g_{HWW}=\cos(\beta-\alpha) \to 0$. And of course, there is no $W$ contribution in the $A \to \gamma \gamma$ case as the $AWW$ coupling is absent as a result of CP--invariance.  

Turning to the decays of the $\Phi$ state, the main modes in a 2HDM would be the fermionic 
decays whose partial widths are given by \cite{Anatomy}
 \begin{eqnarray}
\Gamma(\Phi \to f \bar{f} ) =  N_c \frac{ G_F m_f^2}{4\sqrt{2} \pi}
\, \hat g_{\Phi ff}^2 \, M_{\Phi} \, \beta^{p_\Phi}_f \, ,
\end{eqnarray}
where  the power in the velocity of the final fermion $\beta_f=(1-4m_f^2/M_{\Phi}^2)^{1/2}$
is $p_\Phi =3\,(1)$ for the CP--even (odd) Higgs boson. Hence, the only relevant decays at 
low $\tb$ values are those into $t\bar t$ pairs, whereas the modes $\Phi \to b \bar b, \tau^+ \tau^-$ are relevant only at high $\tb$. All other decay modes, including those to vector boson pairs or to the lighter Higgs and a gauge boson, are strongly suppressed in the alignment/decoupling limits of 2HDMs such as the MSSM \cite{hMSSM}. In  addition, for the mass range $M_A \approx  M_H \approx M_{H^\pm}$ assumed in our analysis, the decays $H/A \to A/H\!+\!Z$ or $H^\pm W^\mp$ are kinematically forbidden at the two--body level and, hence, strongly suppressed. 

As for the $\Phi = H, A$ total decay widths, they are almost the same as the $\Gamma(\Phi \to t \bar{t} )$ partial widths in the low $\tb$ regime and, for $\tb =1$, they are $\Gamma_A=36$ GeV and $\Gamma_H=33$ GeV for the CP--odd and CP--even states with masses of $M_A=750$ GeV and $M_H=766$ GeV \cite{hdecay}.  The branching fractions for the photonic decays $\Phi \to \gamma \gamma$ are extremely small in this case, BR($\Phi \to \gamma \gamma) \approx 0.7 \! \cdot \! 10^{-5}$ \cite{hdecay}, so large contributions of vector-like fermions would be needed to enhance it to a level compatible with the apparent cross section times $\gamma \gamma$ branching ratio of the diphoton state at the LHC, i.e., of order a few fb. 

In the case of a singlet $\Phi$ resonance, the total decay width may be very small, of order 1 GeV or below,  if there are only loop-induced decays into gauge bosons. 
However, a large total width could be generated from the mode $\Phi \to t \bar t$ if the $\Phi t\bar t$~Yukawa coupling is strong enough, or by allowing $\Phi$ to decay into pairs of vector--like leptons with  masses $m_L \lsim 375$ GeV. Such masses for vector--like leptons are still allowed by collider constraints, in contrast to vector--like quarks, which negative LHC searches require to be heavier than about 700~GeV~\cite{PDG}. 

\section{Interference in the $\mathbf{\gamma \gamma}$ Spectrum}

\subsection{Formulation}

\begin{figure}[!h]
\vspace*{-3.cm}
\centerline{ \includegraphics[scale=0.86]{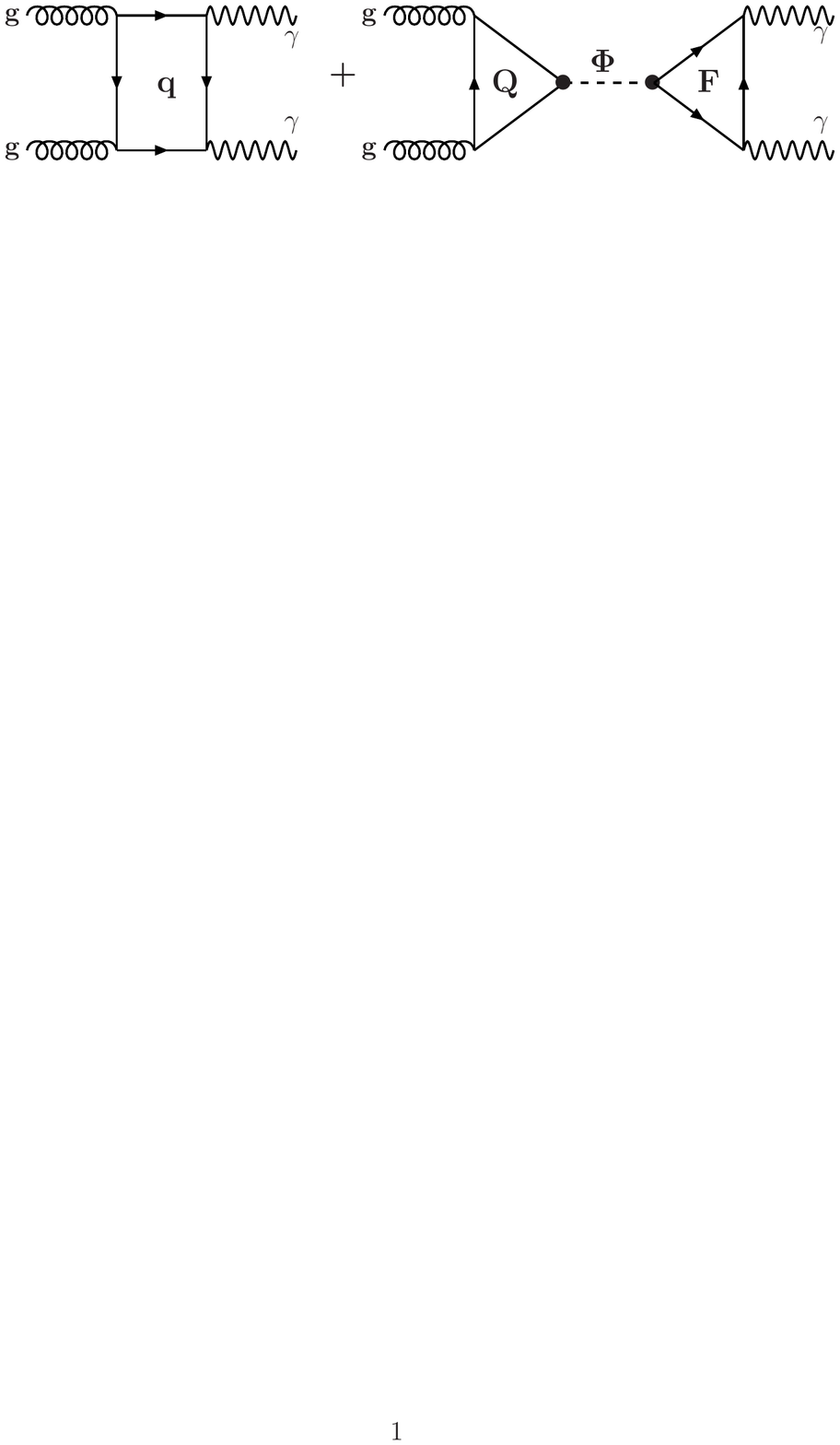} }
\vspace*{-20.1cm}
\caption{\it Feynman diagrams for the continuum background (left) and the $\Phi$ signal (right)
in the process $gg\to \gamma\gamma$ at leading order (LO). The internal particles are light quarks $q$ in the background diagram and heavy fermions $Q, F$ in the signal diagram.} 
\label{fig:feynmanpp}
\vspace*{-2mm}
\end{figure}

At leading order (LO), the process $gg \rightarrow \gamma\gamma$ receives contributions from the two diagrams shown in Fig.~\ref{fig:feynmanpp}: a box diagram in which the two photons are radiated from the internal quark lines, that we call the background or continuum, and a product of two triangle diagrams with circulating heavy fermions linked by the exchange of one or more $\Phi (750)$ states that we call the resonant contribution or signal. We make some simplifying assumptions in our analysis. We neglect the contributions of the 125 GeV Standard Model-like Higgs exchange as well as $q {\bar q} \to \gamma\gamma$ diagrams, which do not contribute to the interference. When calculating the background we also neglect $gg \to \gamma \gamma$ amplitudes that do not interfere with the $\Phi$ signal. Finally, we neglect possible bosonic contributions to the $\Phi \to\gamma \gamma$ amplitude that are small in the alignment limit of the 2HDMs (or the decoupling limit of the MSSM) that we study here, as discussed in  the previous Section.

Averaging/summing over the polarisations of the incoming gluons/outgoing photons and  adding the continuum and the  resonant contributions, the total amplitude of the process can be written as 
\beq
{\cal A} = - \sum_\Phi \frac{{\cal A}_{gg \Phi} {\cal A}_{\gamma\gamma \Phi}}{\hat s - M_\Phi^2 + i M_\Phi \Gamma_\Phi} + {\cal A}_{gg\gamma \gamma} \, .
\label{eq:ggpp}
\eeq
The sum in the first term may run over more than one state, e.g., $\Phi=H,A$ in a 2HDM, and, in the second term, there is a sum that runs over the six standard quark flavors, $q=u,d,s,c,b,t$, as the contributions from heavy (vector--like) quarks decouple as $\hat s /4m_Q^2 \to 0$. At LO, the couplings of the $\Phi$ states to gluons and photons induced by loops of a heavy fermion $F$ are given by \cite{venerable,Anatomy}
\beq
\label{A12}
{\cal A}_{gg \Phi} &= &\frac{\alpha_s}{8 \pi v} \hat s \sum_{Q} \hat g_{\Phi QQ} A_{1/2}^\Phi (\hat \tau_Q) \, , \\  
{\cal A}_{\gamma\gamma \Phi } &= &\frac{\alpha}{4 \pi v} \hat s \sum_{F}  N_c^F e_F^2 \, \hat g_{\Phi FF} \, A_{1/2}^\Phi (\hat \tau_F) \, ,
\eeq
where the  form factors for the contributions of spin--$\frac12$ fermions $A_{1/2}^{\Phi}$
are given in eq.~(\ref{eq:formfactors}) in the CP--even and CP--odd cases but where the loop variable is  now given by $\hat \tau_F \equiv \hat s / 4 m_F^2$ with $\hat s$ the partonic centre-of-mass energy-squared.

As for the continuum contribution, the matrix elements $A_q$ for the one--loop box diagram contribution of a given quark $q$ in the massless limit $\hat s \gg 4m_q^2$, which holds very well for 
the five light quarks $q=u,d,s,c,b$ and is also a good approximation for $q=t$, are given by~\cite{int-gamma,Martin,1loop} 
\beq
A_q  &=&  z \ln \left( \frac{1+z}{1-z}\right ) - \frac{1 + z^2}{4} 
\bigg[ \ln^2 \left (\frac{1+z}{1-z} \right ) + \pi^2 \bigg] \, ,
\label{eq:Mpppp}
\eeq
where $z = \cos\theta$, $\theta$ being the scattering angle in the diphoton centre-of-mass frame, and we have retained only the helicity configurations that give non-vanishing interference with the $\Phi$ amplitudes.  The total amplitude of the continuum is then
\beq
{\cal A}_{gg\gamma\gamma} = 2 \alpha_s \alpha \sum_q e_q^2 A_q \, . 
\label{eq:Aggaa}
\eeq
We note that for light quarks, $m_q \ll \sqrt{\hat s}$, the continuum amplitudes above have only a small absorptive part that is suppressed by powers of $1/ \hat \tau_q = 4m_q^2/ \hat s$. However, the top quark loop induces a relevant contribution, since the $m_t^2$ effects (that yield more complicated expressions for the amplitudes~\cite{1loop}) are not insignificant. In addition, an imaginary contribution with no quark mass suppression occurs at the two--loop level \cite{2loop}. We neglect both contributions in this rather exploratory analysis of interference effects. We note also that heavy quarks decouple as $\hat s /4 m_Q^2$ in the background amplitudes, and we thus neglect their possible contribution in the box diagrams. 

The cross section for the $gg\to \gamma \gamma$ background falls steeply with the square of the centre-of-mass energy $\sqrt{\hat s}$, i.e., the invariant mass of the diphoton pair. For a complete description of the background,  the contribution of the  $q \bar q \to \gamma \gamma$ final state, summing the contributions of all light quarks $q=u,d,s,c,b$ in the initial state, should also be included,  but it does not interfere with the $\Phi$ signal. The partonic cross section $\hat \sigma(q \bar q \to \gamma \gamma)$ is much larger than that for the $gg$-initiated component, as the process occurs at tree level and, unlike $gg \to \gamma \gamma$, is not suppressed by two powers of $\alpha_s$. Nevertheless, at the hadronic level when folding with the parton luminosities,  the  difference between the rates of the two subprocesses  becomes smaller, an order of magnitude only, due to the large compensation arising from the much higher gluon-gluon luminosity at the energies involved at the LHC. 


In contrast to the $gg\to \gamma\gamma$ background amplitude, the form factors $A^\Phi_{1/2}$ develop important imaginary components when the fermions circulating in the $\Phi gg$ and $\Phi \gamma\gamma$ loops have masses below the kinematical threshold, $\hat s = M_\Phi = 4m_F^2$, as seen in Fig.~~\ref{fig:A12}. The imaginary parts are maximal slightly above threshold;  Im($A^A_{1/2}) \approx 2.8$ for $\tau \approx 1.5$--2.5 and Im($A^H_{1/2}) \approx 1.6$ for $\tau \approx 2$--5, remaining significant far above this threshold, as one still has Im($A^A_{1/2}) \approx {\rm Im}(A^H_{1/2}) \approx 1$ for $\tau \approx 10$. On the other hand, for $\tau <1$,  the amplitudes are real and are maximal near threshold, where one has Re($A^H_{1/2}) \approx 2 $ and  Re($A^A_{1/2}) \approx \frac12 \pi^2 \approx 5$. Finally, we remark that for $\tau \approx 4.7$, which corresponds to the  case of the top quark with $m_t=173$ GeV, the form factors are still sizeable, with the real parts being rather smaller that the imaginary ones:  Re($A^H_{1/2}) \approx 0.6 $ and  Re($A^A_{1/2}) \approx 0.2$ versus Im($A^H_{1/2}) \approx 1.5 $ and  Im($A^A_{1/2}) \approx 1.8$, so that $| A^A_{1/2}/A_{1/2}^H|^2  \approx 2$. The $b$--quark contributions are very small in the cases of interest to us, and we neglect them in our analysis.


At the hadronic level, when convoluting with the parton luminosity function
\begin{equation}
G_{gg} (\hat s) \; = \; \int_{\hat s/s}^1 {\rm d}x/(sx) \! \times \! g(x) g(\hat s/sx) \, , 
\label{eq:cgg}
\end{equation}
the cross section for the $pp\to (\Phi \to) ~ \gamma\gamma$ process including the pure signal and its interference with the continuum background is given by
\beq
\frac{{\rm d}^2\sigma } {{\rm d}\sqrt{\hat s}  {\rm d}z} (pp \! \to  \! \gamma\gamma) 
= \frac{G_{gg}(\hat s)}{256\pi \sqrt{\hat s} } \bigg[ \sum_\Phi \frac {N_S^\Phi + N_\Phi^{\rm IRe} + N_\Phi ^{\rm IIm} } { (\hat s - M_\Phi^2)^2 + M_\Phi^2 \Gamma^2_\Phi} + N_B \bigg] \, ,
\label{eq:sigmall}
\eeq
where the various components, except for the pure background $N_B$ that has been discussed previously,  are given by
\beq
N_\Phi^S &=& |{\cal A}_{gg \Phi } \, {\cal A}_{\gamma\gamma \Phi}|^2 \, ,  \\
N_\Phi^{\rm IRe} &=&   -2 {\rm Re} [ {\cal A}_{gg \Phi} \, {\cal A}_{\gamma\gamma \Phi} \,  
{\cal A}_{gg\gamma\gamma}^* ] \times (\hat s - M_\Phi^2) \, ,  \\
N_\Phi^{\rm IIm} &=&  -  2 {\rm Im} [ {\cal A}_{gg \Phi} \, {\cal A}_{\gamma\gamma \Phi} \, 
{\cal A}_{gg\gamma\gamma}^* ] \times M_\Phi \Gamma_\Phi \, . 
\eeq
Since the first component of the interference, $N_\Phi^{\rm IRe}$, is proportional to  $\hat s - M_\Phi^2$, it does not contribute to the total cross section when one integrates over $\hat s$, to the extent that $G_{gg}(\hat s)$ varies slowly over the width of the $\Phi$ state(s). However, it distorts the resonance shape and shifts the position of the peak, changing the apparent mass of the observed resonance. On the other hand, the second interference term, $N_\Phi^{\rm IIm}$, contributes to the total cross section, and its contribution is more significant  if the total width $\Gamma_\Phi$ is large. 

We recall that the results above are only at LO, and higher-order corrections must be taken  into account. The QCD corrections to the signal cross section, $gg \to \Phi$, are known up to N$^3$LO \cite{ggH-NNLO} in the approximation in which the internal quark is much heavier than the Higgs boson, which is a good approximation below the  $Q\bar Q$ threshold,  $M_\Phi \lsim 2m_Q$ where the amplitudes have no imaginary parts. However, above this kinematical threshold, the QCD corrections for both the real and imaginary parts are known only to NLO \cite{ggH-NLO}. 

It is a good approximation at NLO to incorporate these corrections in the limit of infinite loop mass even for $M_\Phi \gsim 2m_Q$, provided that  the Born term contains the full quark mass dependence \cite{ggH-NLO}.  At the LHC with $\sqrt s \approx 13$ TeV, the corrections up to N$^3$LO lead to a $K$--factor~\footnote{The $K$--factor is defined as the ratio of the cross section at the  higher order to the LO cross section, with the coupling $\alpha_s$ and the parton distribution functions (PDFs) taken consistently  at the respective perturbative orders. For the latter, we use always the MSTW set \cite{MSTW}. }  $K_{\rm N3LO}^{\rm gg \to \Phi} \approx 2$ in both the CP--even and CP--odd cases. We note that, for convenience, we make the  choice $\mu_R=\mu_F=M_\Phi$  for the renormalization and factorization scales,  which is different from the standard choice for the SM Higgs boson, namely $\mu_F=\mu_R=\frac12 M_\Phi$ \cite{LHCXS}, which leads to a slightly smaller $K$--factor than our choice (but the same total cross section at the N$^3$LO). 

The NNLO  corrections to the background processes are also known~\cite{2loop}, but the higher-order corrections have not yet been calculated for the interference between the signal and background amplitudes. We assume here, following a standard choice (see for instance Ref.~\cite{Martin}), that the interference has the same $K$--factor as the signal amplitude.  The QCD corrections to the $\Phi \to \gamma \gamma$ decay (known only at NLO) and  the electroweak corrections to $gg\to \Phi$ (which are not completely known in the cases of interest) are or should be rather small \cite{Anatomy,LHCXS}, and can safely be ignored in a first approximation. 
 
In order to fix ideas, we recall the case of the Standard Model Higgs boson $h$ \cite{int-gamma,Martin}.  The main contribution to the dominant $gg\to h$ production mechanism is due to the top quark loop with the $W$ boson loop dominating the $h \to \gamma \gamma$ decay amplitude. Since $M_h <2 M_W, 2m_t$, the amplitudes are real: the sole imaginary component present in the process is that due to the bottom quark loop, which is extremely small. The amplitude from the $gg\to \gamma \gamma$ box diagram that generates the interference with the signal, and which is mediated by the five light quarks only (the contribution of the the top quark decouples as $\hat s/4 m_t^2$ for $\hat s = M_h^2$), is also mostly real at one loop, but the small two--loop contribution 
has an imaginary part that generates a negative interference of few percent at most. The interference between the real parts of the Higgs signal and continuum background has been shown to lead to a downwards shift of the Higgs mass by about 150~MeV at the 8-TeV LHC \cite{Martin}. 
The situation is completely different for the 750 GeV $\Phi$ resonance as we discuss in the next Subsection.

\subsection{Numerical Results}

We study now the effects of interference with the background in various models for the $gg\to \Phi \to \gamma \gamma$ signal. As already mentioned, for simplicity we do not include the $q {\bar q} \to \gamma \gamma$ background, which would not interfere with the $gg \to \Phi \to \gamma \gamma$ amplitude, nor do we include the loop-induced $gg \to \gamma \gamma$ backgrounds in partial waves that would not interfere with the signal. Thus, we underestimate the total $\gamma \gamma$ background but this is not a problem as  our main objective is to study the line-shape and the possible enhancement of any signal by interference effects rather than to compare with data.

We consider initially cases where only Standard Model fermions in the background $gg \to \gamma \gamma$ loops and in the $gg \to \Phi$ and $\Phi \to \gamma \gamma$ amplitudes. The contributions of the light-quark loops in the background calculation are essentially real.  
In principle, one should also include all Standard Model fermion loops in the signal processes
\footnote{As discussed previously, we do not include $W$ boson loops in the $\Phi \to \gamma \gamma$ 
decay amplitude, as we are working in the alignment limit in which the $HWW$ coupling vanishes, and the $AWW$ coupling is absent in CP--invariant theories.}
 $gg \to \Phi$ and $\Phi \to \gamma \gamma$. However, as we assume that their couplings are proportional to those in a Type--II 2HDM with $\tan \beta = 1$, their contributions are negligible and only the top quark loop contributions need be taken into account. In this case, if the $\Phi$ is assumed to be a scalar $H$, the $H t {\bar t}$ coupling has the the opposite sign to that of the $h t {\bar t}$ coupling in the Standard Model, whereas if the $\Phi$ is a pseudoscalar state, the $A t {\bar t}$ coupling has the same sign as the standard $h t {\bar t}$ coupling. As one can see from Fig.~\ref{fig:A12} where the form factors that describe the fermionic contributions to the $\Phi gg$ and $\Phi \gamma\gamma$ vertices, the real and imaginary parts of the top loop contributions are significant in both the scalar $H$ and the pseudoscalar $A$ case, though the imaginary parts are much larger.

If the $\Phi t {\bar t}$ coupling were to have the same magnitude as the Standard Model $h t {\bar t}$ coupling, we would find $\Gamma (H \to t {\bar t}) = 30$~GeV and $\Gamma (A \to t {\bar t}) = 36$~GeV for $M_\Phi = 750$~GeV, the difference being due to the difference between p- and s-wave phase space. In the following we consider these benchmark options, as well as options in which the fermion couplings found in the Type--II 2HDM are modified by universal factors (0.18 and 0.16, respectively) chosen to obtain $\Gamma (H \to t {\bar t}), \Gamma (A \to t {\bar t}) = 1$~GeV for $M_\Phi = 750$~GeV in order to describe also the interference effects in the case of a narrow resonance. In a later stage we will also include loops of heavy fermions in the $gg \to \Phi$ and $\Phi \to \gamma \gamma$ amplitudes in addition to the Standard Model loops. 
As specific models, we consider first minimal scenarios in which the $\Phi$ is either a single scalar $H$ or a pseudoscalar $A$, as was discussed in~\cite{EEQSY,DEGQ}, with the broad and narrow total decay widths given above. We then consider a non-minimal scenario with a pair of near-degenerate states $H$ and $A$, with the couplings and mass difference $M_H - M_A = 16$~GeV found in a supersymmetric version of the Type--II 2HDM with $\tan \beta = 1$ \cite{hMSSM}.

\begin{figure}[!h]
\centerline{\hspace*{-.5cm}
\includegraphics[scale=0.38]{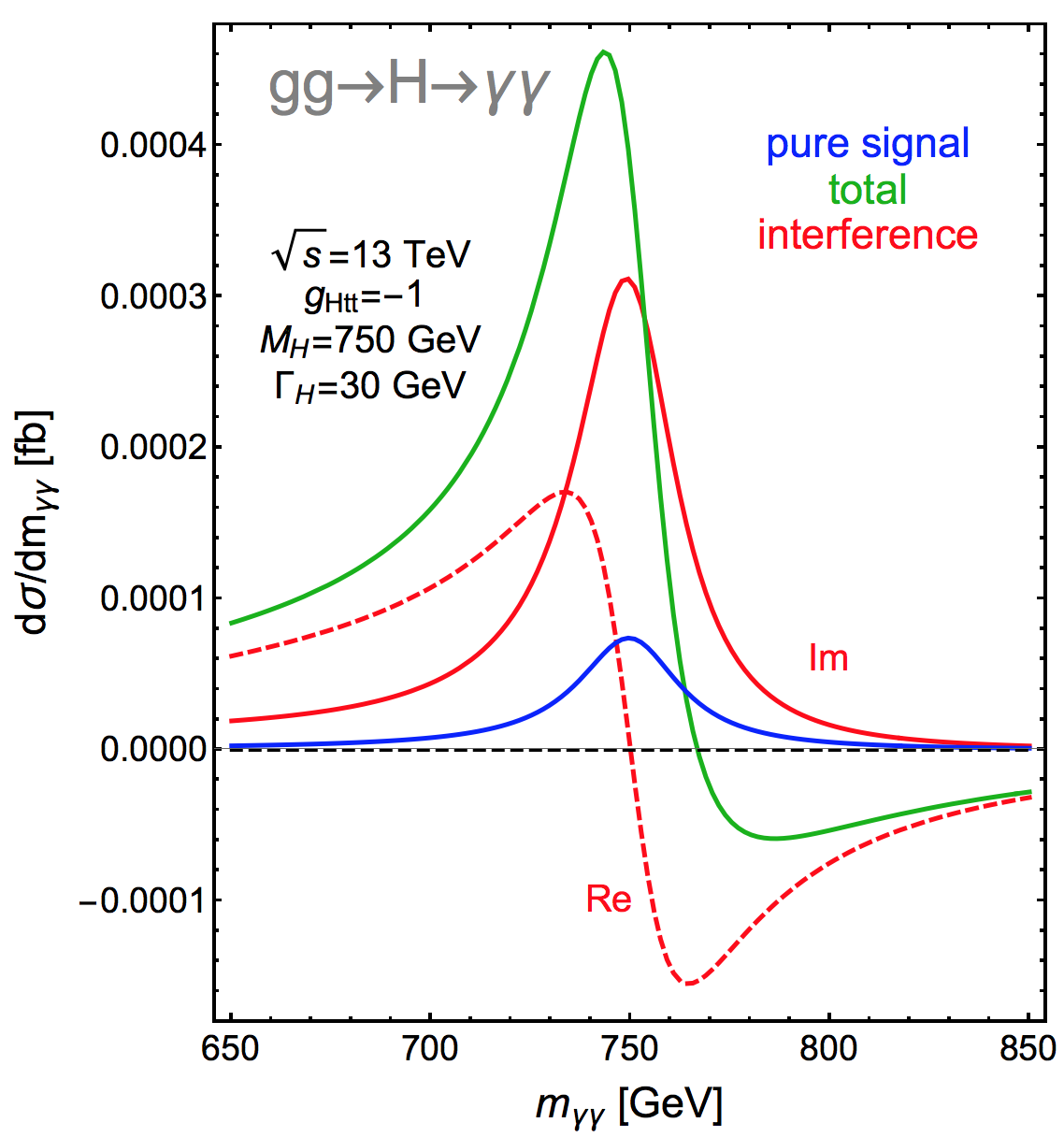}~~~ 
\includegraphics[scale=0.38]{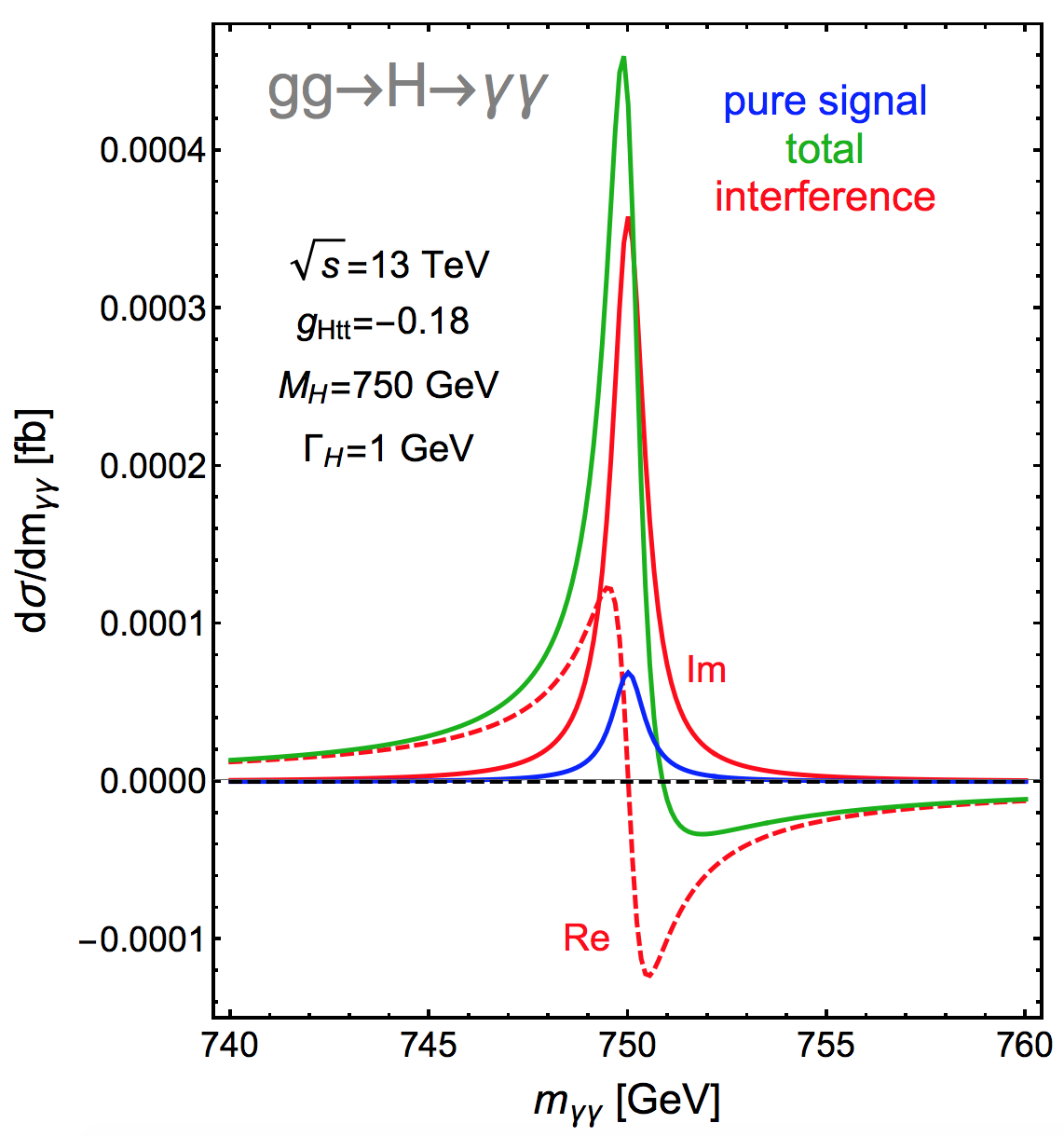}} 
\caption{\it The contributions to the line-shapes of a CP--even $H \to \gamma \gamma$ with mass 750~GeV and total width $\Gamma_H \to t = 30$~GeV (left panel) and $\Gamma_H = 1$~GeV (right panel), as functions of $m_{\gamma \gamma}$, showing the line-shape neglecting interference (solid blue lines), the contributions of interferences in the real and
imaginary parts of the $gg \to H \to \gamma \gamma$ amplitude (dashed and solid lines) and the overall combination including both interferences (solid green lines). These plots were calculated including only Standard Model fermion loops in the $gg \to H$ and $H \to \gamma \gamma$ couplings.}
\label{fig:Hgaga}
\end{figure}

Fig.~\ref{fig:Hgaga} displays contributions to the line-shape of a CP--even $H \to \gamma \gamma$ with mass 750~GeV, assuming a total width $\Gamma_H \approx \Gamma(H \to t {\bar t}) = 30$~GeV (left panel) or $\Gamma_H \approx \Gamma(H \to t {\bar t})= 1$~GeV (right panel), assuming only only Standard Model fermion loops in the $gg \to H$ and $H \to \gamma \gamma$ couplings. 
(Here and in subsequent plots, we use the MSTW set of parton distributions~\cite{MSTW}.)
In each case, the line-shape calculated neglecting interference is shown as a solid blue line, the contributions of interferences in the real and imaginary parts of the $gg \to H \to \gamma \gamma$ amplitude are shown as dashed and solid red lines, and the total line-shape including both interferences is shown as a solid green line. We see that, in both cases, the interference in the imaginary part of the amplitude is much larger than the line-shape calculated neglecting interference, and is symmetric about the nominal $H$ mass. The interference in the real part of the amplitude is also relatively large, and changes sign at the nominal $H$ mass. The overall combination exhibits a peak slightly below the nominal mass and a more modest dip just above the nominal mass. The magnitudes of these features are much greater than in the calculation without the interferences. However, we emphasize that the magnitude of the signal is still far smaller than that reported by ATLAS and CMS, despite the large overall enhancement of the peak, necessitating the introduction of loops of heavy vector-like fermions, 
which, as we discuss later, make the interference effects much less pronounced. 

\begin{figure}[!h]
\centerline{\hspace*{-.5cm}
\includegraphics[scale=0.097]{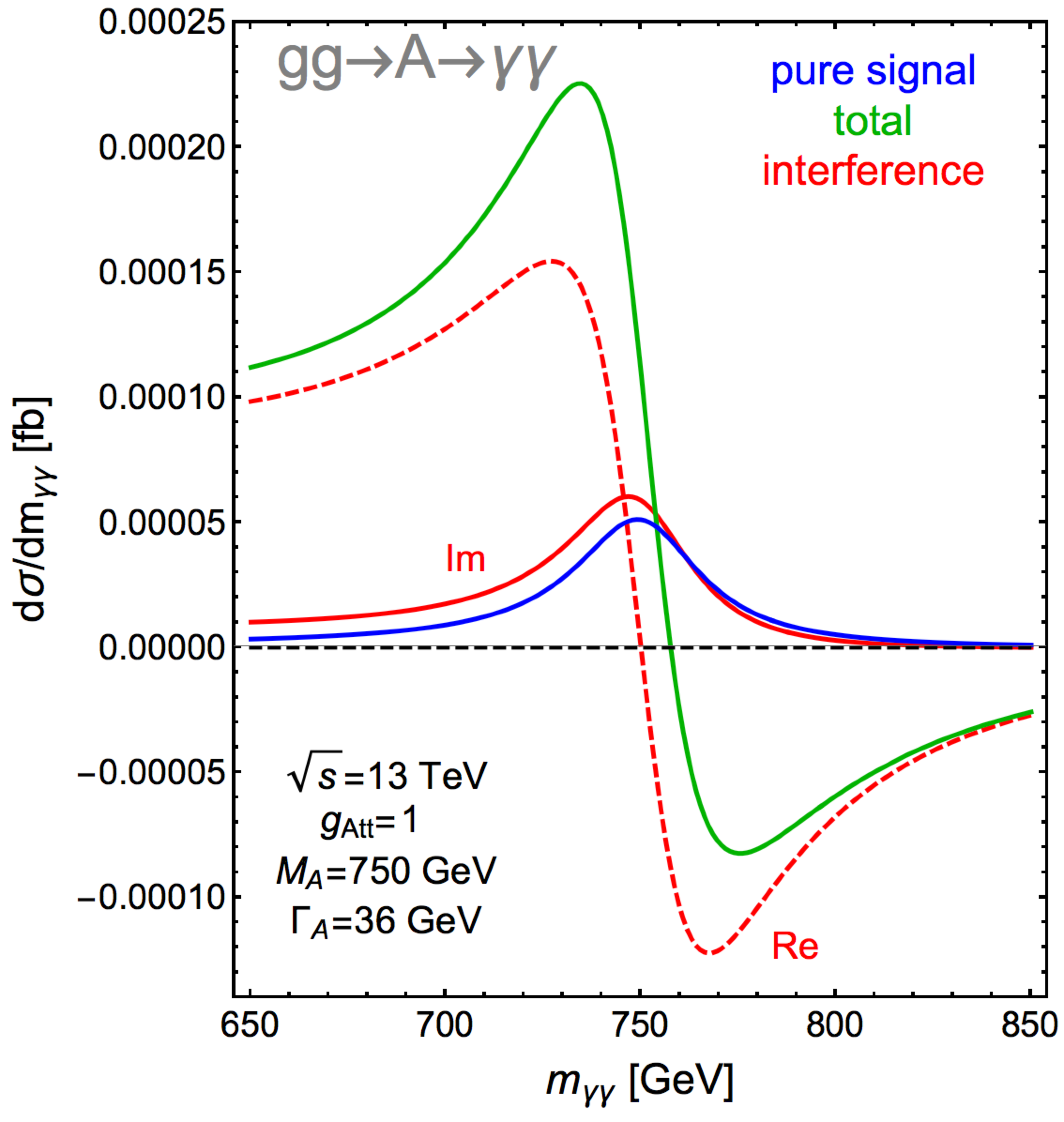}~~~ 
\includegraphics[scale=0.38]{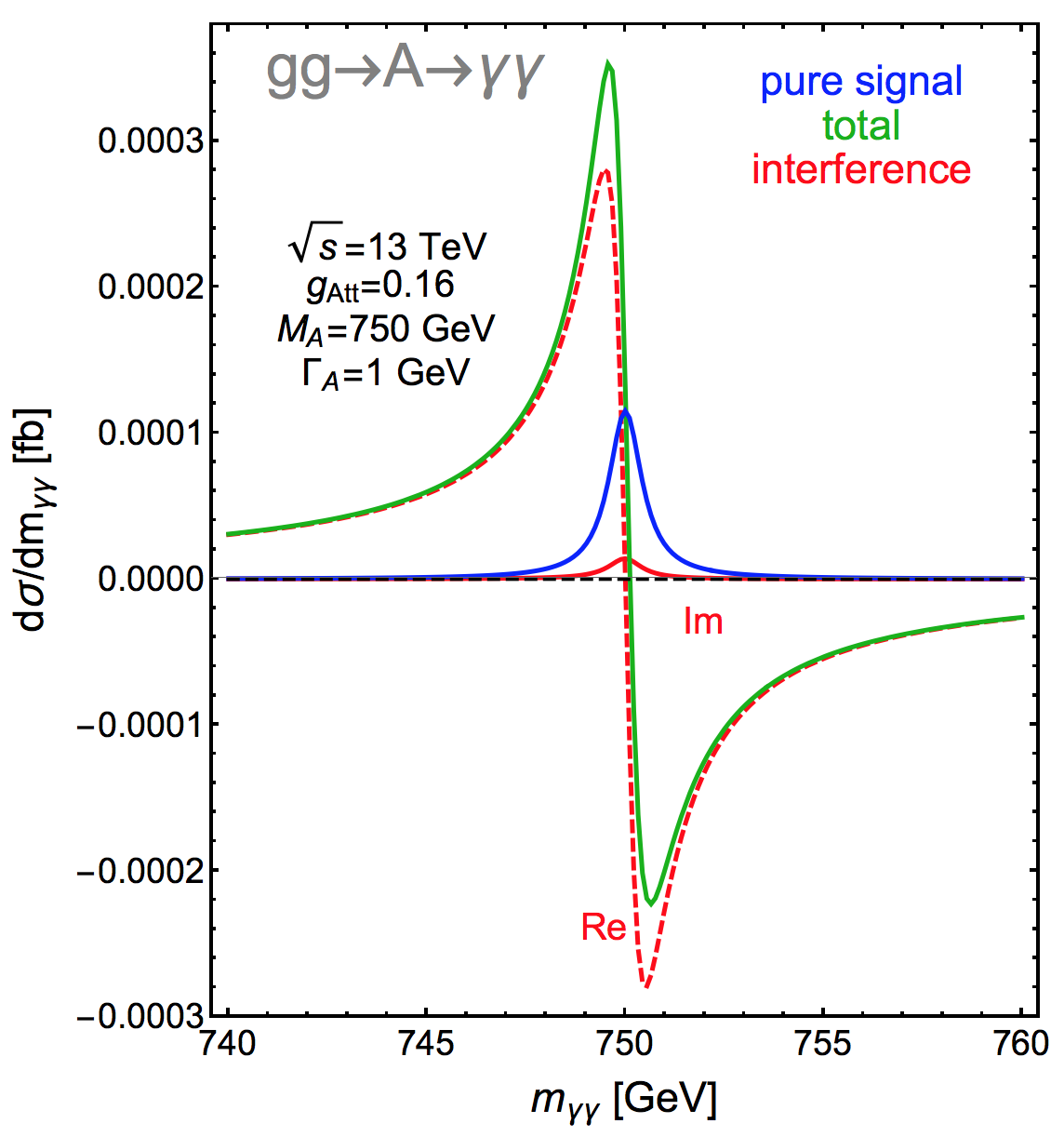}} 
\caption{\it The contributions to the line-shapes of a CP--odd $A \to \gamma \gamma$ with mass 750~GeV and total width $\Gamma_A = 36$~GeV (left panel) and $\Gamma_A = 1$~GeV (right panel), as functions of $m_{\gamma \gamma}$, showing the line-shape neglecting interference (solid blue lines), the contributions of interferences in the real and
imaginary parts of the $gg \to A \to \gamma \gamma$ amplitude (dashed and solid lines) and the overall combination including both interferences (green lines). These plots were calculated including only Standard Model fermion loops in the $gg \to A$ and $A \to \gamma \gamma$ couplings.}
\label{fig:Agaga}
\vspace*{-2mm}
\end{figure}

Fig.~\ref{fig:Agaga} shows the corresponding cases of the line-shapes of a CP--odd $A \to \gamma \gamma$ with nominal mass 750~GeV, assuming a total width $\Gamma_A \approx \Gamma(A \to t {\bar t}) = 36$~GeV (left panel) and $\Gamma_A \approx \Gamma(A \to t {\bar t}) = 1$~GeV (right panel). The overall results are qualitatively similar to those for the CP--even $H$ case in Fig.~\ref{fig:Hgaga}, though in the CP--odd $A$ case the interferences in the imaginary parts of the $gg \to (A \to) ~ \gamma \gamma$ amplitude are less important, and those in the real parts more important. As in the CP--even $H$ case, there are large enhancements of the line-shape compared to the calculation neglecting interference, 
but the overall magnitude is again much smaller than suggested by the 750-GeV data.

\begin{figure}[!h]
\centerline{\hspace*{-.5cm}
\includegraphics[scale=0.38]{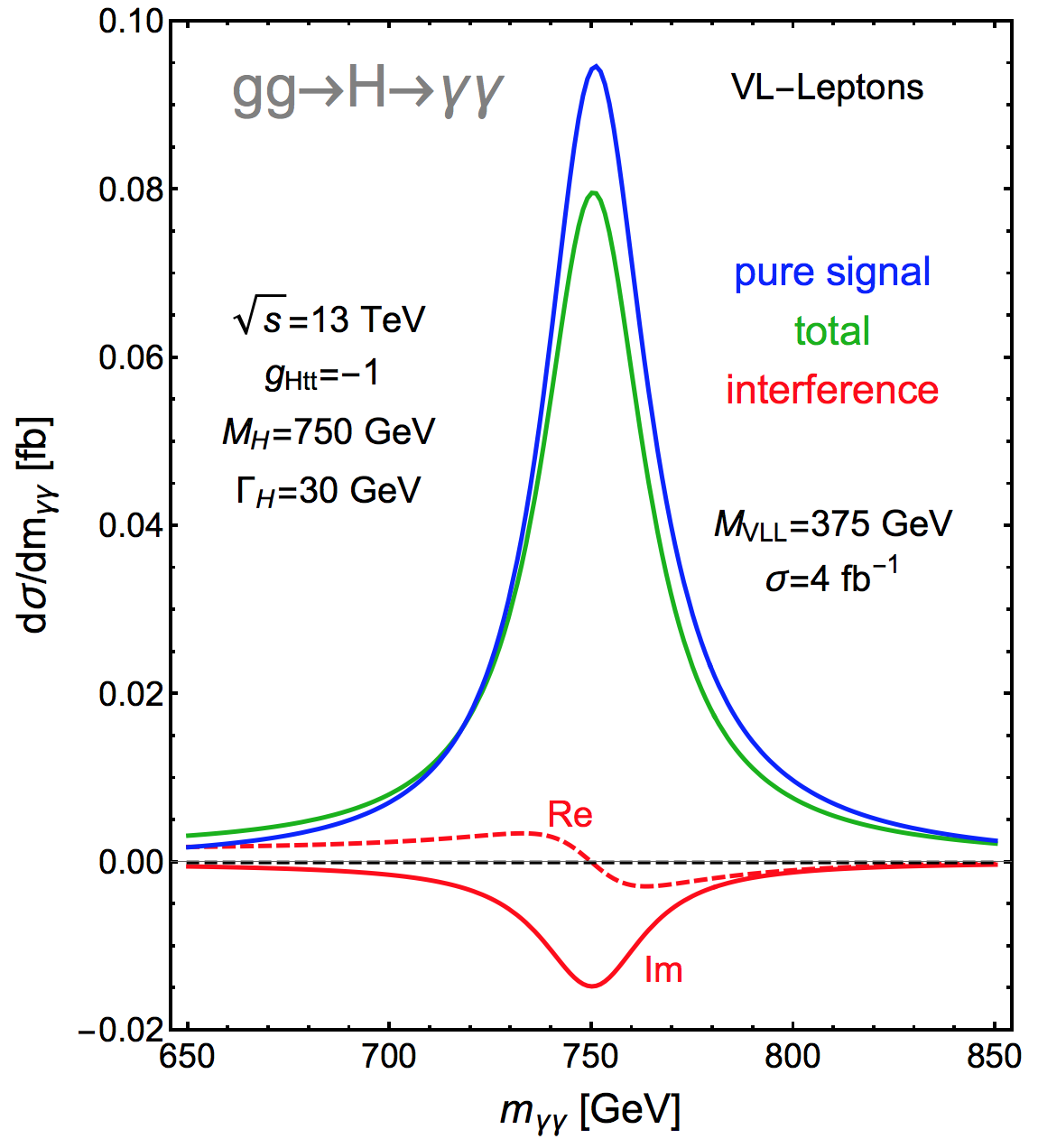}~~~ 
\includegraphics[scale=0.365]{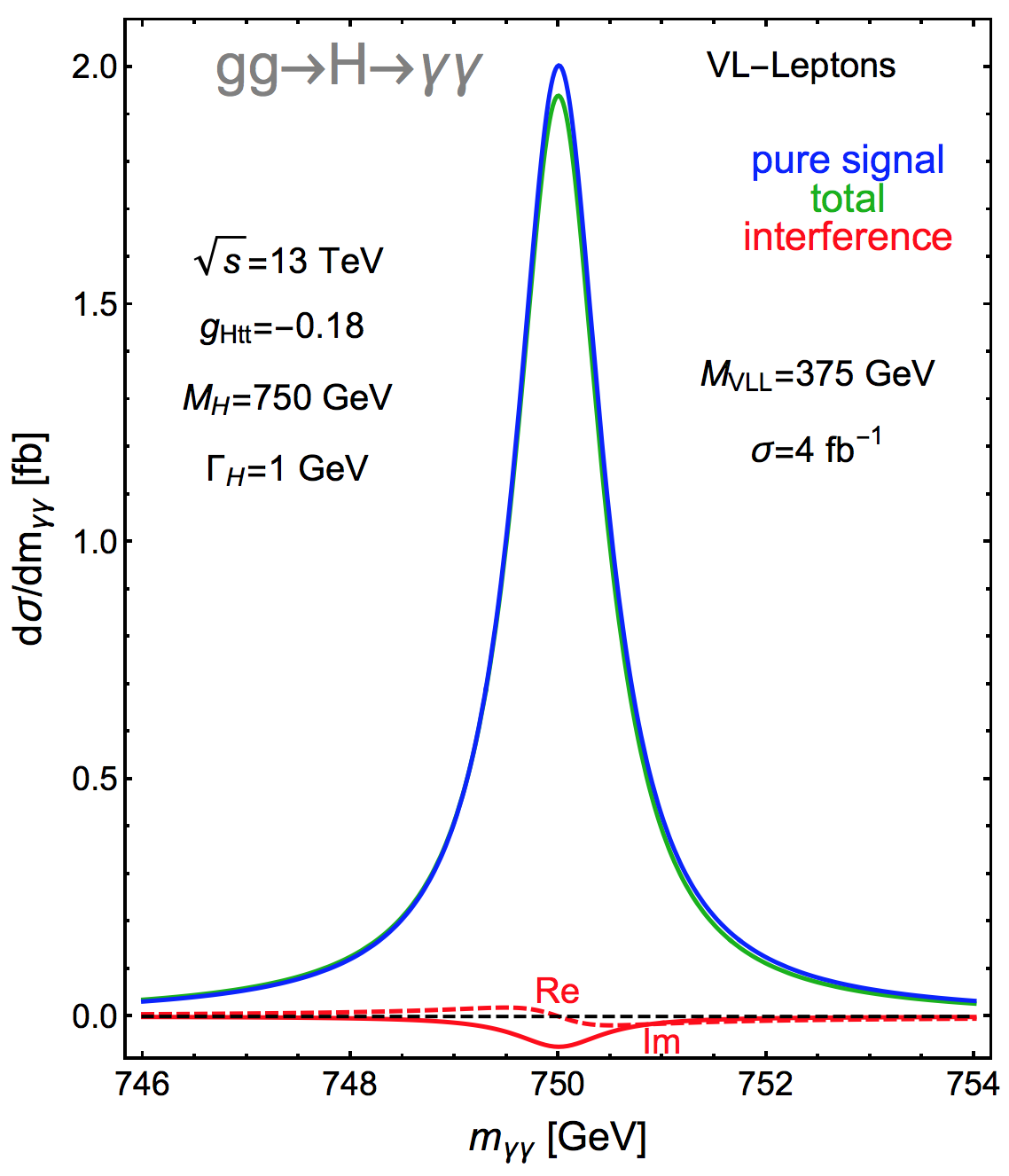}} 
\caption{\it The contributions to the line-shapes of a CP--even $H \to \gamma \gamma$ with mass 750~GeV and total width $\Gamma_H = 30$~GeV (left panel) and $\Gamma_H = 1$~GeV (right panel), as functions of $m_{\gamma \gamma}$, showing the line-shape neglecting interference (solid blue lines), the contributions of interferences in the real and imaginary parts of the $gg \to H \to \gamma \gamma$ amplitude (dashed and solid lines) and the overall combination including both interferences (green lines). These plots were calculated assuming sufficient vector-like leptons to give $\sigma(gg \to H) \times {\rm BR}(H \to \gamma \gamma) = 4$~fb.}
\label{fig:VLLH}
\end{figure}

In a next step, we consider the inclusion of massive vector-like fermions in the signal loop diagrams, in order to enhance the possible $gg \to \Phi \to \gamma \gamma$ signal
to the level where the diphoton cross section reaches the level of $\sigma( gg \to
\Phi) \times {\rm BR}(\Phi \to \gamma \gamma) =4$ fb as suggested by the data at the LHC with $\sqrt s=13$ TeV. In the case of a scalar state $H$ with total width $\Gamma_H=30$~GeV, in order to obtain $\sigma(gg \to \Phi) \times {\rm BR}(\Phi \to \gamma \gamma) \simeq 4$~fb, one needs an enhancement by a factor $\approx  90$ in the product of the $gg\to H$ and 
$H\to \gamma \gamma$ amplitudes given  in eq.~(\ref{A12}), compared to the contribution of the top quark alone. The corresponding enhancement for $\Gamma_H = 1$~GeV would be about 
factor 75, relative to the reduced $H t {\bar t}$ coupling required in this case. One minimal possibility would be to postulate extra vector-like leptons $L$, whose effects are maximized if their masses $M_L \simeq \frac12 M_\Phi$  as 
can be seen from Fig.~1 where the loop factors are shown. We consider this possibility in Fig.~\ref{fig:VLLH}: similar results would be found if responsibility for the enhancement were shared between vector-like quarks and leptons. The effect of such vector-like leptons, assumed to 
be heavier than $\frac12 M_\Phi$ in order not to contribute to the total width, is to increase by a large factor the real part of the product of amplitudes, leaving the imaginary part unchanged. 

However, the dominant contribution to $\sigma(gg \to \Phi) \times {\rm BR}(\Phi \to \gamma \gamma)$ is now provided by the square of the real part of the amplitude, and the interference between this real part and the background is relatively less important, as is the interference in the imaginary part. Note that the different sign of the interferences between the $H$ and 
$A$  cases is simply due to the different signs of the $\Phi t \bar t$ couplings (this might change if new quarks are included in the $\Phi gg$ loop). The net result for $\Gamma_H = 30$~GeV, shown in the left panel of Fig.~\ref{fig:VLLH}, is that the signal strength is reduced by $\sim 20$\% compared to the value that would be found neglecting interference. There would be an analogous, but much smaller, reduction in the case of a narrow total width $\Gamma_H = 1$~GeV, shown in the right of Fig.~\ref{fig:VLLH}.

\begin{figure}[!h]
\centerline{ \hspace*{-.5cm}
\includegraphics[scale=0.38]{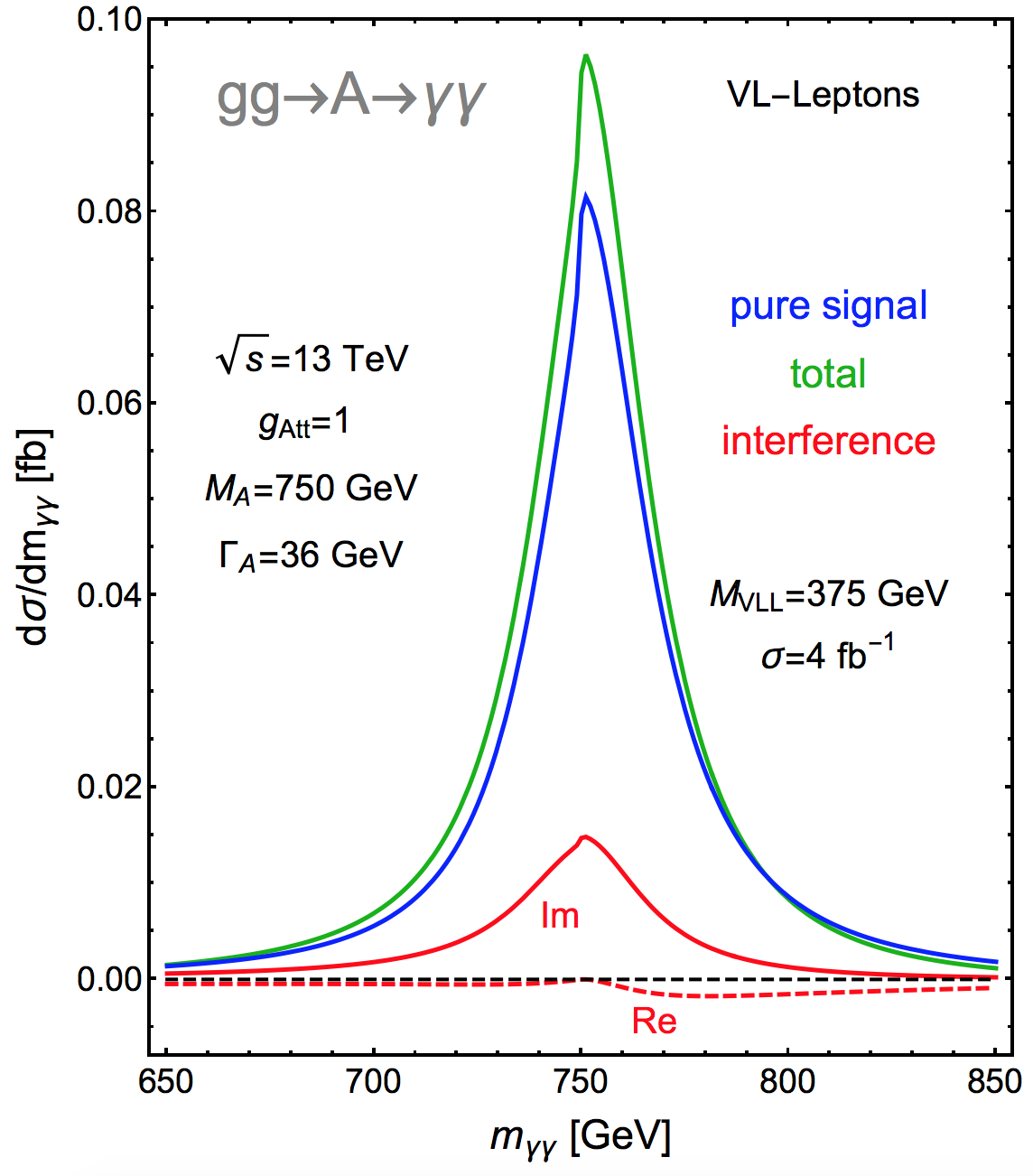}~~ 
\includegraphics[scale=0.186]{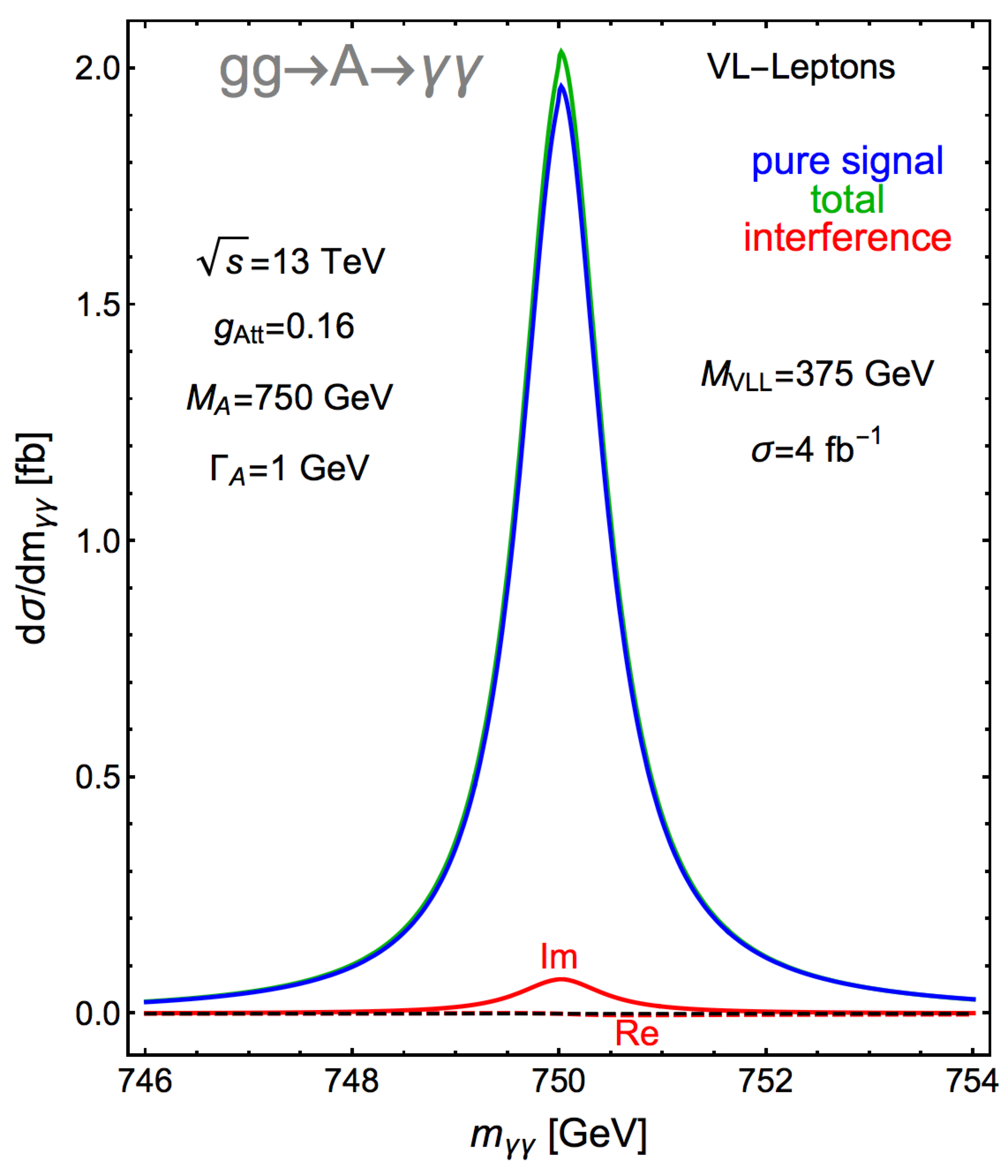}} 
\vspace*{-.2cm}
\caption{\it As in Fig.~\protect\ref{fig:VLLH}, but for the cases of a CP--odd $A \to \gamma \gamma$ with mass 750~GeV and total width $\Gamma_A = 30$~GeV (left panel) and $\Gamma_A = 1$~GeV (right panel). These plots were calculated assuming sufficient vector-like leptons to give $\sigma(gg \to A) \times {\rm BR}(A \to \gamma \gamma) = 4$~fb.}
\label{fig:VLLA}
\end{figure}

Analogous results for a pseudoscalar state $A$ with mass 750~GeV and in the same conditions than the previous CP-even $H$ case are shown in Fig.~\ref{fig:VLLA}. We see that the interference in the imaginary part is positive in this case, leading to an enhancement of the total cross section by $\sim 20$\% for a  wide state with a total width $\Gamma_A = 30$~GeV (left panel).
There is an analogous but much smaller enhancement in the narrow width case with $\Gamma_A = 1$~GeV (right panel).

\begin{figure}[!h]
\centerline{\hspace*{-.5cm}
\includegraphics[scale=0.38]{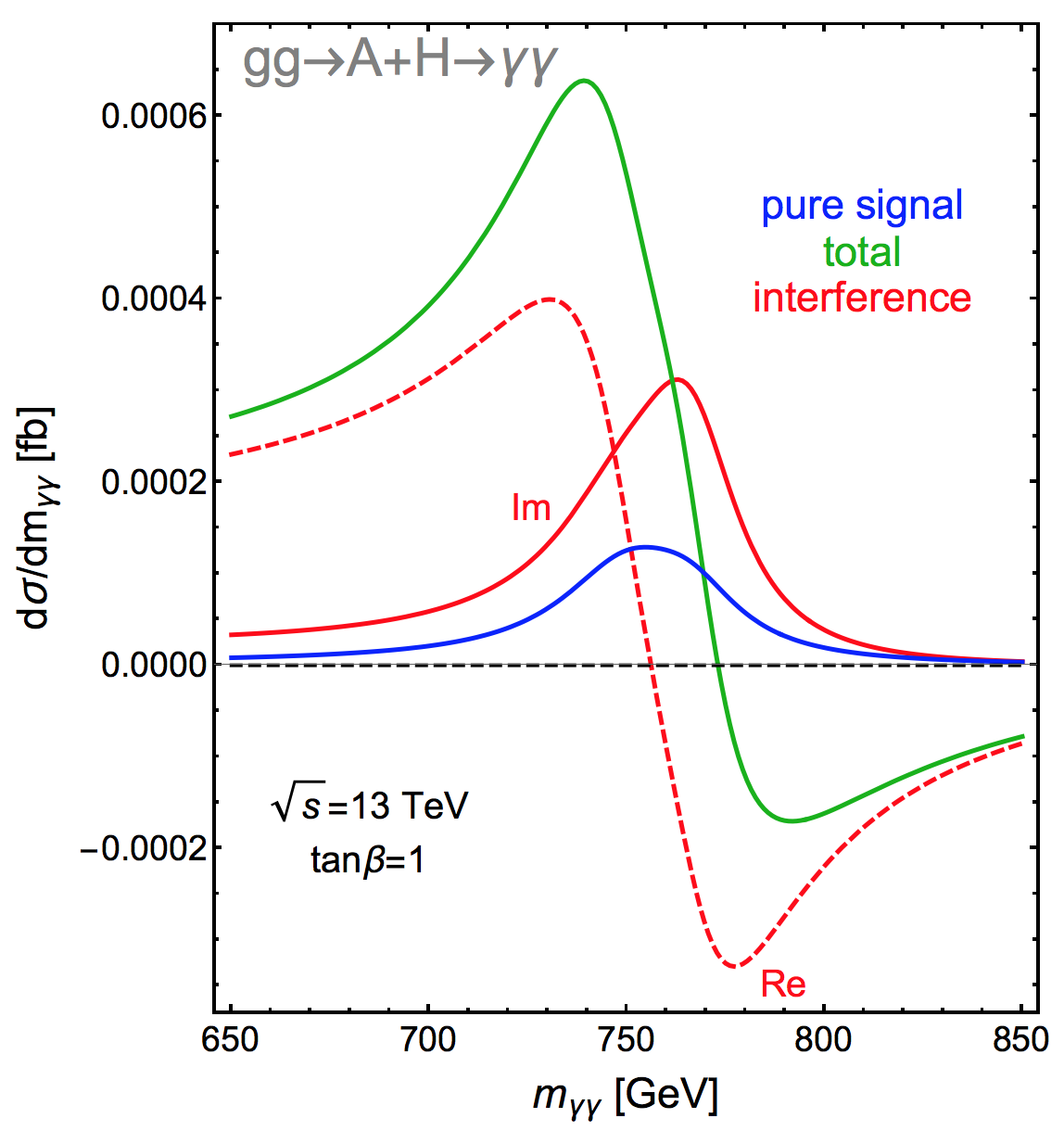}~~~ 
\includegraphics[scale=0.356]{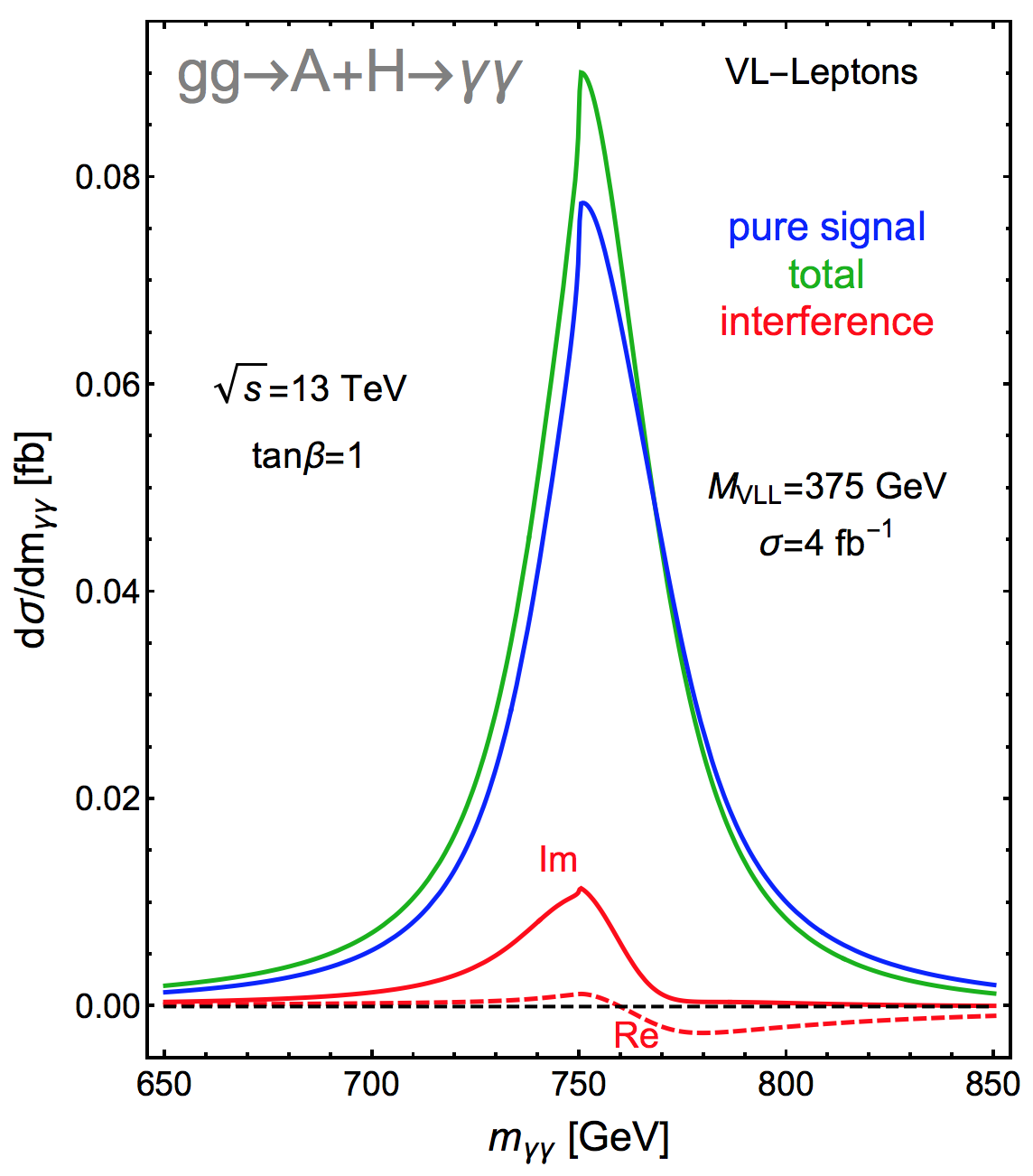}} 
\vspace*{-.2cm}
\caption{\it The contributions to the combined $H+A \to \gamma \gamma$ line-shape in the 2HDM with $M_A = 750$~GeV, $\Gamma_A= 36$~GeV and $M_H = 766$~GeV, $\Gamma_H = 33$~GeV, as functions of $m_{\gamma \gamma}$, showing the line-shapes neglecting interference (solid blue lines), the contributions of interferences in the real and imaginary parts of the $gg \to H \to \gamma \gamma$ amplitudes (dashed and solid red lines) and the overall combinations including both interferences (green lines). The left panel is when only standard fermions are included in the $gg\Phi$ and $\Phi \gamma \gamma$ couplings, whereas the right panel includes vector-like leptons to give $\sigma(gg \! \to \! \Phi) \! \times \! {\rm BR}(\Phi \! \to \! \gamma \gamma) \! = \! 4$~fb.}
\label{fig:VLLHA}
\vspace*{-5mm}
\end{figure}

Finally, our results for the $gg \to \Phi \to \gamma \gamma$ mass spectrum in the 2HDM 
with $\tan \beta = 1$ are shown in Fig.~\ref{fig:VLLHA} when the combined effects of 
the $H$ and $A$ states are considered. We see that, if only Standard Model fermion loops are included in the $gg \to \Phi$ and $\Phi \to \gamma \gamma$ couplings (left panel), there is a significant enhancement in the peak, which is shifted below 750~GeV,  accompanied by a (smaller) dip above 750~GeV. However, as in previous cases with only Standard Model fermion loops, the peak is still much smaller than the reported signal. On the other hand, there are sufficient vector-like fermions to enhance the signal to 4~fb as reported by ATLAS and CMS (right panel), the enhancement is much smaller, namely about 20\%.

\subsection{Extension to the $\mathbf{Z\gamma}$ Process}

Before closing this Section, we make a few remarks on the other diboson channels
that are possible for the $\Phi$ state(s), namely $\Phi \to \gamma Z, ZZ$ and $WW$. For 
the specific case of the decay $\Phi \to Z\gamma$, the situation is very similar to that of the  $\Phi \to \gamma \gamma$ decay, in particular if the $Z$ boson mass in the final state is neglected compared to the invariant mass $m_{Z\gamma}$,
which is justified for the range of interest close to $M_\Phi \approx 750$~GeV, 
where $M_Z^2/M_\Phi^2 \approx 0.015 \ll 1$.  In this case, the total amplitude of the  $gg \to (\Phi \to ) Z\gamma$ process, including the continuum and the  resonant  contributions, can be simply written as 
 \beq
{\cal A} = - \sum_\Phi \frac{{\cal A}_{gg \Phi} {\cal A}_{\gamma Z \Phi}}{\hat s - M_\Phi^2 + i M_\Phi \Gamma_\Phi} + {\cal A}_{gg\gamma Z} \, , 
\label{eq:ggpZ}
\eeq
similarly to eq.~(\ref{eq:ggpp}) for the $gg\to (\Phi \to ) \gamma \gamma$ process. Here again, the sum in the first term runs over the $\Phi=H,A$ states and the second term describes the box diagram contribution of the $gg\to Z\gamma$ QCD background which is given by an amplitude similar to that of  eq.~(\ref{eq:Aggaa}) \cite{ZgammaB}
\beq
{\cal A}_{gg\gamma Z} = 2 \alpha_s  \alpha   \sum_q e_q v_q A_q \, . 
\label{eq:Aggaa}
\eeq
where the sum that runs over the six standard quark flavors, $q=u,d,s,c,b,t$ and the 
amplitude $A_q$ is given in eq.~(\ref{eq:Mpppp}) in the massless $Z$ boson limit. The only difference with the $\gamma \gamma$ case is that now, one of the charges $e_q$ has to be replaced by the vector part of the $Z q\bar q$ coupling given, in the general case of a fermion $F$ with a third component of the left--and right--handed isospin $I_f^{3L,3R}$ and and electric charge $e_F$, by  
\beq
v_F^Z \equiv v_F = (2I_{3L}^F+2I_{3R}^F- 4e_F s_W^2)/ ( 4s_W c_W)
\label{eq:ZFF}
\eeq
where  $s_W^2=1-c_W^2 \equiv\sin^2 \theta_W$. The axial--vector couplings of the $Z q\bar 
q$ coupling do not contribute in the box diagrams.  Hence, the relative weight of 
the $gg \to Z\gamma$ box contribution at the amplitude level, compared to the $gg\to \gamma\gamma$ case is simply given by $\sum_q e_q v_q / \sum_q e_q^2 \approx 1/2$.

Turning to the signal process $gg\to \Phi \to \gamma Z$,  the $\Phi \to  Z \gamma$ decay amplitude in the triangle diagrams should also contain the vectorial part  of the $Z\bar F F$ coupling of the vector--like fermions to the $Z$ boson (here also the axial--vector couplings do not contribute, and they are anyway absent in the case of vector--like fermions) 
\beq 
{\cal A}_{\gamma Z \Phi } = \frac{\alpha}{4 \pi v} \hat s \sum_{F}  N_c^F e_F v_F \, \hat g_{\Phi FF} \, A_{1/2}^\Phi (\hat \tau_F) \, ,
\eeq
where the  form factors for the contributions of spin--$\frac12$ fermions $A_{1/2}^{\Phi}$
can be found in Ref.~\cite{ZgammaS}. In the massless $Z$ boson approximation $M_Z^2/M_\Phi^2 \to 0$ it reduces to the expression of eq.~(\ref{eq:formfactors}) of the $\Phi \to \gamma
\gamma$ case. Here again, one has in general $ |v_F | < |e_F|$ for the vector--like 
fermions (for instance $v_E \approx 0.64\, e_E$ for a vector-like lepton with a charge $-e$ and isospin $-1/2$) and hence, the signal amplitude is suppressed by a factor that
is  similar to the one suppressing the background amplitude. This makes the situation for the signal/background interference quite similar to the previously discussed $gg \to \gamma\gamma$ case. 

\begin{figure}[!h]
\centerline{\hspace*{-.5cm}
\includegraphics[scale=0.38]{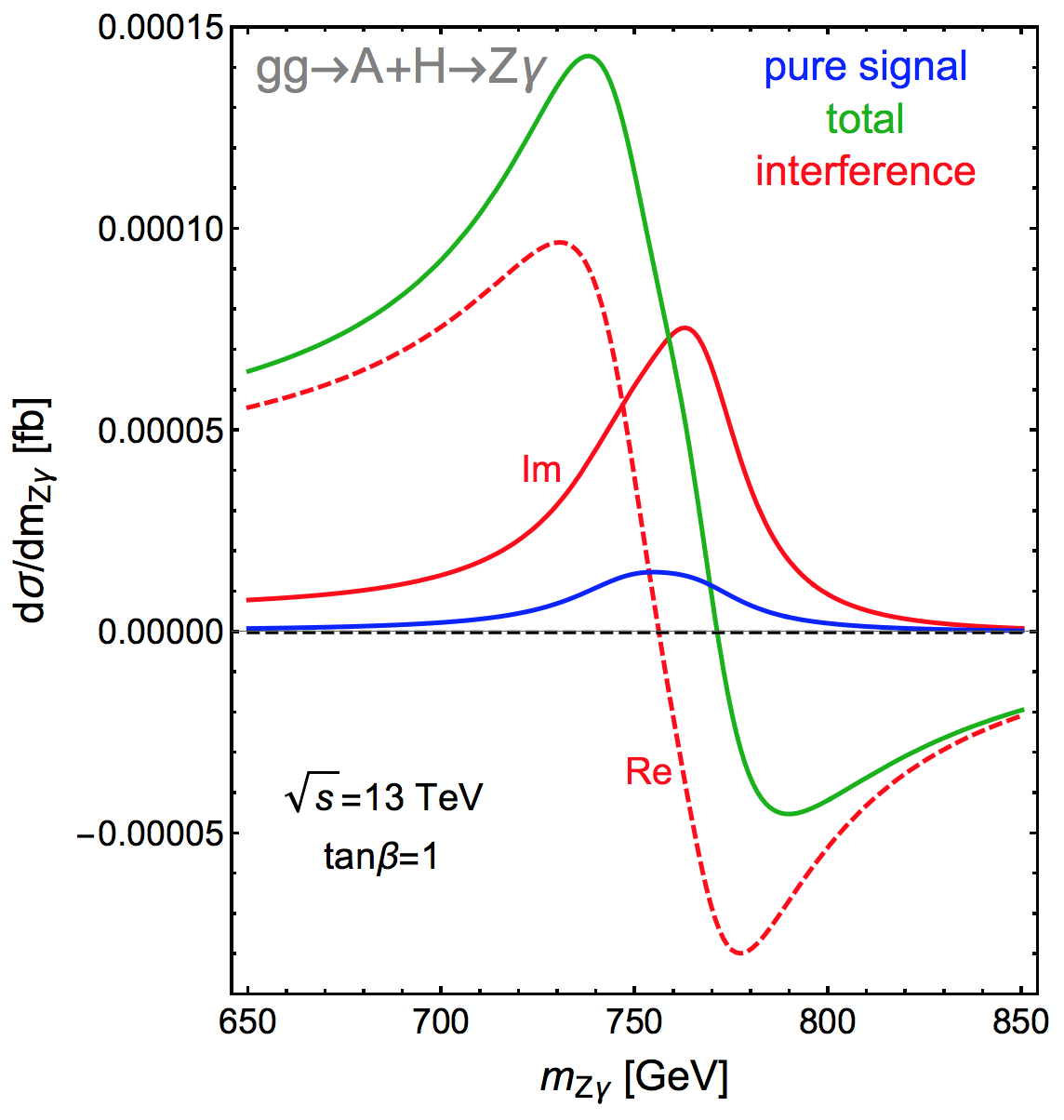}~~~ 
\includegraphics[scale=0.35]{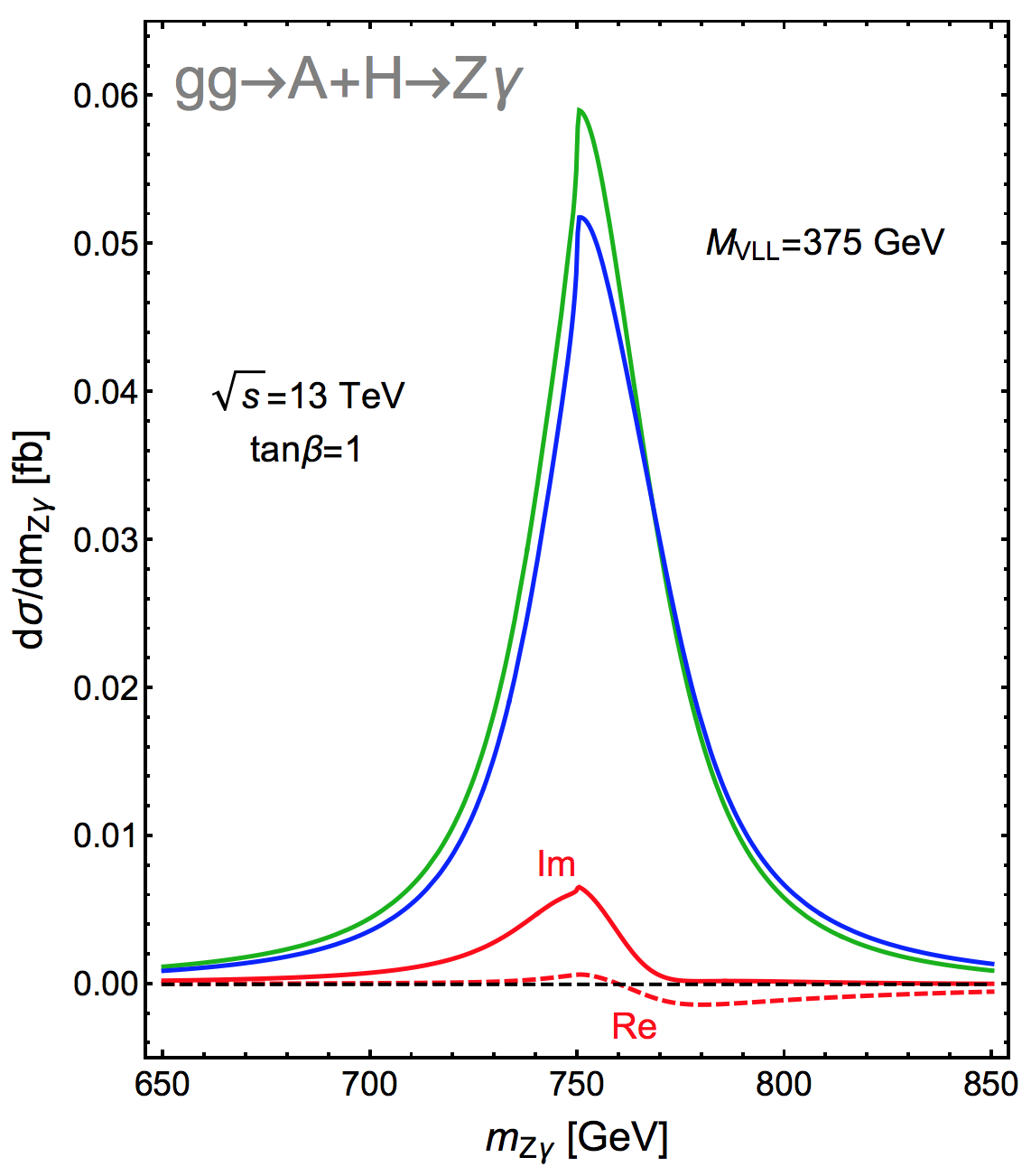}} 
\caption{\it In the $gg \to (\Phi \to) ~ Z\gamma$ process, the contributions to the combined $H+A \to Z \gamma$ line-shape in the 2HDM with $M_A = 750$~GeV, $\Gamma_A= 36$~GeV and $M_H = 766$~GeV, $\Gamma_H = 33$~GeV, as functions of $m_{Z\gamma}$, showing the line-shapes neglecting interference, the contributions of interferences in the $gg \to A/H \to \gamma \gamma$ amplitudes and the overall combinations including both interferences. The left panel is when only standard fermions are included in the $gg\Phi$ and $\Phi \gamma Z$ couplings, whereas the right panel includes the vector-like leptons that are needed to give $\sigma(gg \! \to \! \Phi) \! \times \! {\rm BR}(\Phi \! \to \! \gamma \gamma) \! = \! 4$~fb at the 13 TeV LHC.}
\label{fig:Zgam}
\vspace*{-2mm}
\end{figure}

This is exemplified in Fig.~\ref{fig:Zgam} where the contributions to the combined $H+A \to Z \gamma$ line-shapes in our usual 2HDM scenario with $M_A = 750$~GeV, $\Gamma_A= 36$~GeV and $M_H = 766$~GeV, $\Gamma_H = 33$~GeV, as shown as functions of $m_{Z \gamma}$
and where the signal, background and interference are displayed in two cases: 
when only standard fermions are included in the $gg\Phi$ and $\Phi \gamma Z$ loops (left)
and when the vector-like leptons that are needed to reproduce the LHC diphoton
data are included in the $\Phi \to \gamma Z$ decay (right). As can be seen, compared to
the corresponding $gg \to \Phi \to \gamma  \gamma$ case shown in Fig.~\ref{fig:VLLHA},
the trend is very similar except for the overall normalisation. Hence, as expected,
interference effects in the $\Phi \to Z\gamma$ channel have similar impact  as in the 
$\Phi \to \gamma \gamma$ mode.

This statement can be generalized to the two other possible decay channels
of the $\Phi$ state, namely $\Phi \to ZZ,WW$. This is true not only for a singlet
resonance but also for a $\Phi$ state of a 2HDM in the alignment limit as, in both 
cases, the $\Phi WW$ and $\Phi ZZ$ amplitudes are loop-induced  (there are no tree--level $HWW,HZZ$ couplings) by the same fermions that generate the $\Phi \gamma \gamma$ and $\Phi Z\gamma$ couplings. Also
in these cases, one can neglect the $W$ and $Z$ masses compared to that of the $\Phi$ state,
$M_{W,Z}^2/M_Z^2$ so that the same formalism introduced in the previous subsections
also applies here. Hence, qualitatively the situation should be similar to the one
discussed here. The study of the possible  numerical differences is postponed to a future publication~\footnote{We should note that, for instance, the interference in $gg\! \to\! 
\Phi \! \to \! ZZ, WW$ will affect the analyses that attempt to determine the total decay width of the standard--like $h$ state in these channels \cite{combo}.}  \cite{DEQ2}.  


\section{Interference in $\mathbf{gg \to (\Phi \to) ~ t\bar t}$}

\subsection{Formulation}

\begin{figure}[!h]
\vspace*{-3.cm}
\centerline{ \includegraphics[scale=0.86]{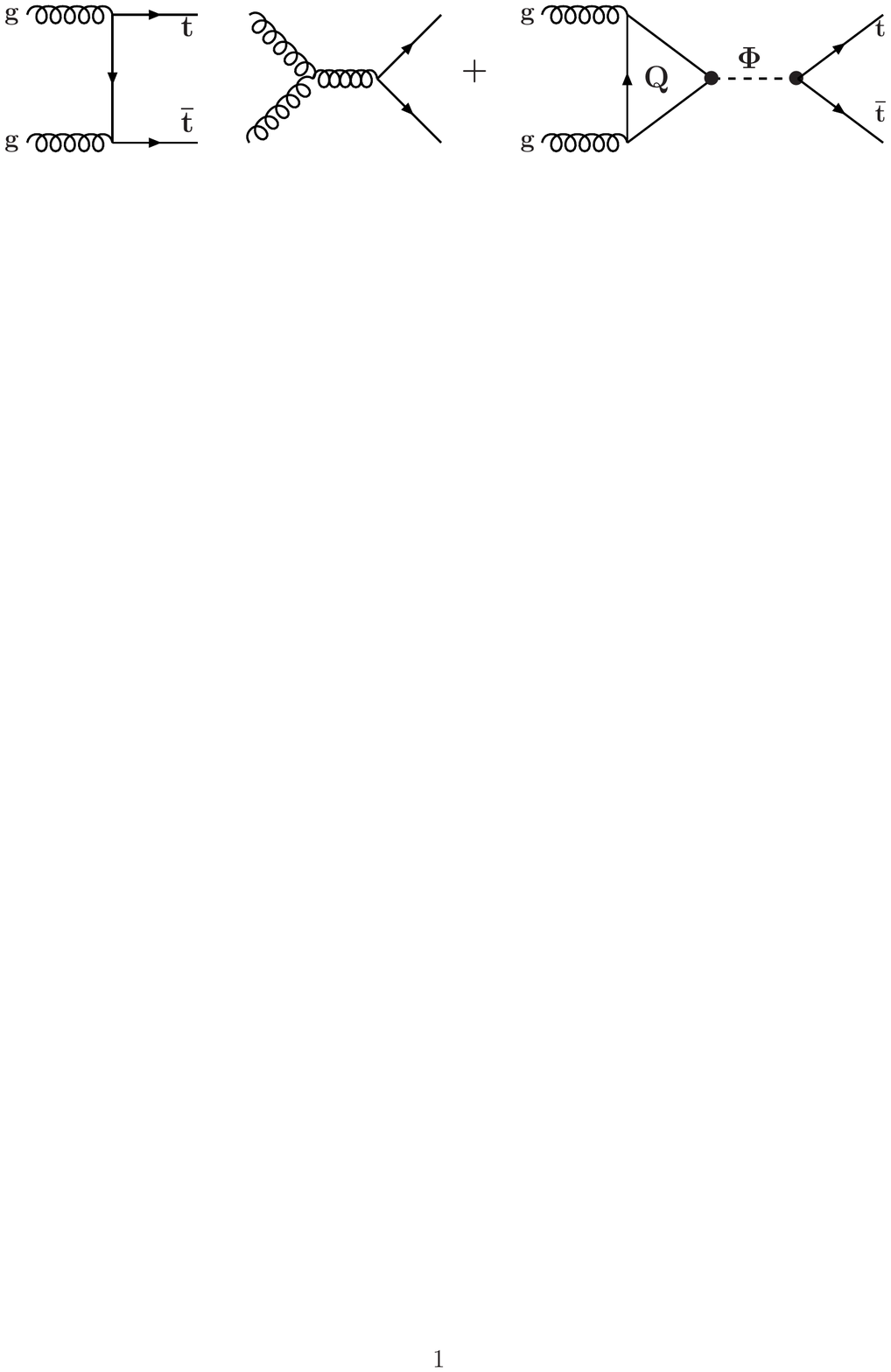} }
\vspace*{-20.1cm}
\caption{\it Leading-order Feynman diagrams for the continuum QCD background  (left) and the resonant $\Phi$ signal (right) in the process $gg \to (\Phi \to) ~ t\bar t$.} 
\label{fig:feynmantt}
\vspace*{-2mm}
\end{figure}

We turn now to $t\bar t$ pair production, for which the leading-order Feynman diagrams for the signal $gg \to \Phi \to t\bar t$ and the QCD background $gg \to t\bar t$ are shown  in Fig.~\ref{fig:feynmantt}. In this case, the situation is completely different from the $gg \to \gamma\gamma$ process in which both the signal and the background were loop-induced and hence comparable in magnitude. For $t\bar t$ production, whereas the $\Phi$ production mechanism $gg\to \Phi$ is the same as in the previous case, 
the background process occurs already at tree--level and has a rate that is much larger than the signal rate. In fact, at the LHC with $\sqrt s= 13$ TeV, 
the $pp \to t\bar t$ process has a cross section  of about 820 pb \cite{tt-NNLO} for a mass $m_t=173$ GeV, 
using the MSTW set of PDFs \cite{MSTW} that we adopt here. The rate is mainly generated by the $gg$-initiated subprocess, 
the contribution of the $q\bar q \to t\bar t $ component being only about 15\% at the above energy. Instead, the signal cross section in the 2HDM is 
$\sigma(gg\to\! H\! +\! A \! \to \! t \bar t ) = 2$~pb at $\sqrt s=13$ TeV, in the optimal case where $\tb=1$ and both $H$ and $A$ have 
masses of about 750 GeV and branching ratios close to unity for their decays into $t \bar t$ final states. 

Hence, although only a small fraction of the background occurs at an invariant mass around  $M_{t \bar t} \approx 750$ GeV, it a formidable task to discriminate between the signal and the background. This is particularly true as, contrary to the previous $pp\to \gamma\gamma$ case, the experimental resolution for $t\bar t$ final states is large and is comparable to the maximal total width expected for the $\Phi$ signal in the 2HDM, $\Gamma_\Phi \approx 45$ GeV. 
Nevertheless, searches for resonances decaying into $t\bar t$ final states have been conducted by ATLAS \cite{ATLAS-tt} and CMS \cite{CMS-tt} and interpreted in various scenarios, although mainly for spin--1 and spin--2 resonances where interference effects do not occur~\footnote{In these cases, the cross sections come from the $q\bar q $ initial state and, because one is dealing with electroweak particles, there is no interference with the colored $q\bar q \to t \bar t$ background. Therefore, in the cases of such resonances, one simply expects an excess or a peak on top of the continuum background.}. They set strong constraints on the cross sections of the resonances that need to be taken into account.

Coming to the description of the process and following the discussion of Section~3, the amplitude in the case of the $gg (\to \Phi ) \to t\bar t$ process, when the contributions of 
resonant signal process and the continuum backgrounds are added, is given by
\beq
{\cal A}^{\Phi }_{gg \to t\bar t} &=&    
- \sum_\Phi   \frac{ {\cal A}_{gg\Phi}   \, \hat s {\cal A}_{\Phi tt} } { \hat s -M_\Phi^2 +i 
M_\Phi \Gamma_\Phi }  +  {\cal A}_{ggtt} \, .
\eeq
The amplitude ${\cal A}_{gg\Phi}$ for the production $gg\to \Phi$ has been given before 
in (\ref{A12}). While one can write the relevant helicity  amplitudes for the signal, the background and their interference in a way similar to the $gg \to \gamma\gamma$ case, 
the partonic differential cross section can be written in a more convenient way as 
\beq 
\frac{ {\rm d}\hat \sigma }{{\rm d} z} = \frac{ {\rm d}\hat \sigma_B }{{\rm d} z}
+ \frac{ {\rm d}\hat \sigma_S }{{\rm d} z} + \frac{ {\rm d}\hat \sigma_I }{{\rm d} z} \, ,
\eeq
where again $z=\cos\theta$ with $\theta$ the scattering angle. The various components, in terms 
of  the velocity of the final top quark at the partonic level $\hat \beta_t = \sqrt{1-4m_t^2 /\hat{s}}$ read \cite{Htt0,Htt1,tt-LO}
\beq
 \frac{ {\rm d}\hat \sigma_B }{{\rm d} z} &=&  \frac{\pi \alpha_s^2}{6 \hat s }  \hat \beta_t \left( \frac {1}{1- \hat \beta_t^2 z^2} -\frac{9}{16}  \right) \bigg[ 3+ \hat \beta_t^2 z^2 - 2 \hat \beta_t^2  - \frac{ 2 (1- \hat \beta_t^2)^2}{1- \hat \beta_t^2 z^2} \bigg] \, ,
\nonumber \\
 \frac{ {\rm d}\hat \sigma_S }{{\rm d} z} &=&  \frac{3 \alpha_s^2 G_F^2 m_t^2}{8192 \pi^3} 
\hat s^2  \sum_\Phi  \frac{ | \hat \beta_t^{p_\Phi} \hat g_{\Phi tt} \sum_Q \hat g_{\Phi QQ} 
A_{1/2}^\Phi (\hat \tau_Q) |^2} {(s- M_\Phi^2) ^2+ \Gamma_\Phi^2 M_\Phi^2}  \, , \nonumber \\
 \frac{ {\rm d}\hat \sigma_I }{{\rm d} z} &=&  - \frac{\alpha_s^2 G_F m_t^2}{64\sqrt 2  \pi} 
\frac {1}{1- \hat \beta_t^2 z^2} {\rm Re} \bigg[ \sum_\Phi \frac{ \hat \beta_t^{p_\Phi} 
\hat g_{\Phi tt} \sum_Q \hat g_{\Phi QQ} A_{1/2}^\Phi (\hat \tau_Q) }
{s- M_\Phi^2+i \Gamma_\Phi M_\Phi} \bigg] \, ,
\eeq
where $p_\Phi =3(1)$ for the CP--even (CP--odd) Higgs boson. The total cross sections for the signal, the background and interference are then obtained by integrating partonic cross sections over the scattering angle $\theta$ and folding them with the $gg$ luminosity, cf, (\ref{eq:cgg}). 

In this case too, the higher-order effects need to be included. The QCD corrections to the $gg\to \Phi$ production cross section were also discussed above, and lead to a $K$--factor of 1.8 at NLO, while those to the decay $\Phi \to t\bar t$ are known  to NNLO \cite{tt-NNLO,tt-NLO} and are $\sim 1.35$. At NLO, the $K$--factor in the case of the $pp\to t\bar t$ QCD background process $K_{\rm NLO}^{\rm QCD} \approx 1.3$ \cite{tt-NLO},  i.e., significantly smaller than that for the Higgs signal process. The NNLO QCD corrections to the $pp\to t\bar t$ process have been completed recently \cite{tt-NNLO}, and increase the total cross section only slightly beyond the NLO value.  The electroweak corrections are rather small in both the signal and background processes, and can be ignored to first approximation. As in the $gg\to \gamma\gamma$ case, we take account of the QCD corrections simply by rescaling the Higgs signal cross section as well as the interference term by the same NNLO correction factor, $K_{\rm NNLO}=2$.

We start our considerations of interference effects in $gg \to t {\bar t}$ by considering the case of a single state $\Phi$, which may be either a scalar $H$ or a pseudoscalar $A$. We note that there is, in principle, an ambiguity in the sign of the $t {\bar t} H$  ($t {\bar t} A$) coupling. These are fixed to be negative (positive) in the 2HDM, but either sign is possible for either coupling, in general. There is also the magnitude of the coupling to be considered. In the 2HDM with $\tan \beta = 1$, as discussed in Section 2, the magnitudes of the couplings are both unity when normalized relative to that in the Standard Model, and the decay widths are $\Gamma_H \approx \Gamma (H \to t {\bar t}) = 30$~GeV, $\Gamma_A \approx \Gamma (A \to t {\bar t}) = 36$~GeV for a nominal mass $M_{H, A} = 750$~GeV, the difference being due to the difference between p-wave and s-wave phase space, respectively. These are two of the benchmark cases for singlet models that we consider in the following. However, in a general singlet model, $\Gamma_H$ or $\Gamma_A$ is arbitrary, and we also consider alternative benchmark scenarios with $
\Gamma_H \approx \Gamma (H \to t {\bar t})$, $\Gamma_A \approx \Gamma (A \to t {\bar t}) = 1$~GeV, which require $|g_{H t \bar t}| = 0.18$, $|g_{H t \bar t}| = 0.16$, respectively.


\subsection{Numerical Results}

In the following Figures we show the results of calculations of the ratios 

\centerline{ (S + B)/B = (signal + background)/(background alone),} 

\noindent for each of these singlet scenarios, $H, A$, both broad with $\Gamma_{H,A} = 30, 36$~GeV and narrow with $\Gamma_{H, A} = 1$~GeV,  as well as analogous results for the 2HDM case with $\tan \beta = 1$ and $M_A = 750$~GeV, for which $\Gamma_A = 36$~GeV, $M_H = 766$~GeV and $\Gamma_H = 33$~GeV. Both ATLAS and CMS have published measurements of the $t {\bar t}$ cross section as a function of $M_{t {\bar t}}$, providing also values of the ratio of the data to smoothed fits to the background. The ATLAS 8-TeV data~\cite{ATLAS-tt} are more constraining for our purposes, so we focus on them in these and subsequent Figures. Their results  are displayed in our plots as ``Brazil" 1- and 2-$\sigma$ green and yellow bands. The data were used in~\cite{ATLAS-tt} to present upper limits on peaks above the background. However, in the models we study {the data are more significant for the constraints they impose on {\it dips} below the background level.}

\begin{figure}[!h]
\centerline{ 
\includegraphics[scale=0.4]{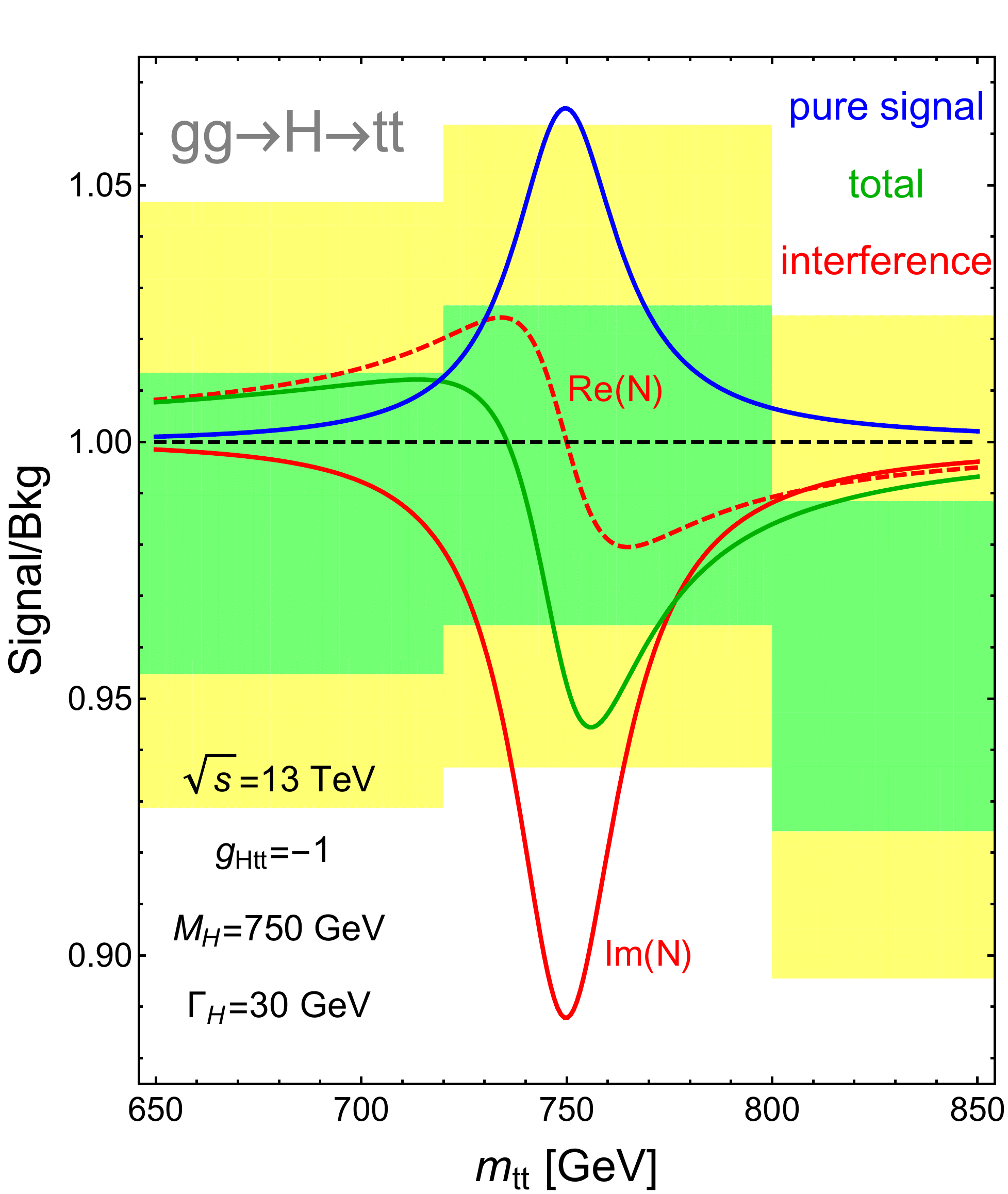} 
\includegraphics[scale=0.4]{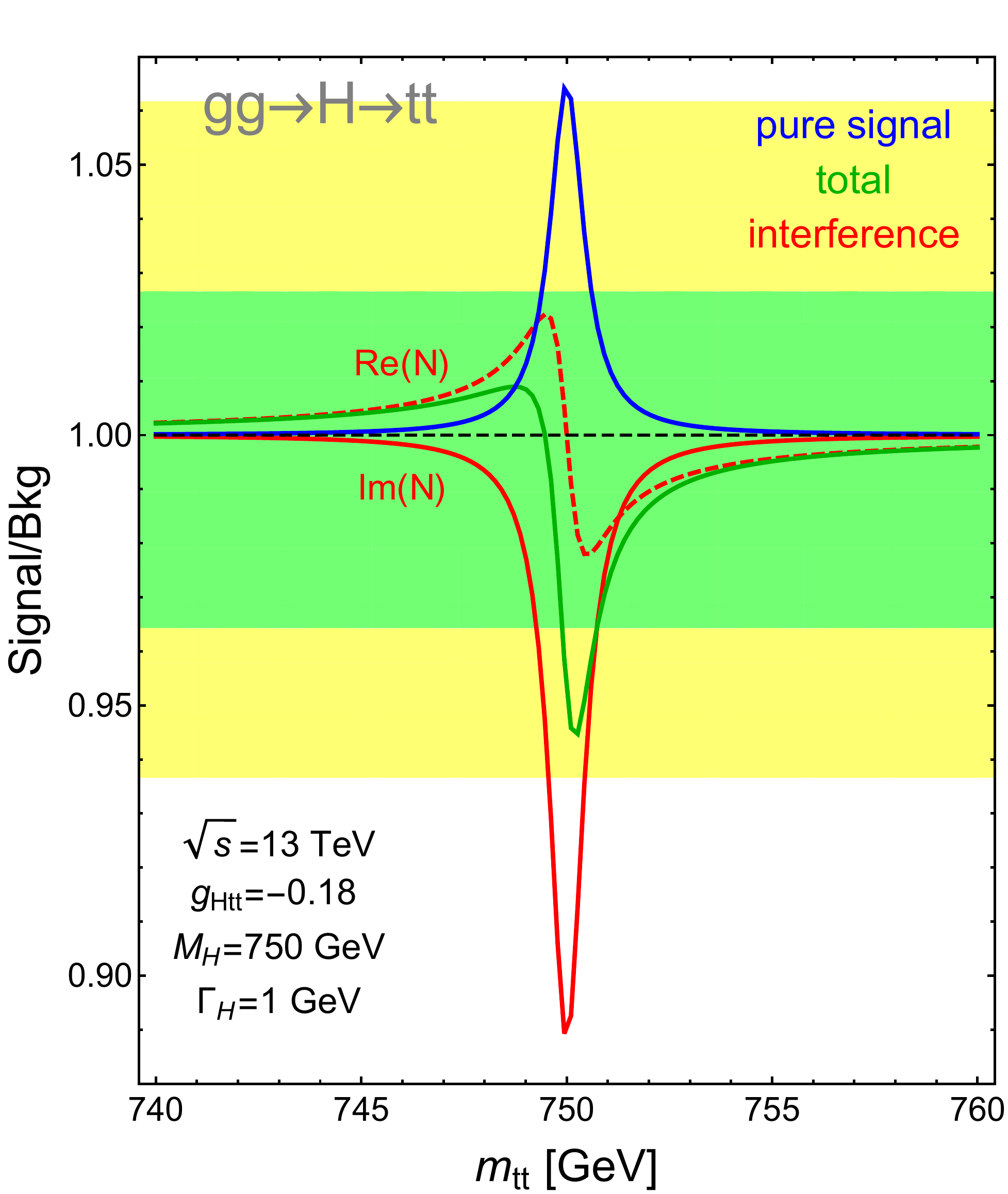}} 
\caption{\it The contributions to the line-shapes of a CP--even $H \to t {\bar t}$ with mass 750~GeV and total width $\Gamma_H = 30$~GeV (left panel) and $\Gamma_H= 1$~GeV (right panel), as functions of $m_{t {\bar t}}$, showing the line-shape neglecting interference (solid blue lines), the contributions of interferences in the real and imaginary parts of the $gg \to H$ amplitude (dashed and solid lines) and the overall combination including both interferences (green lines).}
\label{fig:Httbar}
\end{figure}

Fig.~\ref{fig:Httbar} shows our results for a singlet scalar $H$ with a unit-normalized coupling $g_{H t {\bar t}} = -1$, for which $\Gamma_H \approx \Gamma( H \to t\bar t) = 30$~GeV (left panel), and $g_{H t {\bar t}} = - 0.18$ for which $\Gamma_H = 1$~GeV (right panel)~\footnote{The minus signs are defined relative to the sign of the standard $h t {\bar t}$ coupling. This sign choice corresponds to the sign of the heavy $H t {\bar t}$ coupling in the 2HDM, but has no effect on these plots. However, it will play a role when vector-like quarks are introduced, as we discuss later.}.  We display separately the interference in the imaginary part of the production amplitude (solid red line) and the interference in the real part (dashed red line), as well as the line-shape without interference (solid blue line) and with interference (solid green line).

The interference in the real part changes sign across the nominal $H$ mass, whereas the interference in the imaginary part (which is due to the top quark loop in $gg \to H$ production) is larger in magnitude and always negative. For this reason, the combined interference effect is {\it negative} and overwhelms the putative peak, resulting finally in a {\it dip} in the $m_{t {\bar t}}$ distribution.  In both the $\Gamma_ H = 30$~GeV and 1~GeV cases, the depths of the dips almost reach the ATLAS 2-$\sigma$ lower limit. However, when integrated over the ATLAS $[720, 800]$~GeV bin the net effect would be $< 1 \sigma$, even if $\Gamma_ H = 30$~GeV (left panel). We note that the dip is not symmetric about the nominal mass of 750~GeV, and greater sensitivity to interference effects could be obtained by comparing off-centre bins $[750 - X, 750]$~GeV and $[750, 750 + X]$~GeV, where the choice of $X$ depends on the attainable mass resolution. However, the dip structure in the $\Gamma_ H = 1$~GeV  case (right panel) is unlikely to be unobservable because of the resolution in $M_{t {\bar t}}$.

\begin{figure}[!h]
\centerline{ 
\includegraphics[scale=0.4]{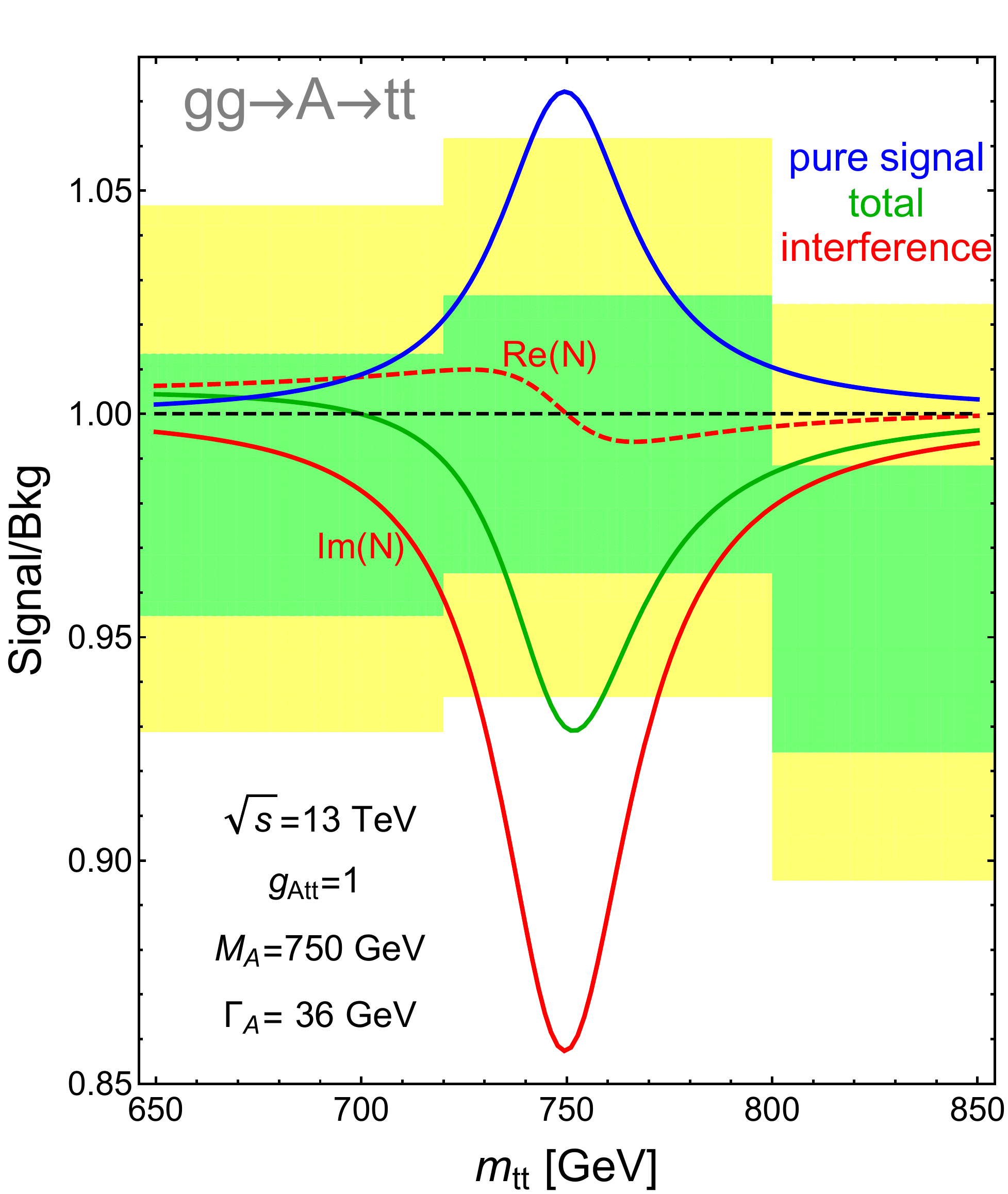} 
\includegraphics[scale=0.4]{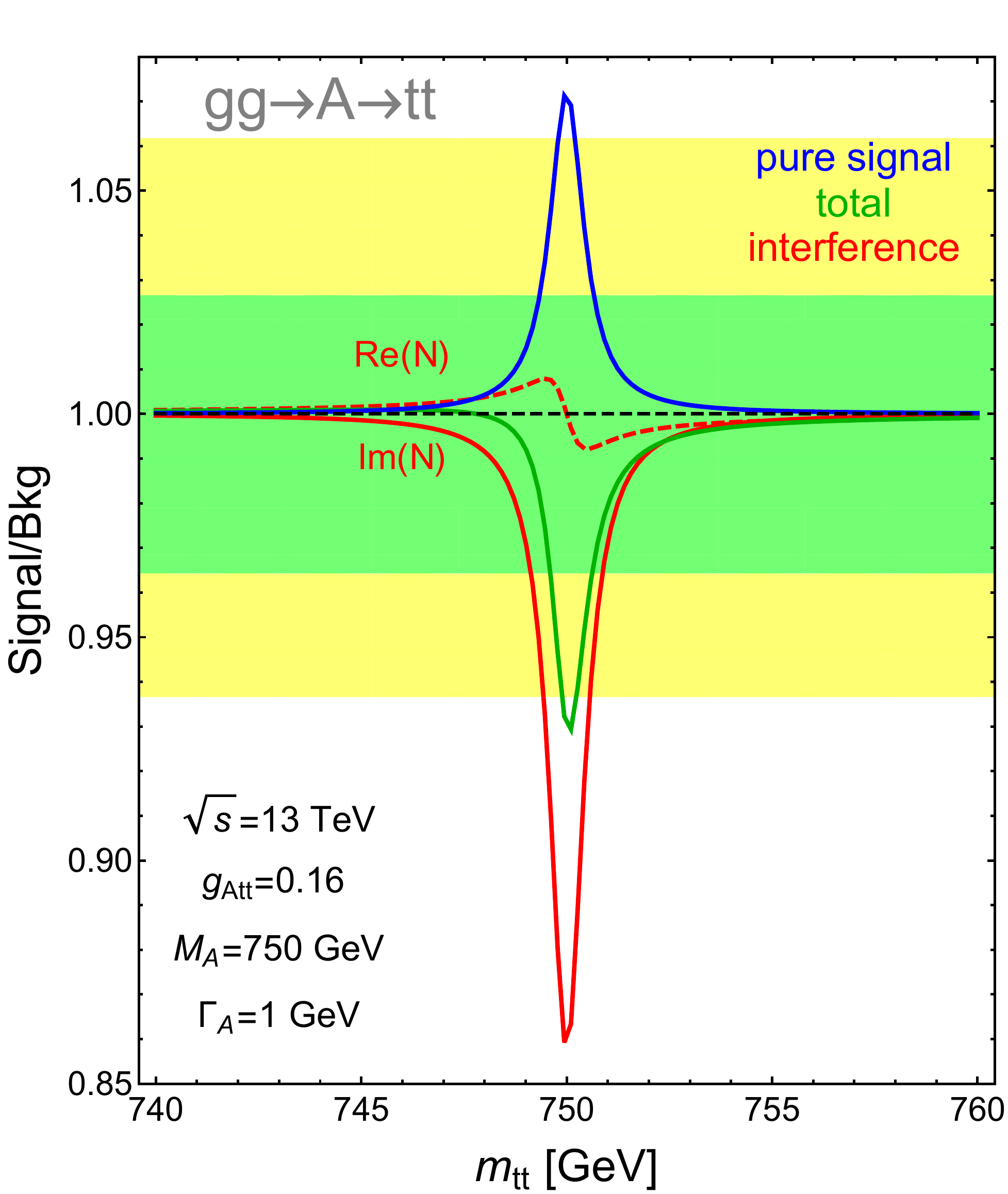}} 
\vspace*{-.1cm}
\caption{\it The contributions to the line-shapes of a CP--odd $A \to t {\bar t}$ with mass 750~GeV and total width $\Gamma_A = 36$~GeV (left panel) and $\Gamma_A = 1$~GeV (right panel), as functions of $m_{t {\bar t}}$, showing the line-shape neglecting interference (solid blue lines), the contributions of interferences alone (red dashed lines) and the overall combination (green lines).}
\label{fig:Attbar}
\vspace*{-3mm}
\end{figure}

Fig.~\ref{fig:Attbar} shows analogous results for a singlet pseudoscalar $A$ with unit-normalized couplings (left panel) and $g_{A t {\bar t}} = 0.16$ for which $\Gamma_A 
\approx \Gamma (A \to t {\bar t}) = 1$~GeV (right panel). We see that the interference is again {\it negative} and overwhelms the putative peak, resulting again in a {\it dip} in the $M_{t {\bar t}}$ distribution, whose depth exceeds the ATLAS 2-$\sigma$ lower limit in this case. However, when integrated over the ATLAS  $[720, 800]$~GeV bin the net effect would again be $< 1 \sigma$, and the sensitivity to interference effects would not be increased greatly by comparing off-centre $[750 - X, 750]$~GeV and $[750, 750 + X]$~GeV bins. As before, the peak and dip effects are very dramatic, but likely unobservable for $\Gamma_A  = 1$~GeV. 

\begin{figure}[!h]
\centerline{ 
\includegraphics[scale=0.4]{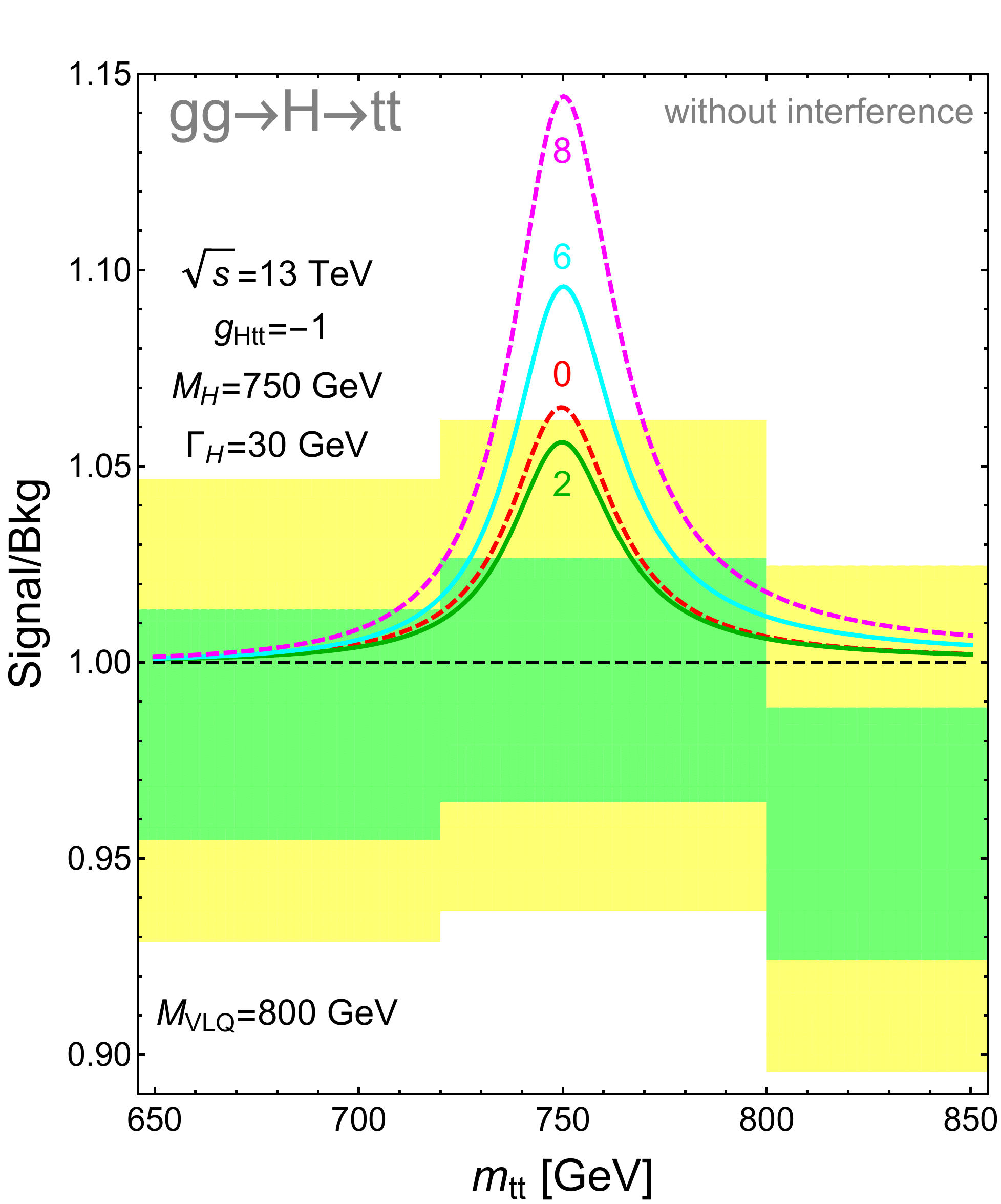} 
\includegraphics[scale=0.4]{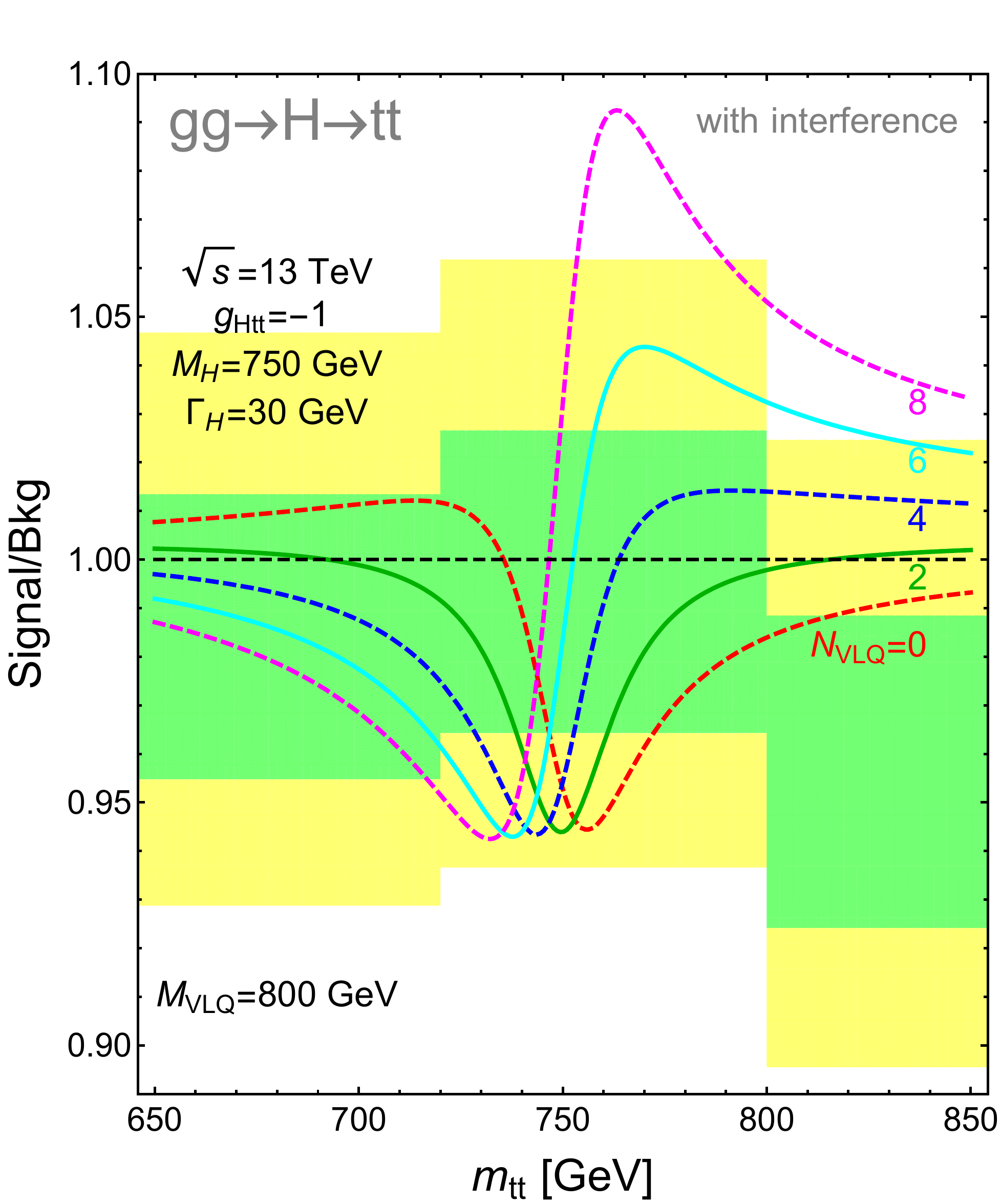}} 
\centerline{
\includegraphics[scale=0.4]{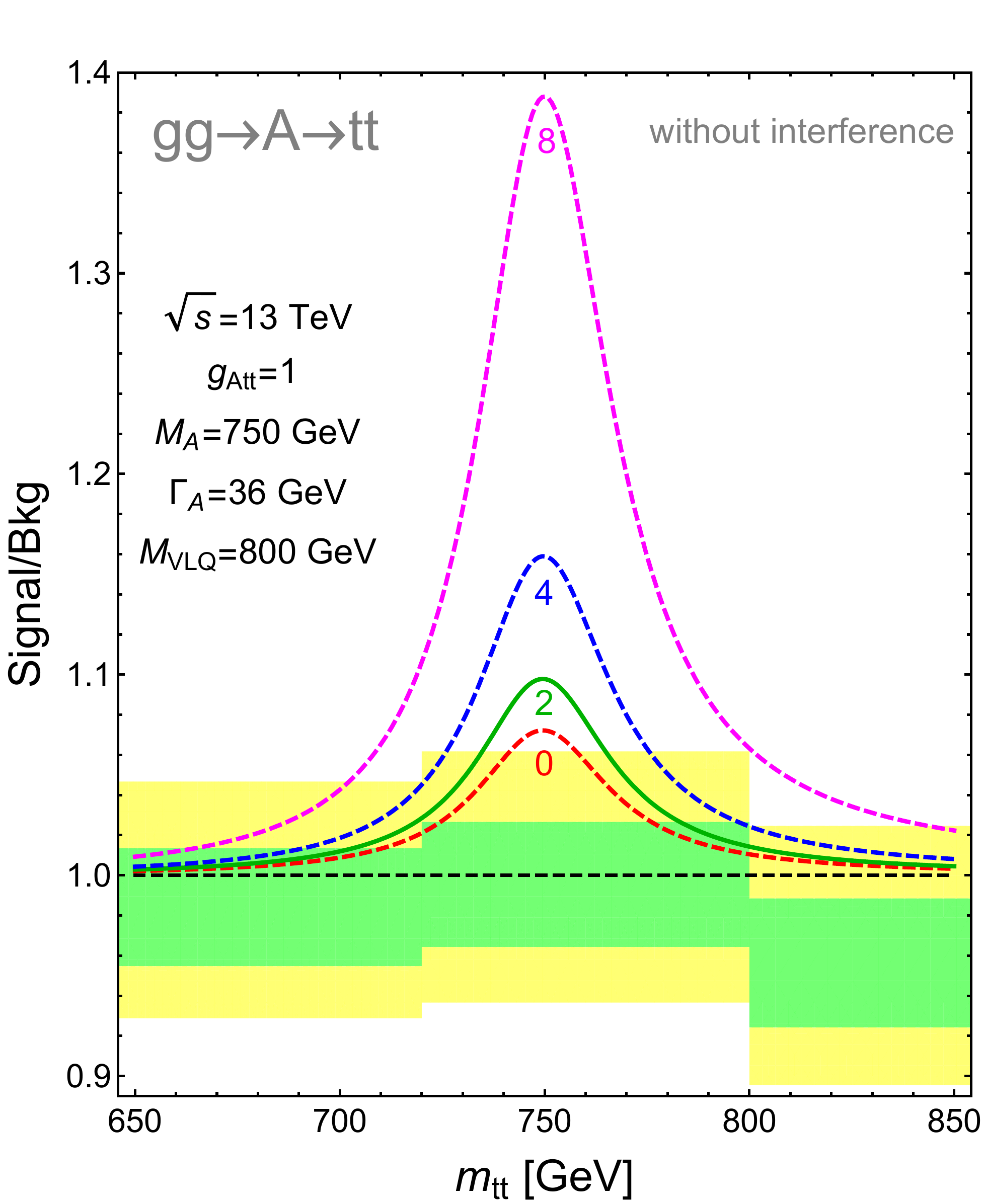} 
\includegraphics[scale=0.4]{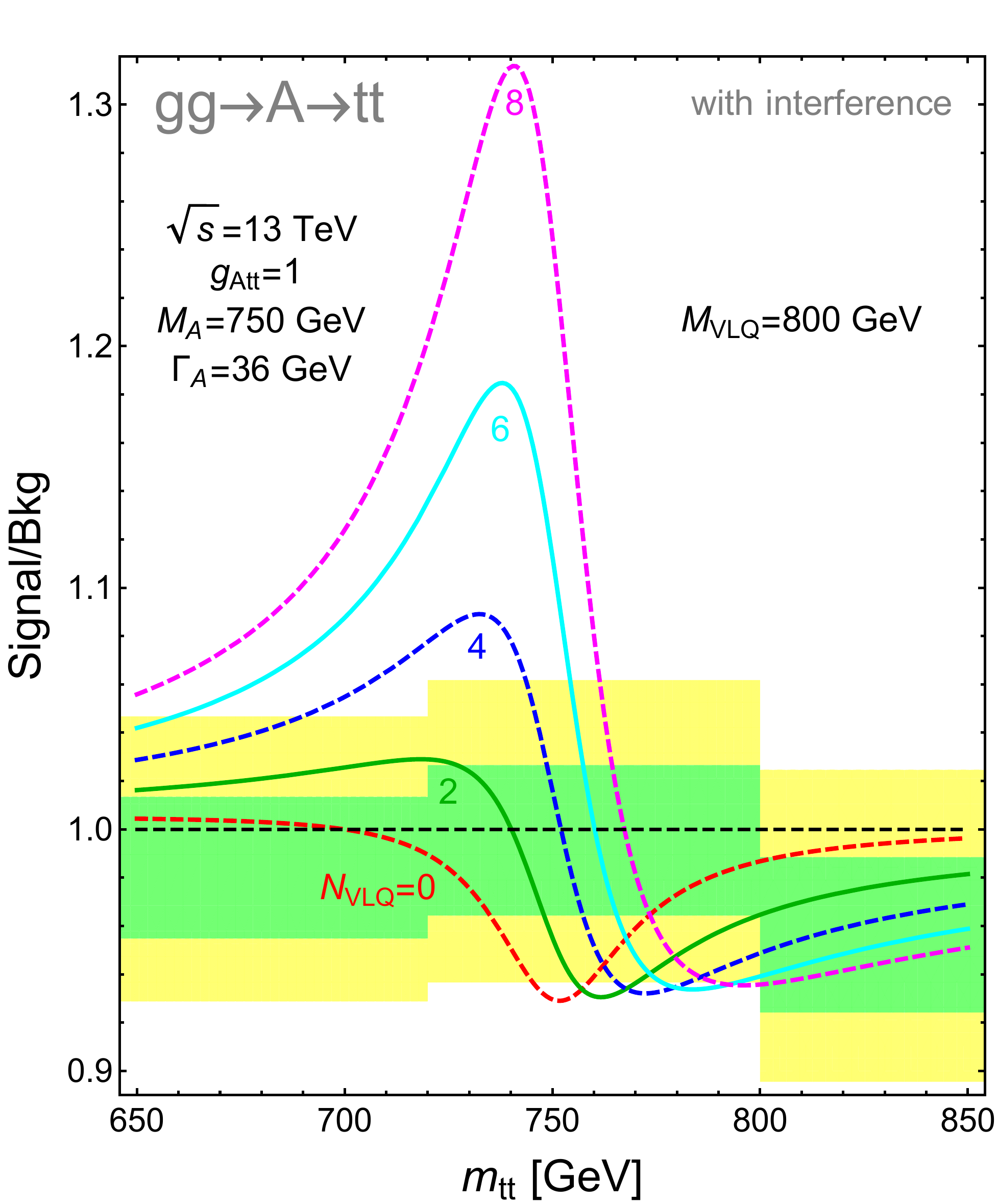} 
}
\caption{\it The contributions to the line-shapes of a CP--even $H \to t {\bar t}$ with mass 750~GeV and total width 30~GeV (upper panels) and a CP--odd $A \to t {\bar t}$ with mass 750~GeV and total width 36~GeV (lower panels), showing the effects of varying numbers of vector-like quarks with masses 800~GeV. The left panels neglect interference, which is included in the right panels.} 
\label{fig:ttbarmore}
\end{figure}

The upper panels of Fig.~\ref{fig:ttbarmore} show the effects of including different numbers of vector-like quarks $Q$ in the loops responsible for $gg \to H$, assuming $\Gamma_H = 30$~GeV and common $Q$ masses of 800~GeV and universal positive, unit-normalized $HQ {\bar Q}$ couplings~\footnote{This is the same sign as the conventional
$h t {\bar t}$ coupling, but opposite to that of the $H t {\bar t}$ coupling in the 2HDM with $\tan \beta = 1$. The interference effects would be larger if the $HQ {\bar Q}$ couplings were negative.}. In the absence of interference (upper left panel)  we see that adding 6 or 8 such heavy vector-like quarks takes the peak outside the 2-$\sigma$ ATLAS range. However, the upper right panel reveals a different picture when interference effects are included. There are {\it dips} for $N = 0, 2, 4, 6, 8$ vector-like quarks, but there are also significant peaks for $N = 6, 8$, in particular. The net result of integrating over the ATLAS $[720, 800]$~GeV bin would lie within the 2-$\sigma$ range. However we see again the potential gain in sensitivity to the antisymmetric interference effect that could be obtained by using off-centre $[750 - X, 750]$~GeV and $[750, 750 + X]$~GeV bins. The lower panels of Fig.~\ref{fig:ttbarmore} show the effects of including 
different numbers of vector-like quarks $Q$ in the loops responsible for $gg \to A$, assuming $\Gamma_ A = 30$~GeV, common $Q$ masses of 800~GeV and universal positive, unit-normalized $AQ {\bar Q}$ couplings.  In this case we see effects that are qualitatively similar to those in the scalar case, but quantitatively more important. It seems likely that a detailed numerical analysis in this case using the present ATLAS binning could exclude $N \ge 6$, but using off-centred bins would again be more sensitive to the interference effects.

\begin{figure}[!h]
\centerline{ 
\includegraphics[scale=0.4]{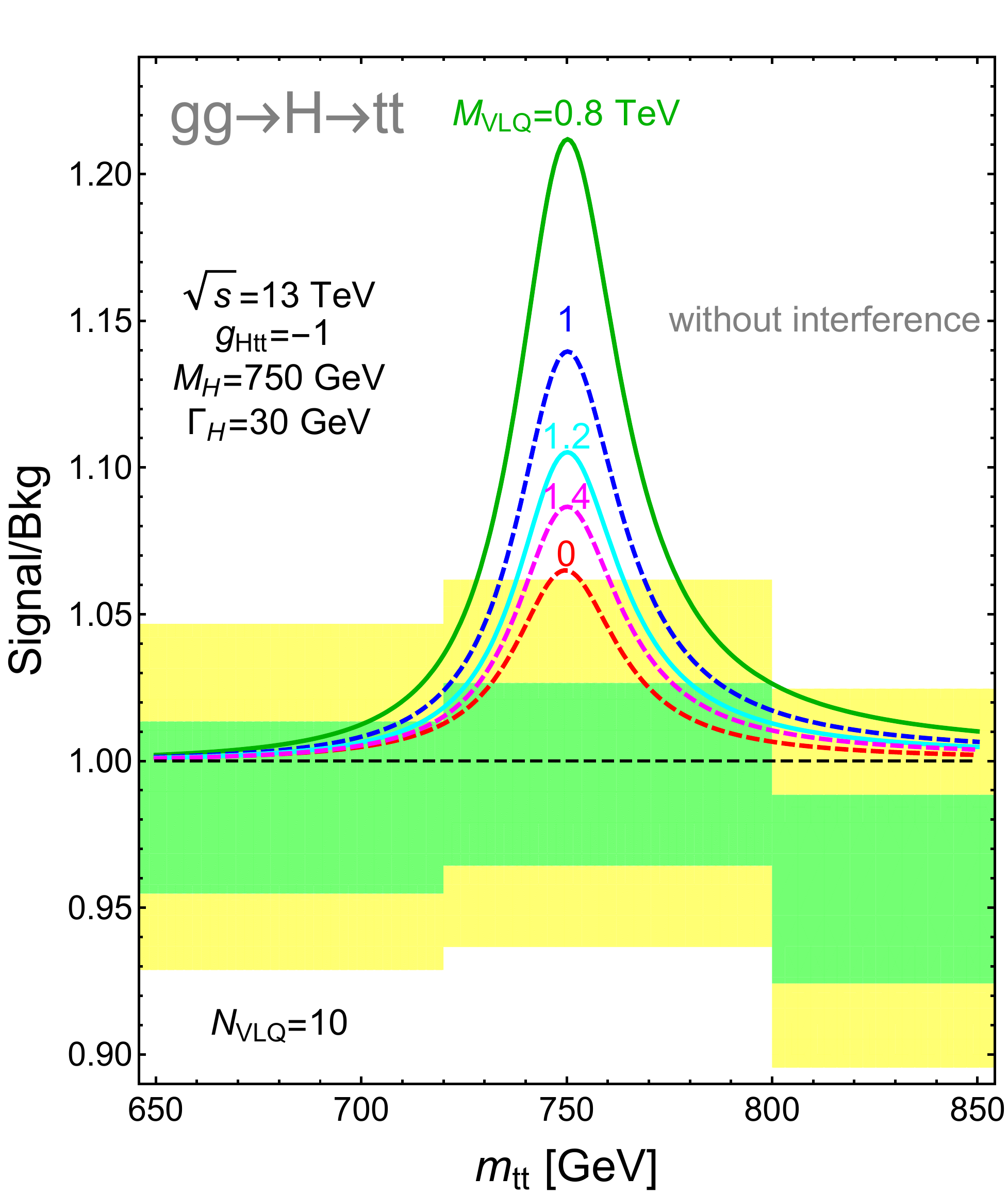} 
\includegraphics[scale=0.4]{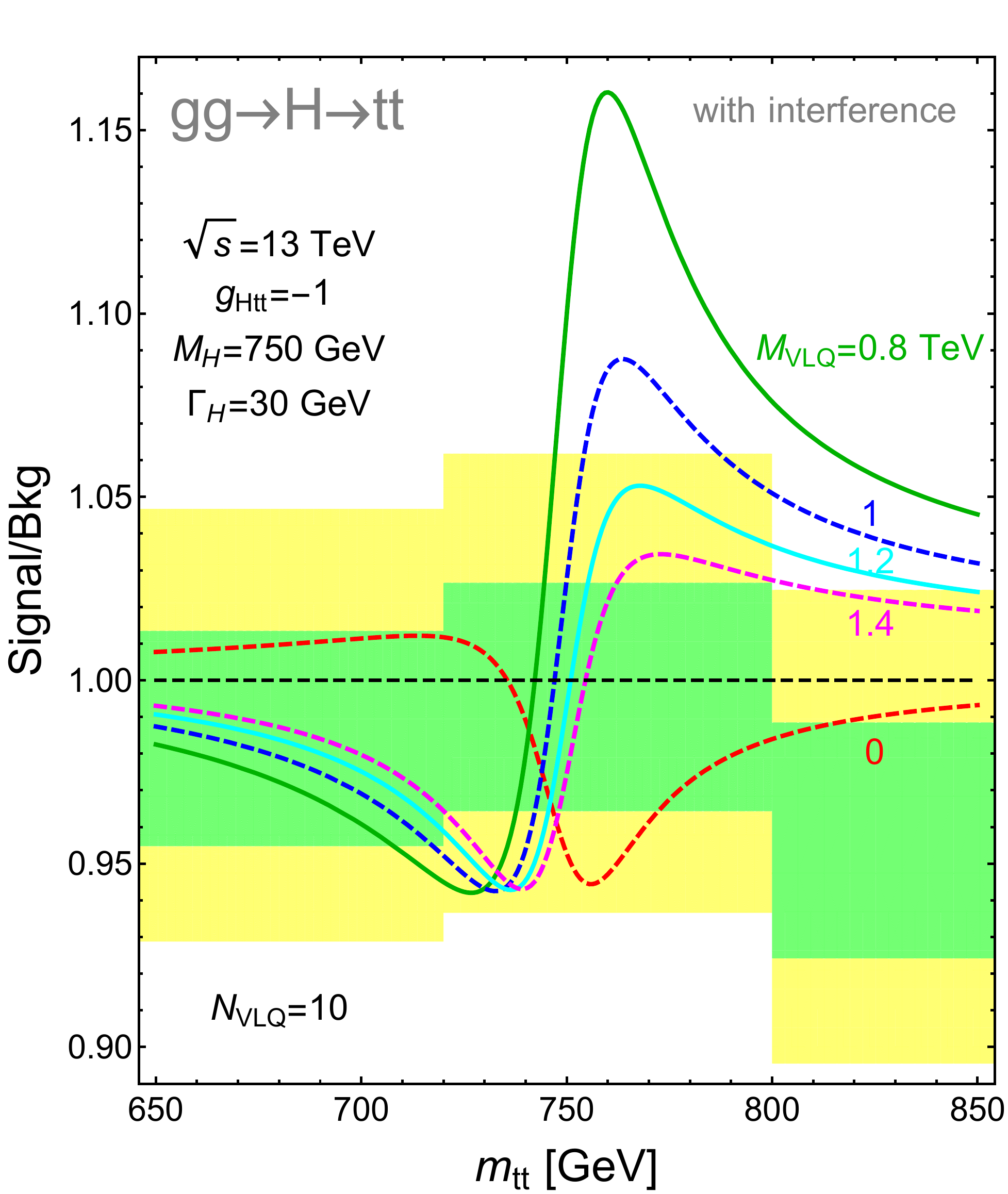}} 
\centerline{
\includegraphics[scale=0.4]{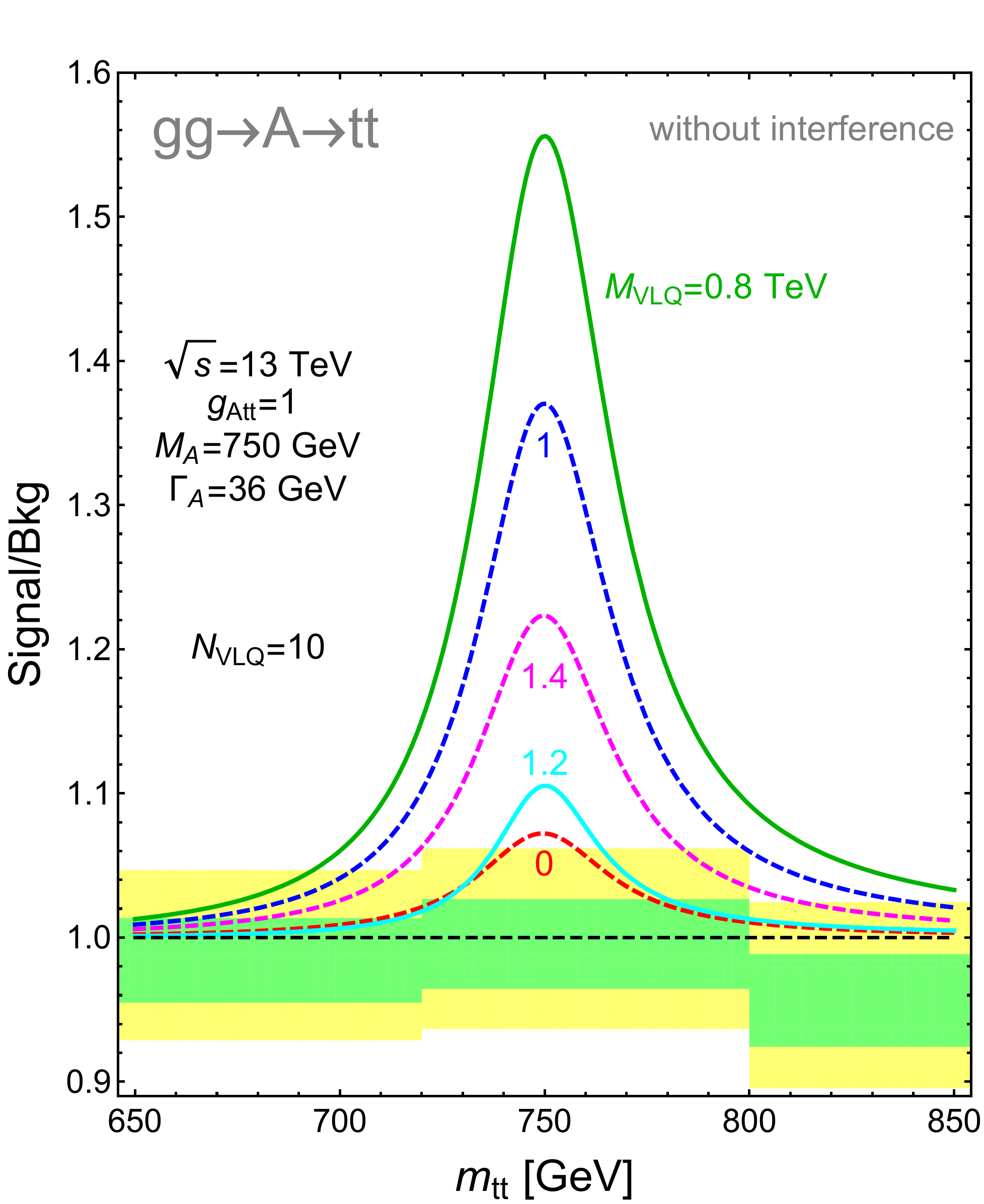} 
\includegraphics[scale=0.4]{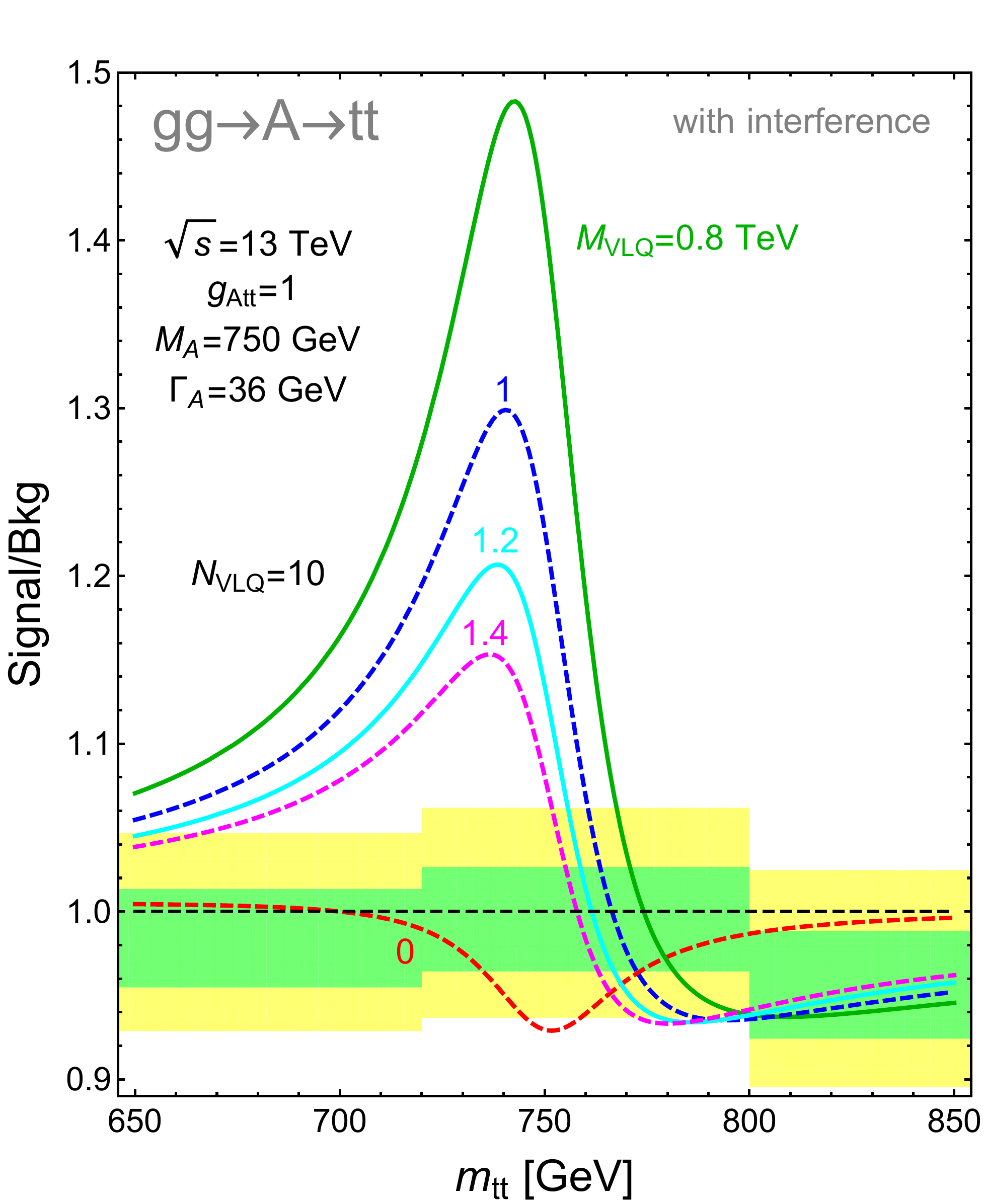} 
}
\caption{\it The line-shapes of a CP--even $H \to t {\bar t}$ with mass 750~GeV and total width 30~GeV (upper panels) and a CP--odd $A \to t {\bar t}$ with mass 750~GeV and total width 36~GeV (lower panels), showing the effects of varying the common mass of the vector-like quarks, assumed here to be 10 in number. The left panels neglect interference, which is included in the right panels.} 
\label{fig:ttbarmass}
\end{figure}

Fig.~\ref{fig:ttbarmass} show analogous results showing the effects of varying the masses of the vector-like quarks $Q$ in the loops responsible for
$gg \to H$ (upper panels) and $gg \to A$ (lower panels), assuming $N = 10$, common masses of 800~GeV, 1~TeV, 1.2~TeV and 1.4~TeV and  universal positive, unit-normalized $HQ {\bar Q}$ couplings, assuming in each case 10 vector-like quarks~\footnote{As discussed above, the interference effects would be larger if the $HQ {\bar Q}$ couplings were negative.}. In the absence of interference (upper left panel) the peak of the $H$ signal would be outside the ATLAS 2-$\sigma$ range for all the displayed values of $M_Q$,
though the value integrated over the ATLAS $[720, 800]$~GeV bin might be allowed for $M_Q > 1$~TeV. However, including interference (upper right panel)
changes drastically the $H \to t {\bar t}$ line-shape. As before, the peak is shifted, there is always a dip, and the integral over the ATLAS $[720, 800]$~GeV bin
is certainly within the allowed range for $M_Q \lesssim 1$~TeV. In the $A$ case (lower panels), both the enhancement in the absence of interference and the effects of interference are greater than in the scalar case, because of the relative (+) sign between the $A t \bar t$ and $A Q \bar Q$ couplings. These plots emphasize once more the increase in sensitivity that could be obtained using off-centre bins.

\begin{figure}[!h]
\centerline{ 
\includegraphics[scale=0.4]{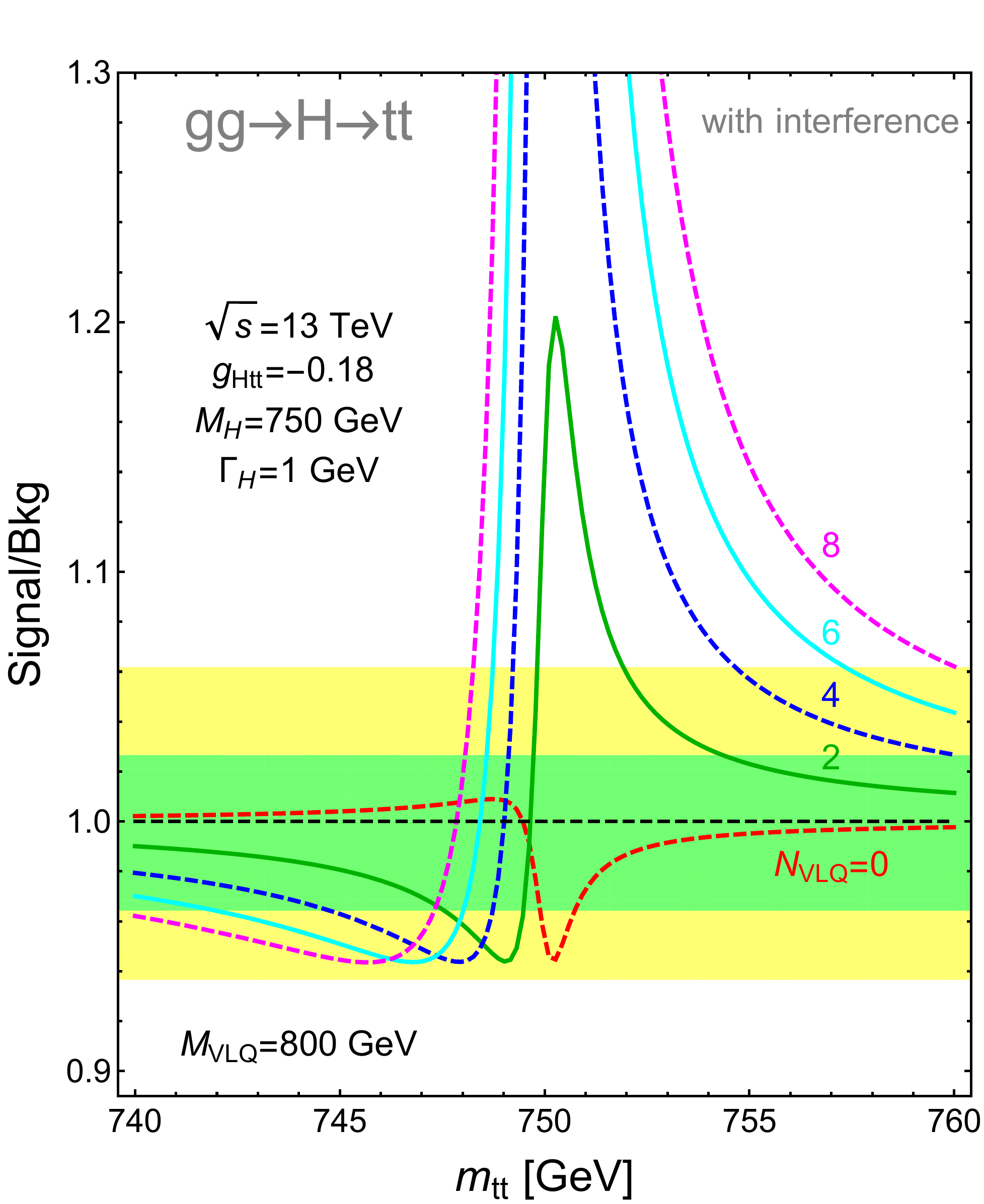} 
\includegraphics[scale=0.4]{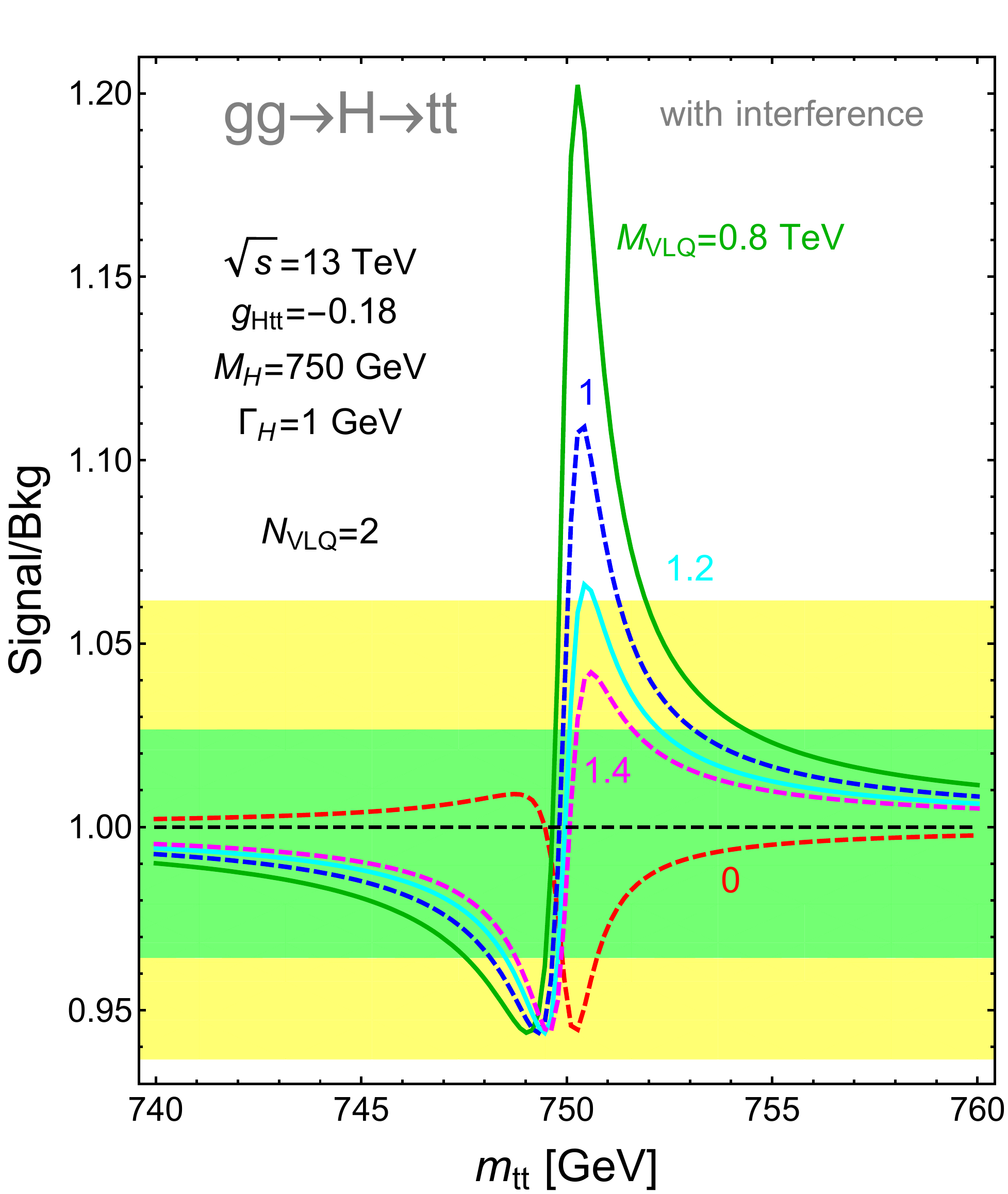}} 
\caption{\it The line-shapes of a CP--even $H \to t {\bar t}$ with mass 750~GeV 
and total width 1~GeV, showing the effects of varying the number of vector-like quarks (left panel) and the common mass of the vector-like quarks, assumed here to be 2 in number, including interference in both cases.} 
\label{fig:ttbarmass1}
\vspace*{-2mm}
\end{figure}

Fig.~\ref{fig:ttbarmass1} shows analogous results for a narrow scalar state with $\Gamma_H = 1$~GeV: varying the number of vector-like quarks with an assumed common mass of 800~GeV (left panel) and varying the common mass assuming just 2 vector-like quarks (right panel). We see that, as in the large-width case shown in Fig.~\ref{fig:ttbarmass}, there are dramatic changes in the interference structure and line-shape that depend
sensitively on the properties and number of vector-like quarks. However, these effects are probably unobservable because of the $t {\bar t}$ mass resolution. For this reason, we do not show the analogous results for a narrow pseudoscalar state, which are very similar.

\begin{figure}[!h]
\centerline{ 
\includegraphics[scale=0.4]{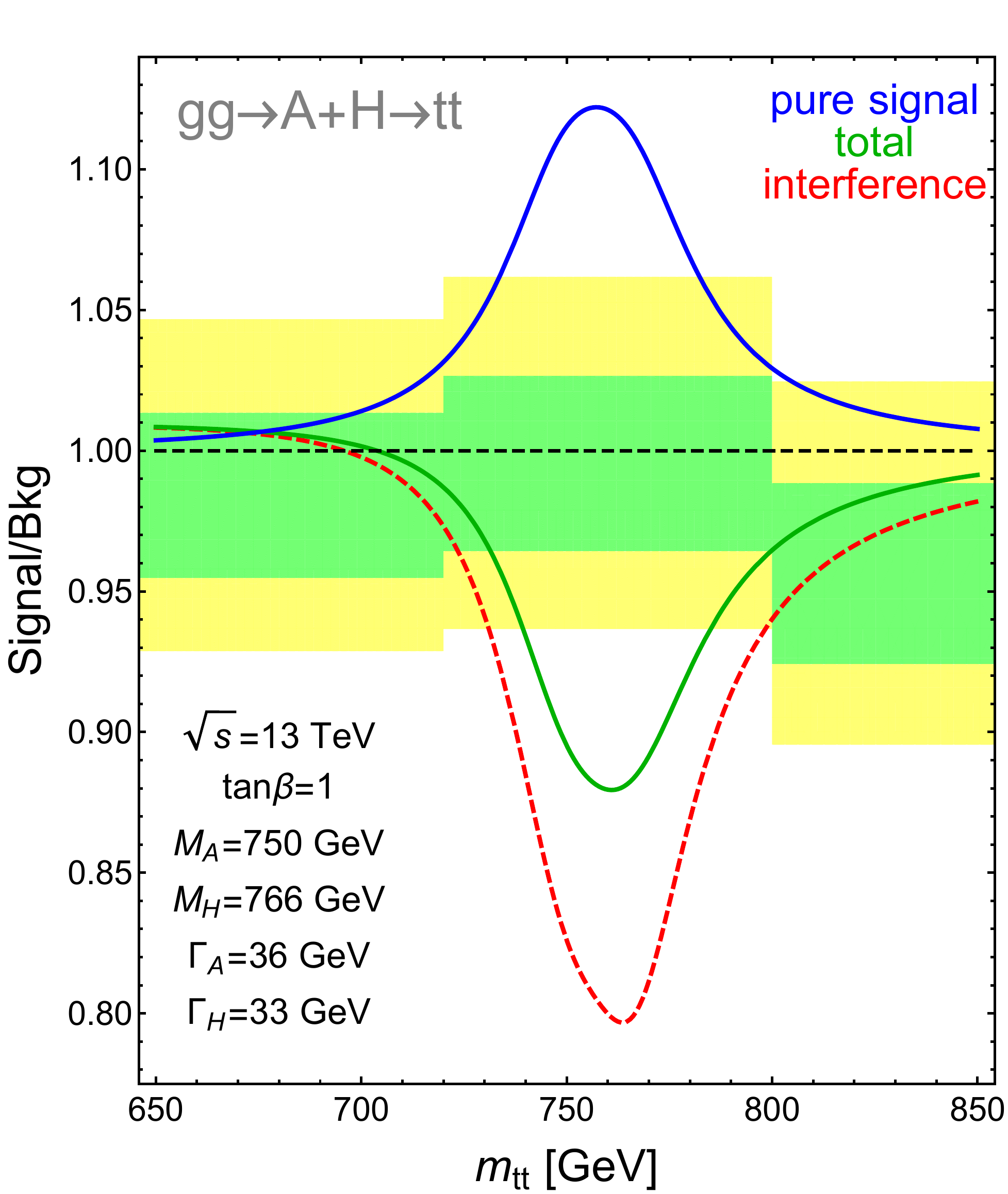} 
}
\caption{\it The contributions to the combined $t \bar t$ line-shape of a CP--odd $A$ with mass 750~GeV and 36~GeV width and a CP--even $H$ with mass 766~GeV and 33~GeV width, with 2HDM  couplings,  neglecting interference (solid blue line), the contribution of interference (dashed red line) and the combination of the two (solid green line).} 
\label{fig:ttbarAH}
\end{figure}

We display in Fig.~\ref{fig:ttbarAH} the combined effects in the 2HDM with nominal masses of 750~GeV for the pseudoscalar $A$ and 766~GeV for the scalar $H$ and corresponding decay widths $\Gamma_A \approx \Gamma (A \to t {\bar t}) = 36$~GeV and $\Gamma_H \approx \Gamma (H \to t {\bar t}) = 33$~GeV. As previously, the solid blue line is the result that would be obtained neglecting interference, the dashed red line is the contribution of the
interference term, and the solid green curve is the combination. We assume here that only the top quark contributes to the $gg \to H, A$ production amplitudes. We see that the interference in this case causes a {\it dip} that is presumably excluded by the ATLAS 8-TeV
data at the 2-$\sigma$ level.

\begin{figure}[!t]
\centerline{ 
\includegraphics[scale=0.35]{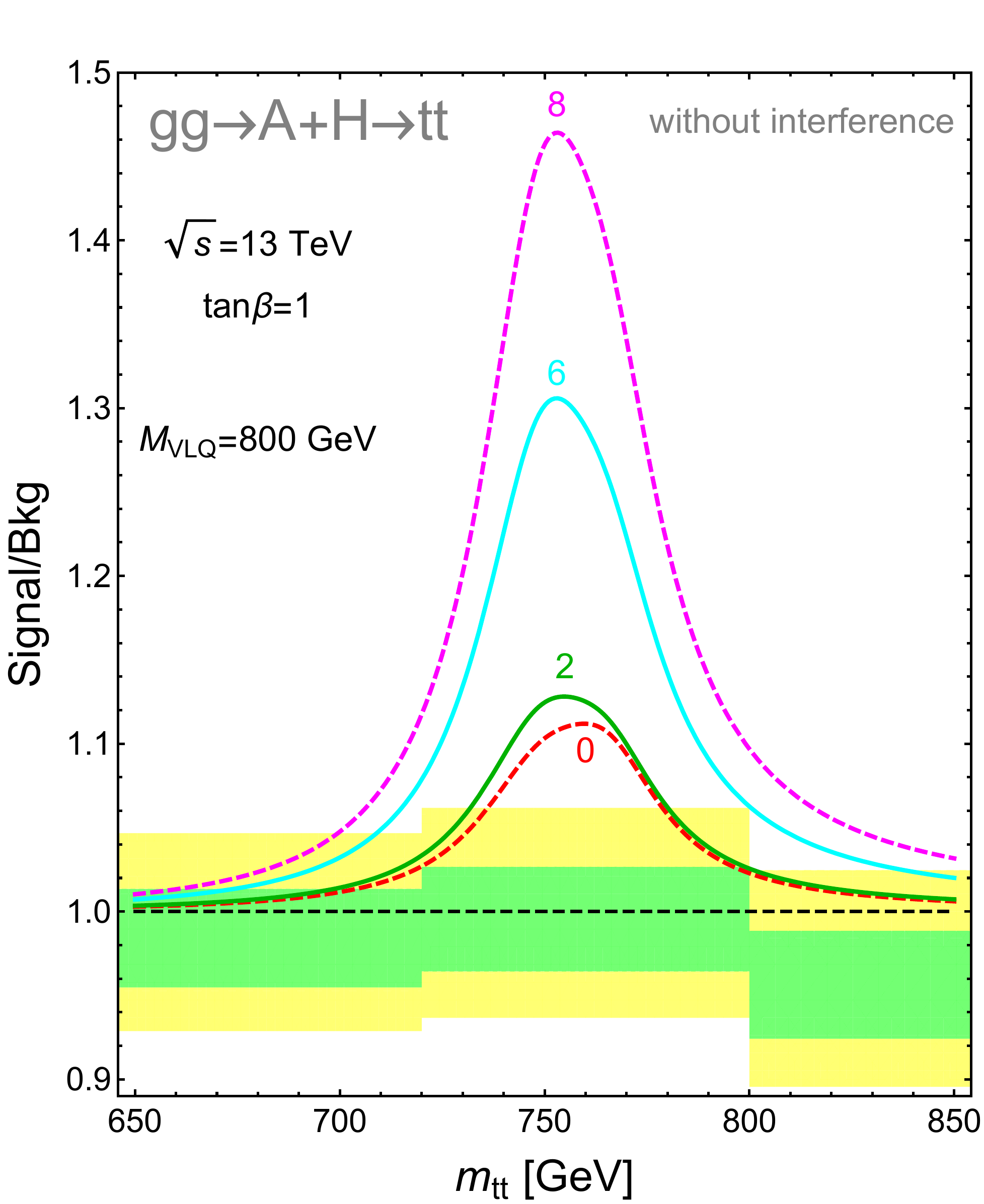}~~~ 
\includegraphics[scale=0.35]{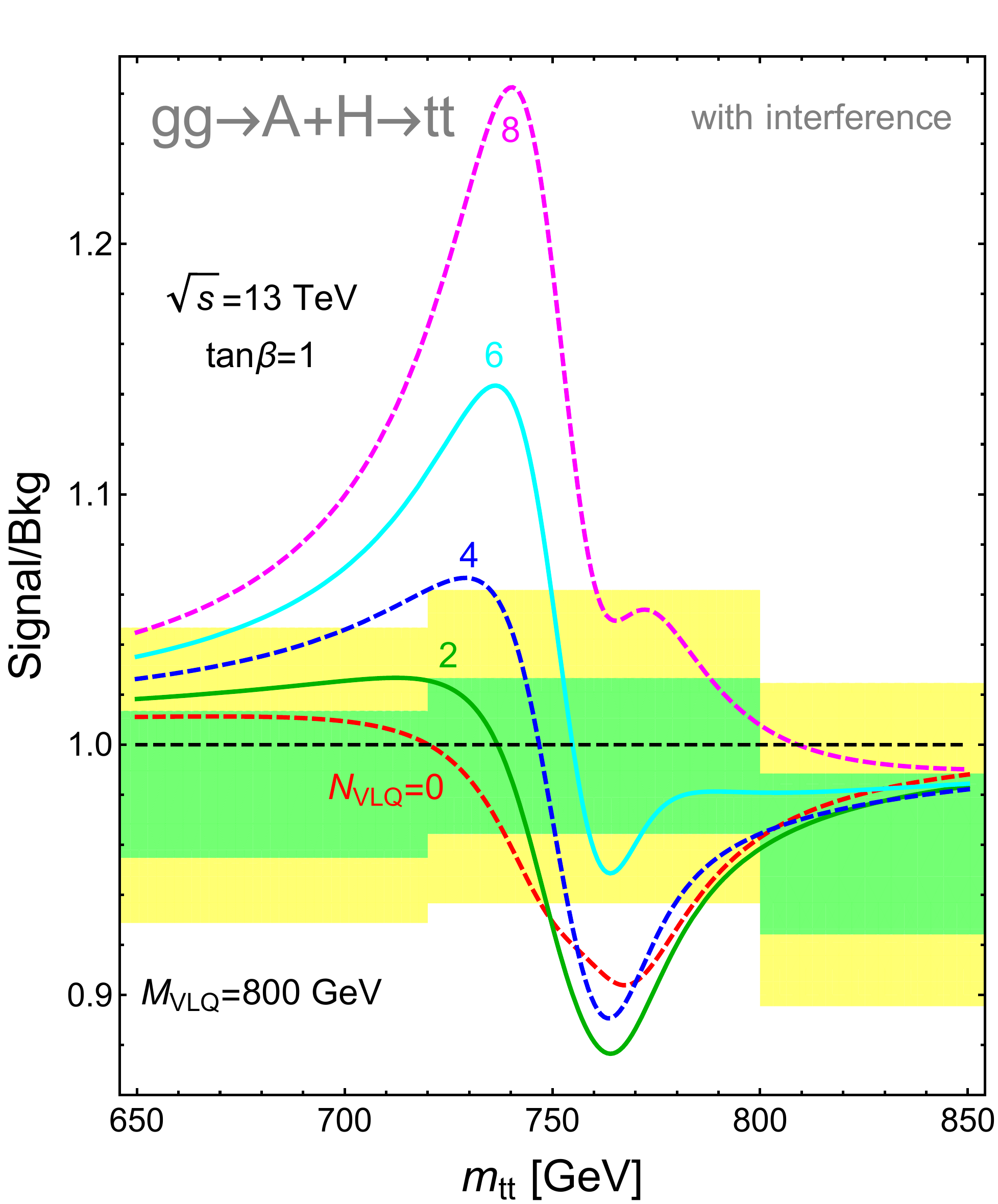}} 
\centerline{
\includegraphics[scale=0.35]{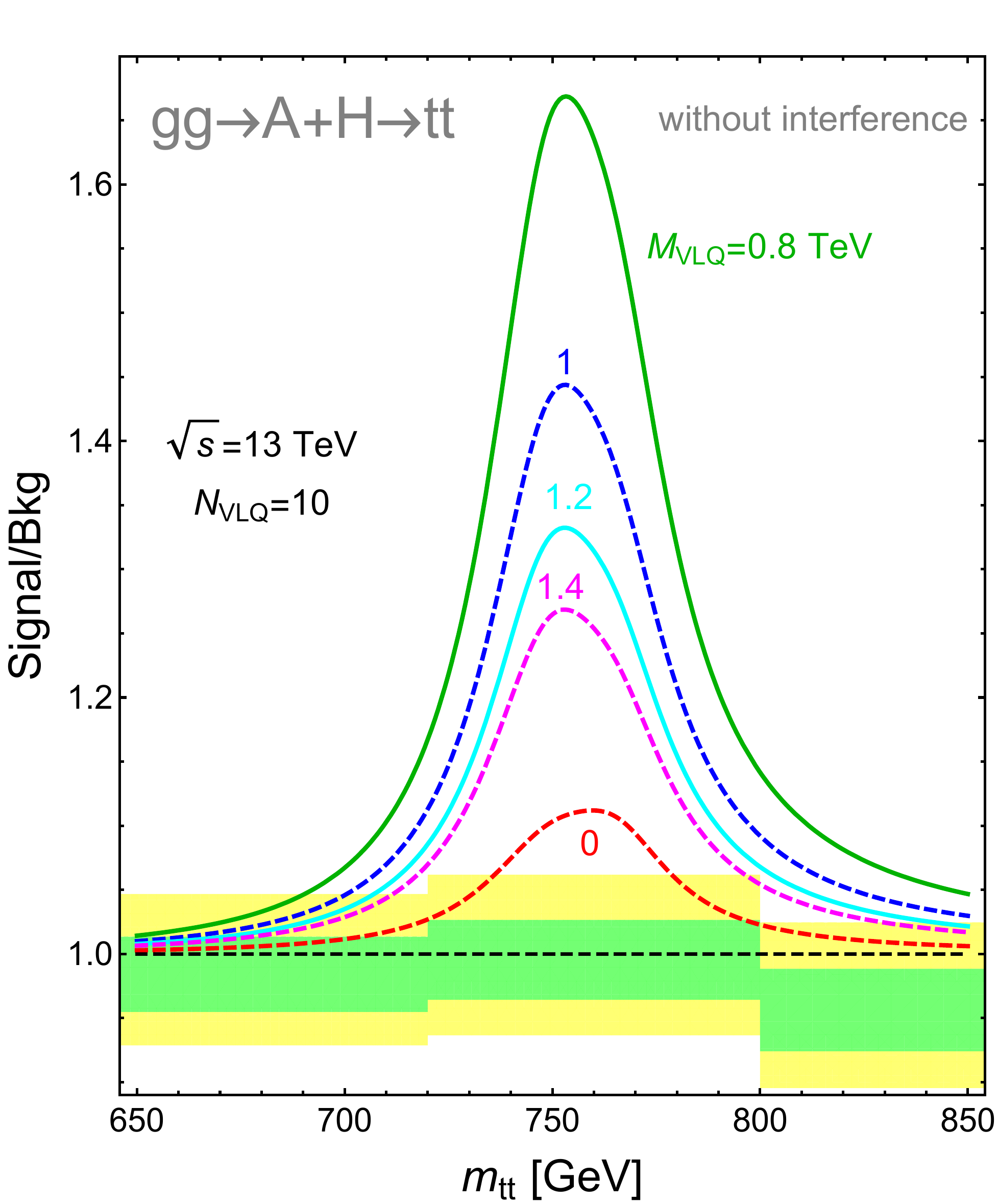}~~~ 
\includegraphics[scale=0.35]{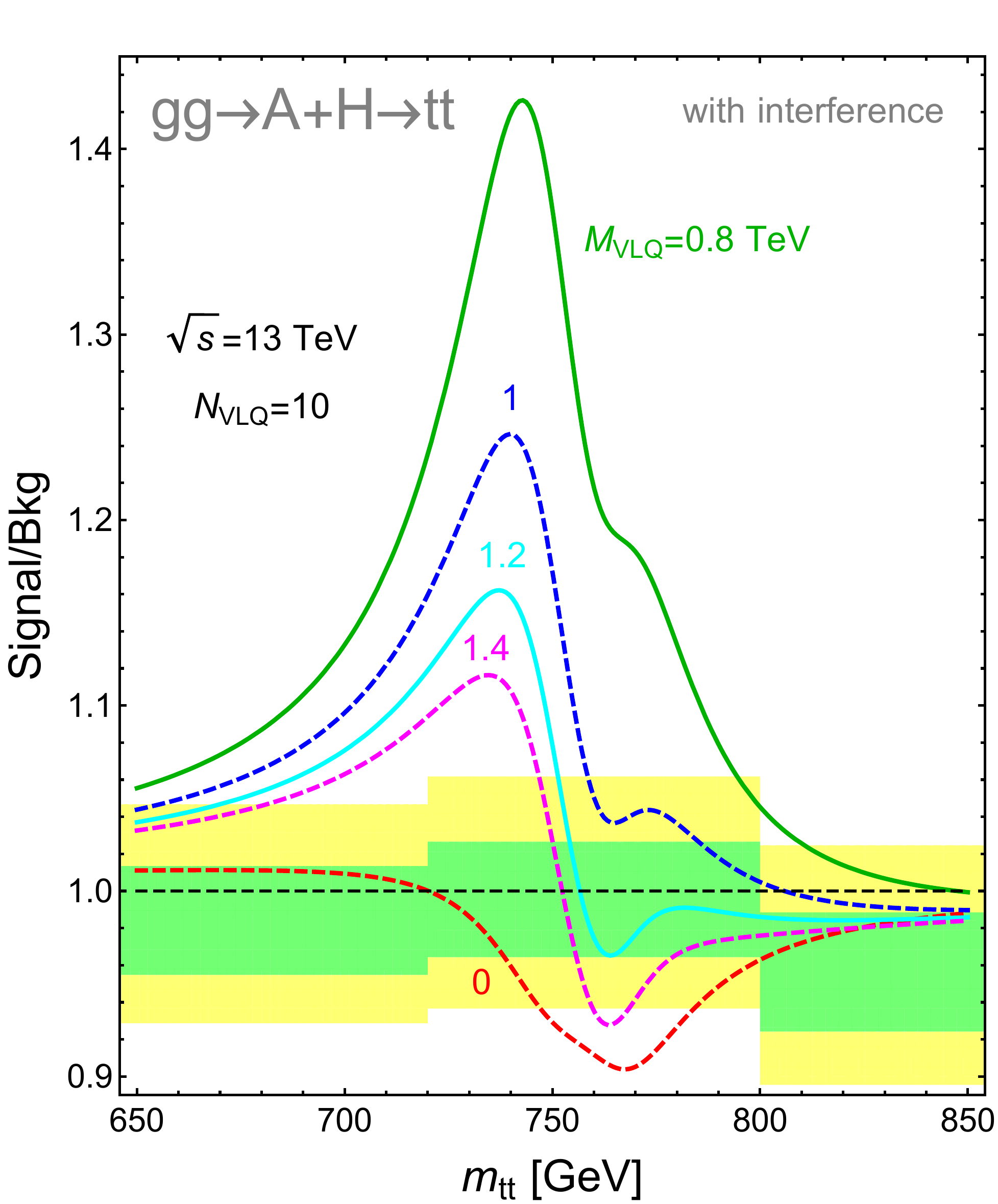} 
}
\caption{\it The combined $t {\bar t}$ line-shape of a CP--odd $A$ with mass 750~GeV and total width 36~GeV and a CP--even $H$ with  mass 766~GeV and total width 33~GeV, with the couplings predicted in the 2HDM, showing the effects of varying the number of vector-like quarks weighing 750~GeV (upper panels) and the common mass of the vector-like quarks, assumed here to be 10 in number (lower panels). The left panels neglect interference, which is included in the right panels..} 
\label{fig:ttbarAHNM}
\vspace*{-3mm}
\end{figure}

Fig.~\ref{fig:ttbarAHNM} shows the effects of including varying numbers $N_{VLQ}$ of heavy vector-like quarks (upper panels) with masses 800~GeV, and varying their masses, assuming $N_{VLQ} = 10$ (lower panels). The former neglect interference effects, which are included in the latter. If one neglected interference, one would conclude from the upper left panel that any number of 800-GeV vector-like quarks is excluded by the absence of a peak. However, we see in the upper right panel that the situation is more nuanced: while the case with no vector-like quarks is presumably excluded at the 2-$\sigma$
level by the absence of a dip, as discussed in connection with Fig.~\ref{fig:ttbarAH}, and presumably also the case with 2 vector-like quarks, the case with 4 vector-like quarks may be compatible with the data because of a change in sign across the $[720, 800]$~GeV bin used by ATLAS. On the other hand, the cases with 6 and 8 quarks
are presumably excluded by the absence of a peak. As before, we note that judicious off-centre binning would increase the sensitivity to interference.

Varying the masses of 10 vector-like quarks, as in the lower panels of Fig.~\ref{fig:ttbarAHNM}, we see that all the masses studied would be excluded if interference were neglected, whereas masses $\ge 1.4$~TeV are probably compatible with the ATLAS data when interference effects are taken into account, again because of the change in sign across the $[720, 800]$~GeV bin used by ATLAS.

\section{Conclusion}

In the context of the indications for a 750 GeV state(s) observed in the early 13 TeV LHC data and that we assume to be due to a new scalar and/or pseudoscalar particle, we have studied in this paper the effects of interferences between the signal and the QCD background in the process $gg \to (\Phi \to) \gamma \gamma$, refining previous calculations~\cite{KIAS}, and in the process $ gg \to (\Phi \to) t {\bar t}$, presenting original results. The interference effects are quite complex (pun intended), and their measurement would provide information on both the real and imaginary parts of the $gg \to \Phi$ amplitude in both processes and, in the first case, also the $\Phi \to \gamma \gamma$ amplitude. We have used two benchmark scenarios to study these effects in the scalar (CP-even) $H$ and pseudoscalar (CP-odd) $A$ cases: a singlet state whose total width may be either 1 or $\approx 30$~GeV, and a 2HDM model in which there are adjacent scalar and pseudoscalar states with total widths of $\approx 30$~GeV, with nominal masses of about $\approx 750$ GeV and eventually differing by about 16~GeV.

The following are some key general features of our analysis.\vspace*{-2mm} 

\begin{itemize}

\item[$i)$] In general, interference effects may change significantly the $gg \to \Phi \to \gamma \gamma $ signal cross section but only if the signal rate is much smaller than the background rate. In this case, peaks before the nominal resonance mass value and dips after this value can be observed and an enhancement of the total rate by a  factor up to about four  can be  obtained. This is particularly true if the resonance is narrow.\vspace*{-2mm}   

\item[$ii)$] In the context of the putative 750 state, the diphoton rate observed at the LHC 
is so large that interference effects are rather small, increasing the rate  by a few 10\% at most and not altering significantly the resonance shape.  This is mainly due to the fact that the new vector--like fermion contributions that are necessary to explain the observed diphoton rate should be real if new decay channels of the $\Phi$ states (which would increase the total width and suppress the $\gamma \gamma$ branching ratios) are not kinematically allowed.\vspace*{-2mm} 

\item[$iii)$] Similar effects are expected in the $\Phi \to Z\gamma$ process that we have 
briefly considered, and we expect that it will also be the case in the two remaining electroweak diboson channels of the $\Phi$ state, namely $\Phi \to ZZ$ and $\Phi \to WW$.\vspace*{-2mm} 

\item[$iv)$] In the $gg \to \Phi \to t{\bar t}$ case, interference effects have a much larger impact.   Negative interference may cause the total cross sections to exhibit a dip instead of a bump, invalidating limits on resonances based on putative bump signatures. This occurs, for instance, in the case where the production of the $\Phi$ states is initiated by the standard
quark (mainly top quark) loops only. The presence of additional vector--like quarks might change the situation though and peaks followed by dips might occur, possibly requiring judicious off-centre binning.\vspace*{-2mm} 

\item[$v)$] On the other hand LHC data probably have similar sensitivity to possible dips as we have illustrated with ATLAS 8-TeV data.  Since interference effects change sign across the nominal $\Phi$ resonance mass, the most sensitive way to search for such effects would be to use off-centre bins.\vspace*{-2mm} 

\end{itemize}

Our analysis has barely scratched the surface of possible interference effects. For example, as commented above, there would be analogous effects in the $Z \gamma$, $ZZ$ and $W^+ W^-$ final states that must be present at some level and we have explicitly discussed only the specific case of the $\Phi \to Z\gamma$ process in the approximation $M_Z^2/M_\Phi^2 \to 0$. While the situation should be qualitatively similar to the $ \gamma\gamma$ case,  numerical differences would arise depending on the quantum numbers of the heavy vector-like fermions circulating in the loops generating the various decays.  Even in the $\gamma\gamma$ and $t\bar t$ cases discussed here, we have not made a systematic exploration of all the effects that might affect the signals, backgrounds and  their interferences and, in particular,  we have not incorporated in a thorough way the higher--order QCD radiative corrections, nor considered the theoretical uncertainties and the systematic experimental errors. Nor we have included in a detailed way  all the ingredients that would be required to interpret the LHC diphoton signal, in particular, the constraints on models of vector-like quarks that could be inferred from present data. Some of these issues will be addressed in future work \cite{DEQ2}. 

In any case, comprehensive analyses of the experimental data may be premature in advance of confirmation that the $\Phi$ enhancement is due to one or more new particles. However, our analysis has relevance even if its existence is not confirmed, since many other searches for massive spin-zero particles are ongoing at the LHC and will continue in the future, there and at any future $pp$ colliders.

\subsection*{Acknowledgements:} 
JQ would like to acknowledge discussions with Roberto Barcel\`o, Anne-Laure Cunrath Pequegnot and Ritesh Singh. The work of AD is supported by the ERC Advanced Investigator Grant Higgs@LHC. The work of JE and JQ is supported partly by the STFC Grant ST/L000326/1.
AD and JE thank the CERN Theoretical Physics Department for its hospitality.

\end{document}